\documentclass[namedreferences]{solarphysics}

\usepackage[hyperref,optionalrh,showbiblabels]{spr-sola-addons} 
\usepackage{graphicx}        
\usepackage{color}           
\usepackage{breakurl}        
\usepackage{gensymb}
\usepackage{multirow}

\newcommand{\etal}{{\it et al.}}
\newcommand{\IE}{{\it i.e.}}
\newcommand{\EG}{{\it e.g.}}
\newcommand{\VERSUS}{{\it versus}}

\newcommand{\longscore}{CorPITA score}
\newcommand{\speedunit}{\,\mbox{km}\mbox{s}^{-1}}
\newcommand{\accunit}{\,\mbox{km}\mbox{s}^{-2}}
\newcommand{\Deg}{\degree}
\newcommand{\arcsec}{^{\prime\prime}}

\newcommand{\rtr}[1]{\textcolor{black}{#1}}

\newcommand{\acceleration}{a}
\newcommand{\ainit}{\acceleration}
\newcommand{\afit}{\acceleration_{\mbox{fit}}}
\newcommand{\atrue}{\acceleration_{\mbox{true}}}

\newcommand{\velocity}{v}
\newcommand{\vinit}{\velocity_{0}}
\newcommand{\vfit}{\velocity_{\mbox{fit}}}
\newcommand{\vtrue}{\velocity_{\mbox{true}}}
\newcommand{\awareversion}{\url{https://github.com/wafels/eitwave/releases/tag/v0.1}}
\newcommand{\RDP}{RDP}

\newcommand{\sdomission}{{\it Solar Dynamics Observatory}}
\newcommand{\aiainstrument}{{\it Atmospheric Imaging Assembly}}

\newcommand{\sohomission}{{\it Solar and Heliospheric Observatory}}
\newcommand{\eitinstrument}{{\it Extreme ultravioiet Imaging
    Telescope}}

\newcommand{\stereomission}{{\it Solar Terrestrial Relations
    Observatory}}
\newcommand{\euviinstrument}{{\it Extreme Ultraviolet Imager}}

\chardef\us=`\_

\begin{document}

\begin{article}
\begin{opening}

\title{AWARE: An Algorithm for The Automated Characterization
  of EUV Waves in The Solar Atmosphere}

\author[addressref={aff2}, corref, email={jack.ireland@nasa.gov}]{\inits{J.}\fnm{Jack}~\lnm{Ireland}}
\author[addressref={aff2,aff3}, email={a.r.inglis@nasa.gov}]{\inits{A.R.}\fnm{Andrew R.}~\lnm{Inglis}}
\author[addressref={aff2}, email={a.y.shih@nasa.gov}]{\inits{A.Y.}\fnm{Albert Y.}~\lnm{Shih}}
\author[addressref={aff2}, email={s.christe@nasa.gov}]{\inits{S.}\fnm{Steven}~\lnm{Christe}}
\author[addressref={aff4}, email={stuart.mumford@sheffield.ac.uk}]{\inits{S. J.}\fnm{Stuart}~\lnm{Mumford}}
\author[addressref={aff5}, email={hayesla@tcd.ie}]{\inits{L. A.}\fnm{Laura A.}~\lnm{Hayes}}
\author[addressref={aff2}, email={barbara.j.thompson@nasa.gov}]{\inits{B. J.}\fnm{Barbara J.}~\lnm{Thompson}}
\author[addressref={aff6}]{\inits{B. J.}\fnm{V. Keith}~\lnm{Hughitt}}

\address[id=aff2]{NASA Goddard Spaceflight Center, Greenbelt, MD 20771, USA.}
\address[id=aff3]{Catholic University of America.}
\address[id=aff4]{University of Sheffield, Sheffield, UK.}
\address[id=aff5]{Trinity College Dublin, Dublin, Ireland.}
\address[id=aff6]{Laboratory of Cancer Biology and Genetics, NCI, NIH, Bethesda, MD.}

\runningauthor{J. Ireland \etal}
\runningtitle{An algorithm for the automated characterization of EUV waves}
\begin{abstract}Extreme ultraviolet (EUV) waves are large-scale
  propagating disturbances observed in the solar corona, frequently
  associated with coronal mass ejections and flares.  They appear as
  faint, extended structures propagating from a source region across
  the structured solar corona. Since their discovery, over two hundred
  papers discussing their properties, causes, and physical nature have
  been published. However, despite this their fundamental properties
  and the physics of their interactions with other solar phenomena are
  still not understood. To further the understanding of EUV waves, we
  have constructed the Automated Wave Analysis and Reduction (AWARE)
  algorithm for the measurement of EUV waves. AWARE is implemented in
  two stages.  In the first stage, we use a new type of running
  difference image, the running difference persistence image, which
  enables the efficient isolation of propagating, brightening
  wavefronts as they propagate across the corona.  In the second
  stage, AWARE detects the presence of a wavefront, and measures the
  distance, velocity, and acceleration of that wavefront across the
  Sun.  The fit of propagation models to the wave progress isolated in
  the first stage is achieved using the Random Sample Consensus
  (RANSAC) algorithm.  AWARE is tested against simulations of EUV wave
  propagation, and is applied to measure EUV waves in observational
  data from the \aiainstrument\ (AIA).  We also comment on unavoidable systematic errors that bias the estimation of wavefront velocity and acceleration.  In addition, the full AWARE software suite comes with a package that creates simulations of waves propagating across the disk from arbitrary starting points.
\end{abstract}

%
%


\end{opening}

\section{Introduction}\label{sec:Intro}
Extreme ultraviolet (EUV) waves are large-scale propagating
disturbances observed in the solar corona. These waves were discovered
through observations made by \eitinstrument\ (EIT) onboard the
\sohomission\ (SOHO) \citep{1997SoPh..175..571M, 1998GeoRL..25.2465T,
  1999ApJ...517L.151T}, and were hence initally dubbed `EIT'
waves. Since those first observations, over two hundred papers
discussing their properties, causes and physics have been
published. EUV waves appear to be strongly associated with coronal
mass ejection (CME) activity \citep{2002ApJ...569.1009B}, and to a
lesser extent with solar flares \citep{2006ApJ...641L.153}. However,
the initiation of EUV waves and their physical nature is not
completely understood.

Recent reviews by \citet{2011SSRv..158..365G},
\cite{2014SoPh..289.3233L}, \cite{2015LRSP...12....3W} and
\cite{2017SoPh..292....7L} present and interpret the EUV wave
literature.  In interpreting EUV wave observations, a variety of
explanations have been put forward. Some studies present evidence
supporting a magnetohydrodynamic (MHD) wave interpretation
\citep{1998GeoRL..25.2465T, 1999ApJ...517L.151T,2000ApJ...543L..89W,
  2001JGR...10625089W, 2002ApJ...574..440O, 2010ApJ...713.1008S},
while others argue for what \cite{2012SoPh..281..187P} call a
“pseudo-wave” due to either the evolving manifestations of a CME
\citep{1999SoPh..190..107D, 2000ApJ...545..512D, 2008SoPh..247..123D,
  2011ApJ...738..167S} or transient localized brightenings
\citep{2007AN....328..760A, 2007ApJ...656L.101A}.  Some authors have
found evidence indicating that the complex brightenings associated
with EUV waves can be due to a combination of both MHD waves and
pseudo-waves \citep{2002ApJ...572L..99C, 2005ApJ...622.1202C,
  2004A&A...427..705Z, 2009ApJ...705..587C}. A unified explanation of
this phenomenon is complicated by the broad range of observed wave
propagation speeds \citep{2009ApJS..183..225T, 2011A&A...532A.151W}
and amplitudes of EUV waves. \rtr{The path of EUV waves have been
  observed to be modified by nearly all major coronal features,
  including active regions \citep{2000ApJ...543L..89W}, filaments
  \citep{2012ApJ...753...52L}, coronal holes
  \citep{2009ApJ...691L.123G}, streamers
  \citep{2013ApJ...766...55K}. \cite{2014SoPh..289.3233L} and
  \cite{2015LRSP...12....3W} review observations of EUV wave-like
  behavior including reflection, transmission and refraction.}

EUV waves are also clearly correlated with other dynamic phenomena in addition to CMEs and flares. For example, investigations such as \cite{2000SoPh..193..161T}, \cite{2004A&A...427..705Z}, and \cite{2010ApJ...709..369P} have indicated that the development of coronal dimmings may be closely linked to the development of EUV waves. Some authors have also explored the potential of EUV waves in
diagnosing properties of the coronal medium that are otherwise hard to
measure, \IE, their use as tools to perform coronal seismology
\citep{1970PASJ...22..341U}. For example, if a fast MHD wave mode
interpretation is assumed, then the wave
propagation speed, coronal density and temperature can all be
estimated from observations, allowing the coronal magnetic field
strength to be derived \citep{2005LRSP....2....3N}.  This value can
also be used to test the accuracy of magnetic field extrapolation
codes \citep{2008ApJ...675.1637S} and other indirect measurements of
the coronal magnetic field strength \citep{2007Sci...317.1192T}.

Recent advances in solar instrumentation have allowed substantial
progress to be made over early SOHO/EIT observations. Data from the
\stereomission\ (STEREO) \euviinstrument\ (EUVI) instruments provided a
significant improvement in both spatial and temporal resolution
\citep[\EG][]{2008ApJ...680L..81L, 2008ApJ...681L.113V}.  Critically,
with the launch of \sdomission\ (SDO) in 2010, highly detailed,
multi-wavelength observations of EUV waves are now possible,
illuminating the complex structure and interactions of these waves
\citep[\EG][]{2012ApJ...753...52L}. With these new data, studies of
individual wave events \citep[\EG][]{2011ApJ...741L..21L} have
augmented earlier kinematic studies \citep{1999SoPh..190..467W,
  2000ApJ...543L..89W}, improving the description of the initiation
and subsequent deceleration of EUV waves. \cite{2013ApJ...776...58N}
catalogs 171 EUV waves identified through visual inspection of AIA
images in the 193 \AA\ channel between April 2010 and January 2013.


In order to answer fundamental questions regarding the physical nature of EUV waves, many waves must be studied using reproducible methods that can scale to perform automated searches of the AIA data. Such studies are too time-intensive in scope to be carried out manually. It is challenging to produce exactly reproducible results using a manual approach applied to large datasets. Automated feature detection algorithms have an advantage over human detections of features because they generate repeatable results for the same input data, \IE\ they enable reproducibility. In addition, their ability to examine large quantities of data quickly makes them a valuable tool and a complement to manual inspection and analysis. The solar physics community already makes use of the Computer Aided CME Tracking \citep[CACTus,][]{2004A&A...425.1097R} and Solar Eruptive Event Detection System \citep[SEEDS,][]{2008SoPh..248..485O}) CME catalogs, both of which are generated from automated feature detection algorithms.

Hence, in this article we present the Automated Wave Analysis and
Reduction in EUV (AWARE) algorithm\footnote{The results shown here
  were derived using SunPy version 0.7.10, with AWARE code available
  at \awareversion.}, a new automated EUV wave detection and
characterization procedure applied to EUV image data. Such a fully
automated procedure is essential in order to unlock the full potential
of the large full-disk image datasets available from SDO and STEREO,
and enables the characterization of EUV waves in large numbers. AWARE
has been developed in the Python programming language, and makes use
of features provided by the SunPy data analysis package
\citep{2015CSD....8a4009S}. AWARE is also a fully open-source and version controlled package, which is freely available (see Section \ref{s:conclusions}) under a BSD-3 clause license.

In Section 2 we discuss existing algorithms for the detection of solar features, including EUV waves, and their current status. In Section 3 we discuss in detail the AWARE algorithm and pipeline. In Section 4 we analyze the performance of AWARE and demonstrate its diagnostic and characterization features.

\section{Existing EUV Wave Detection Algorithms}
\label{sec:existing}
EUV waves appear as faint, extended, enhancements that propagate against the complex background structure of the solar corona.  Their relative faintness and structurally complex corona make them difficult to isolate. There are at least three automated EUV detection
methods currently published, the Novel EUV-wave Machine Observing (NEMO)
algorithm, described by \cite{2005SoPh..228..265P} \citep[see also][]{2012SoPh..276..479P}, the Coronal Pulse Identification and Tracking Algorithm (CorPITA) described in \cite{2014SoPh..289.3279L}, and Solar Demon \citep{2015JSWSC...5A..18K}.

\begin{figure}
\begin{center}
\includegraphics[width=12cm]{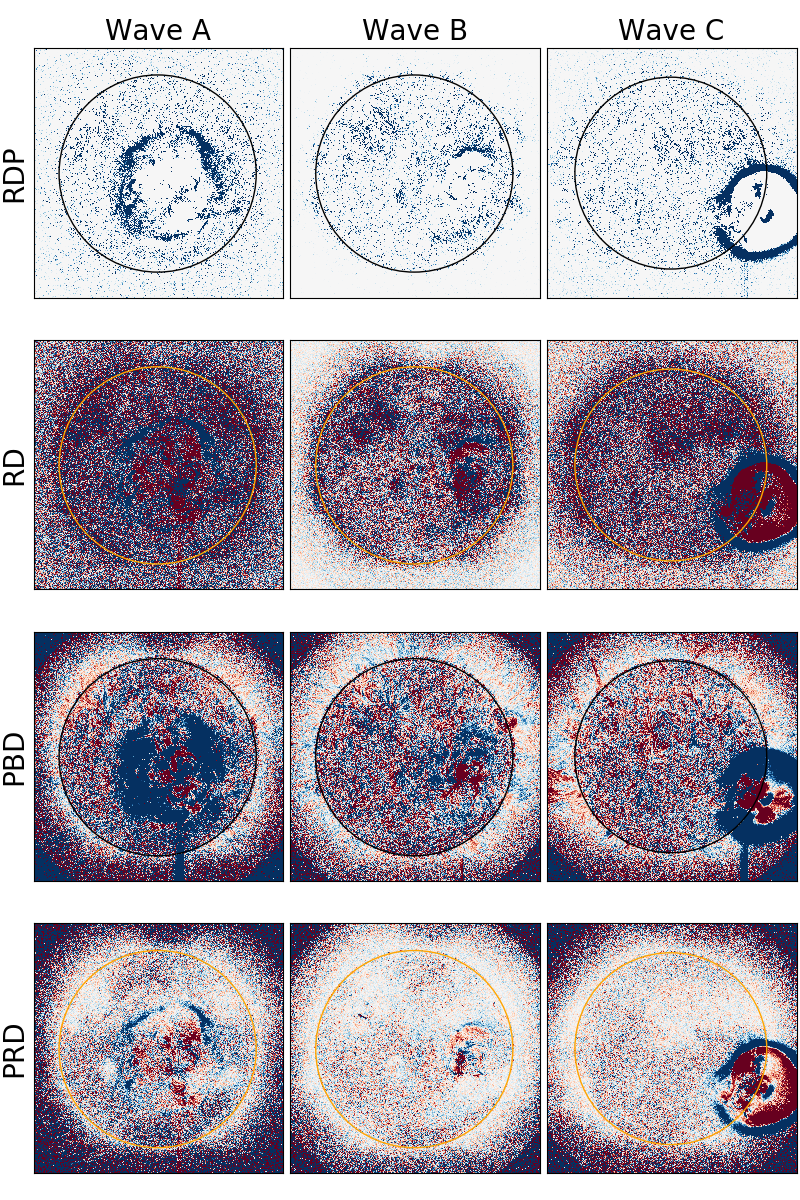}
\caption{Illustration of the effect of four different imaging
  processing techniques applied to three EUV waves.  The columns show
  the result of applying different differencing methods to three
  different EUV wave events as seen in the AIA 211 \AA\ channel.  Wave
  A refers to data 15 February 2011 02:00:24 UT, Wave B refers to 16
  February 2011
  14:35:48 UT and Wave C refers to 07 June 2011 06:29:12 UT (see also
  Figures \ref{fig:longetal8a}, \ref{fig:longetal8e}  and
  \ref{fig:longetal4} respectively). In all plots, blue pixels
  indicate a positive change in emission, white indicates no change in
  emission, and red indicates a negative change in emission. The first
  row shows the running difference persistence (RDP) images used by
  AWARE (Section \ref{sec:aware:segment:persistence}), the second row
  shows the running difference (RD) images (used in NEMO analysis),
  and the third row shows the percentage base difference (PBD, used by
  CorPITA).  The final row shows the percentage running difference
  (PRD) images used by Solar Demon.  The color representations in the RDP and RD images are identical; values are linearly scaled in the range -100\,--\,100.  The colors in the PBD images are linearly scaled in the range -25\%\,--\,25\%. The \RDP\ images show a cleaner separation of the propagating bright wavefront compared to the other three differencing methods.}
\label{rpdm_figure}
\end{center}
\end{figure}

NEMO was originally designed for analysis of SOHO/EIT data, but has since been
modified to analyze STEREO/EUVI images. The original NEMO algorithm
\cite{2005SoPh..228..265P} consists of three components. These are: 1)
source event detection, 2) recognition of eruptive dimmings, 3)
detection and analysis of EUV waves. The event detection component is
based on the higher-order moments of running difference (RD) images. A
RD image is simply the difference between two consecutive images. A
sharp change in the skewness or kurtosis of the distribution of RD
image values is a reliable signature that an impulsive event, such as
a flare or an EUV wave, has been observed in that image.  Results from
NEMO are available at \url{http://sidc.be/nemo/}; however, the
implementation ceased operations in 2010 and so there are no new EUV
detections being provided to the community.
\cite{2012SoPh..276..479P} is concerned with advances to original NEMO
algorithm with respect to source event detection and eruptive
dimmings, and does not explicitly tackle EUV waves.  Example RD images are shown in the second row of Figure \ref{rpdm_figure}.


The second algorithm that has been developed is CorPITA
\citep{2014SoPh..289.3279L}. CorPITA uses percentage base difference
(PBD) images as the foundation for detection (for example images see the third row of Figure \ref{rpdm_figure}).  A PBD
image is formed by taking the difference between a selected base image
and the current image, and then scaling that difference by the base
image, multiplied by 100.  CorPITA is triggered by the occurrence of a
flare.  In CorPITA PBD images, the base image is taken two minutes
before the flare start time. The flare position is used as the origin
of the EUV wave; great circles intersecting this origin are analyzed
to identify whether an EUV wave is present. The intensity profile
along the great circle is fitted for each time-step with a
multi-Gaussian function, based on the observation of
\cite{2006ApJ...645..757W} that cross-sections of EUV wave events have
this approximate form. This assumption allows the wave to be
characterized in terms of its position, velocity, and width.

\rtr{The third algorithm that has been developed is Solar Demon \citep{2015JSWSC...5A..18K}. Solar Demon detects three types of solar phenomena - flares, dimmings, and EUV waves. Whenever Solar Demon detects a flare, its starting location and time are used to initiate Solar Demon EUV wave detection and characterization, using percentage running difference images. The percentage running difference image (PRD, examples are shown on the final row of Figure \ref{rpdm_figure}) is calculated by subtracting an image taken two minutes earlier from the current image, and dividing by the image two minutes earlier. The characterization procedure involves de-rotation, limb brightness correction, and divides the image into 24 sectors. A Hough transform-based scheme is used for identifying and characterizing the EUV wave front in every sector.}

In this context, AWARE provides a new, alternative approach for the
detection and characterization of EUV waves, based on running difference persistence (RDP images, first row of Figure \ref{rpdm_figure}).  In the following section, we describe in detail the image processing and data analysis steps in AWARE, and demonstrate its application to solar data.
\section{The AWARE Wave Characterization Algorithm}
\label{sec:aware}
AWARE is implemented in two stages.  In the first stage, image processing techniques are used to isolate the wavefront.  These are described in Section \ref{sec:aware:segment}.  In the second stage, the location of the propagating wavefront is estimated and a model of the propagation is fit to the wavefront motion.  These are described in Section \ref{sec:aware:dynamics}.

\subsection{Stage 1: Image Processing Steps to Segment the Propagating Wavefront}
\label{sec:aware:segment}
As was noted in Sections 1 and 2, EUV waves are difficult to detect
since they are faint, extensive, and propagate over a complex
background image, the solar corona. This realization has driven past
attempts to enhance and detect EUV waves by making use of running
difference or percentage base difference images. However, these images, while
enhancing potential wavefronts, remain noisy and populated by other
extraneous features (see Figure \ref{rpdm_figure}). AWARE adopts a new, simple, and very
promising strategy for segmenting an EUV wave wavefront from image
data, running differences of persistence images
\citep{2016ApJ...825...27T}.

\subsubsection{Segmentation Using Running Difference Persistence Images}
\label{sec:aware:segment:persistence}
\begin{figure}
\begin{center}
\includegraphics[width=12cm]{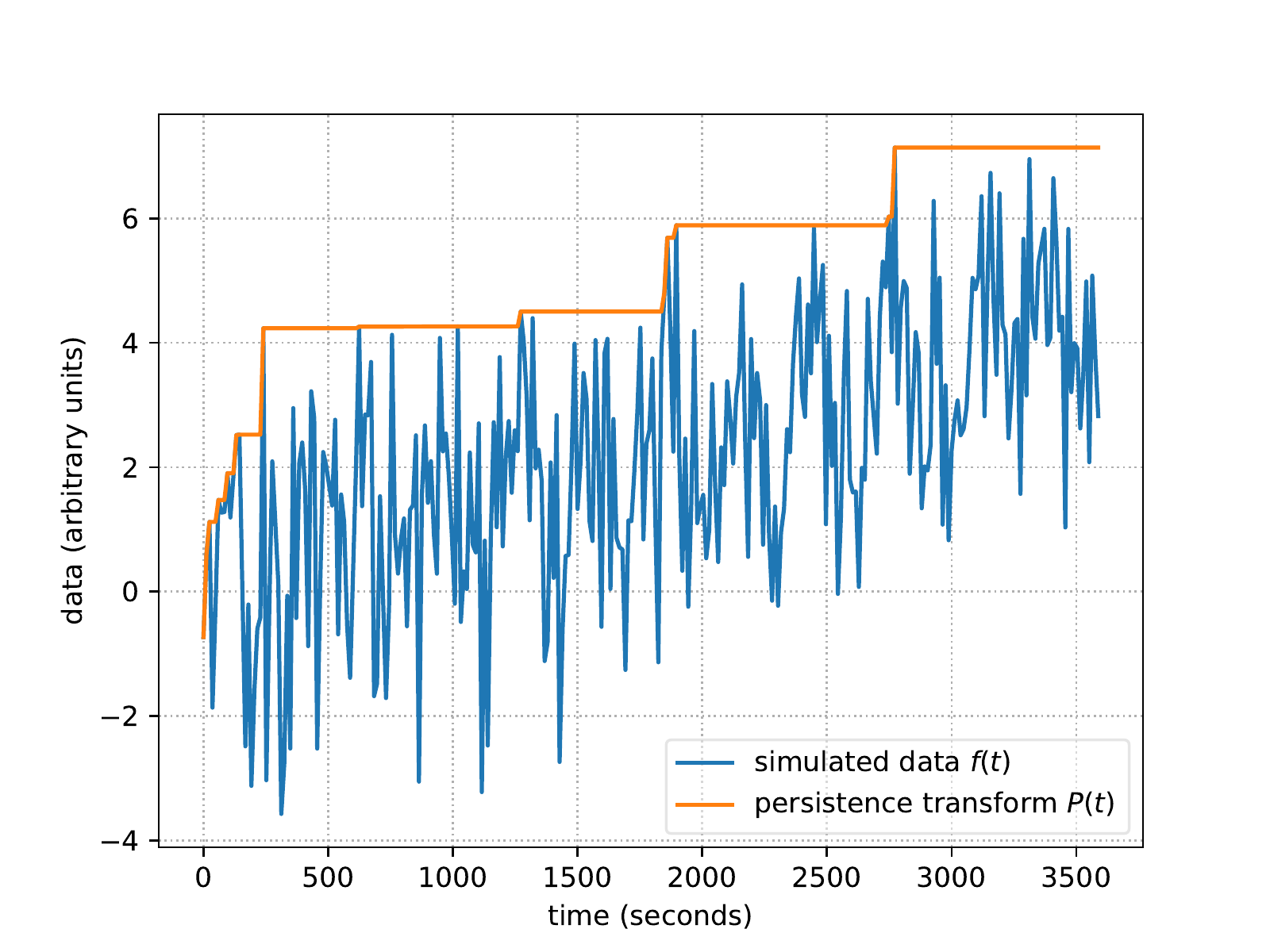}
\caption{Example of the application of the persistence transform.  The
original data $f(t)$ is shown in blue, and its persistence transform
$P(t)$ is shown in red.}
\label{fig:persistence}
\end{center}
\end{figure}

A persistence image is found by calculating the persistence (or running maximum) value of the emission at each pixel at all locations and times.  The
persistence value at time $t$ of a time series $f(t)$ is simply the
maximum value reached by that pixel in the time range $0\rightarrow
t$.  If at later times the pixel value increases, the persistence
value increases accordingly. If the pixel value decreases however, the
persistence value remains unchanged. Hence, a set of persistence maps
constructed from an image sequence will indicate the brightest values
yet achieved in that sequence at each $t$.  The persistence transform
$P(t)$ of the time-series $f(t)$ is defined as
\begin{equation}
\label{eqn:persisttransform}
P(t) = \max_{t'=0}^{t'=t}f(t').
\end{equation}
Figure \ref{fig:persistence} illustrates the persistence transform
$P(t)$ of a time-series of simulated data $f(t)$. In AWARE, the persistence transform is applied on a pixel-by-pixel basis on time-ordered sets of AIA images to obtain a persistence transform of the original AIA images.  The running difference of these images generates the running difference persistence (RDP) images.
Figure \ref{rpdm_figure} illustrates \RDP\ images for three example EUV wave events (these were also analyzed by CorPITA \citep{2014SoPh..289.3279L}).  The first row shows running difference persistence (RDP) images, the basic image type used by AWARE.  The second row shows running difference (RD) images, the basic image type analyzed by the NEMO algorithm \citep{2005SoPh..228..265P}. The final row shows percentage base difference (PBD) images, used by the CorPITA algorithm. Comparison with \RDP\ images shows that in standard RD and PBD images the wavefront is more diffuse, and much coronal structure not associated with the wavefront remains in the image. RD and PBD images also have much denser noise compared to the \RDP\ images of the same data; hence, separating the EUV wave from the noise is substantially easier when using \RDP\ images.

\RDP\ images have two desirable properties when searching for EUV waves.
Firstly, only pixels that brighten over previous values have a
non-zero value in the running difference of persistence images, while
zero-value pixels correspond to areas that did not increase in
brightness. Hence, since an EUV wave brightens neighboring pixels as
it moves across the Sun, the \RDP\ images isolate those brightening
pixels.  In other words, the \RDP\ images isolate the leading part of
the wavefront that brightened new pixels.  Secondly, since much of the
corona does not vary significantly during an EUV wave, \RDP\ images
show very little residual coronal structure distant from the EUV,
greatly simplifying the resulting images.

\subsubsection{Reducing Noise in Running Difference Persistence Images}
\label{sec:aware:segment:noise}
EUV waves are faint against the background corona and so some additional steps are applied in order to improve the chance of their detection.  The image processing steps used to segment the EUV wave from the input data are described below.  In the following, a datacube refers to a three-dimensional array of data ordered as $(x,y,t)$, two spatial directions $(x,y)$ and one temporal direction $t$.  An image is a slice in the datacube at a fixed value of time $t$.

\begin{figure}
\begin{center}
\includegraphics[width=0.48\textwidth]{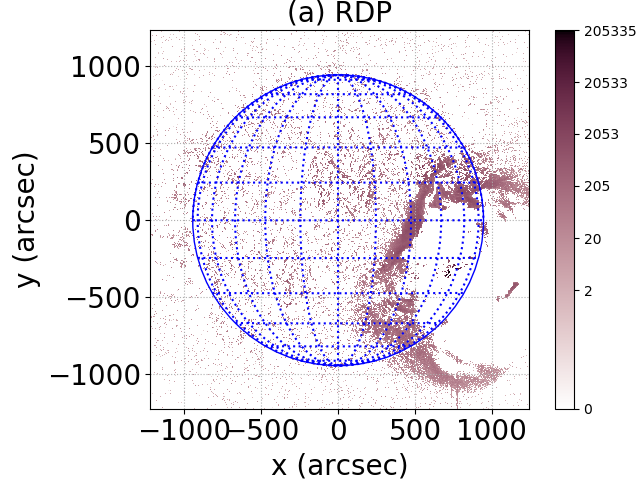}
\includegraphics[width=0.48\textwidth]{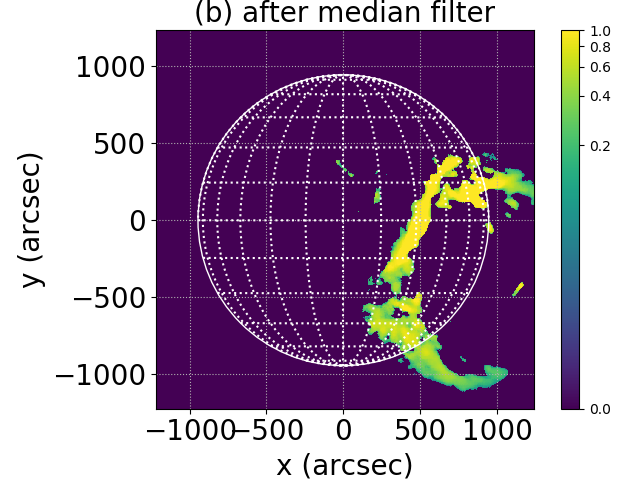}
\end{center}
\begin{center}
\includegraphics[width=0.48\textwidth]{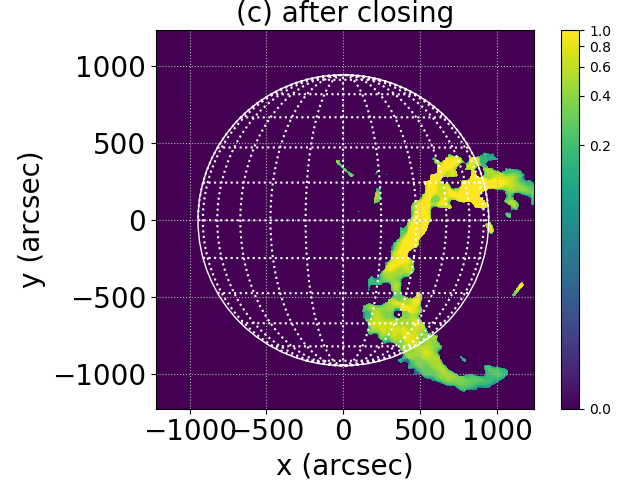}
\end{center}
\caption{(a) Shows the running difference image between two
  consecutive persistence images. The wavefront is already
  evident. (b) Shows the image at the same time after applying
  the median filter (Section \ref{sec:aware:segment}).  (c)
  Shows the resulting image after the morphological closing operation
  is applied.  This has the effect of filling in small gaps in the
  detected wavefront \citep[\EG][]{2002dip..book.....G}.}
\label{method_figure}
\end{figure}

\begin{enumerate}

\item Given a set of time-ordered solar EUV images (\EG\ SDO/AIA or STEREO/EUVI data), data are summed in time and space to increase the signal to noise ratio of the wave against the background. Images may be summed in space as desired, for example an AIA image may be binned using $2\times2$ super-pixels to form $2048\times2048$ pixel images. In the time dimension, images may be summed as required. Typically, pairs or triplets of consecutive images are used. After this step the datacube consists of $N_{D}$ images (the summations in the space and time dimensions are used to determine the EUV wave detections shown in Figures \ref{fig:longetal4}, \ref{fig:longetal6}, \ref{fig:longetal7}, \ref{fig:longetal8a}, and \ref{fig:longetal8e}).

\item The persistence transform is then applied to the resulting image set.  This creates a set of persistence images, showing the brightest values obtained in each pixel (Equation \ref{eqn:persisttransform}) as a function of time.

\item Perform a running difference operation on the persistence images. Hence, only areas that increase in brightness from one time to the next remain.  An example RDP image is given in Figure \ref{method_figure}a.  There are now $N_{D}-1$ images in a datacube $R(x, y, t)$.

\item \rtr{The dynamic range of the datacube $R(x,y,t)$ is reduced by applying a square root transformation. AWARE uses a percentile clipping on the transformed datacube where the top 1\% of the intensity distribution in the datacube are clipped and set to the value at the 99\%th percentile. This filters out spikes in the data caused by cosmic ray strikes, bright flares, etc. The datacube is then normalized to the range zero to 1, yielding a datacube $D(x,y,t)$.  The purpose of this step is to remove the high values so that the intensity distribution better represents the bulk of the image pixels.}

\item 
The above steps extract propagating features from the input datacube.

However, the resulting images $I_{i}$ $(1\le i \le N_{D}-1)$ still show substantial noise.  The following steps are intended to further isolate the wavefront by applying additional filtering steps that analyze each image at multiple length-scales $r_{k}, 1\le k \le N_{r}$.
\begin{enumerate}

\item Apply a noise reduction filter to the data cube of images.  Our demonstration algorithm uses a two-dimensional median filter applied to each image in the datacube.  This replaces every pixel in the image with the median value found in its neighborhood.  The median filter used is a circle $C_{r_{k}}$ in the spatial dimensions with a given radius $r_{k}$ pixels. The median filter is a commonly used and simple method of removing noise from an image \citep{2002dip..book.....G}.  The resulting image is $I_{noise, k}=\mbox{median}(I, C_{r_{k}})$.  An example of the effect of this operation is shown in Figure \ref{method_figure}b.

\item Apply a morphological closing operation to the noise-reduced
  image \citep{2002dip..book.....G}.  This operation helps to close
  small ‘cracks’ in structures.  The structuring element used is the
  same as that used by the median filtering operation, and generates
  an image $I_{close, k} = I_{noise, k}\bullet C_{r_{k}}$, where
  $\bullet$ denotes the morphological closing operator. An example of the effect of this operation is shown in Figure \ref{method_figure}c. 

\end{enumerate}

\item The non-zero locations in the image $F(x,y)=\sum_{k=1}^{N_{r}}I_{close, k}(x,y)$ indicates the location of the EUV wave as determined by the multi-scale operations in the previous step.   \rtr{Note that the temporal summing needs to be large enough so that when the running difference of the persistence image is calculated a significant number of pixels are newly brightened in the data so that the median noise reduction at the specified length scale does not replace those pixels with zeroes (the background value).}

\item Masks indicating the non-zero locations in each image $F$ are created:
\[
    \mbox{mask}(x,y)=\left\{
                \begin{array}{ll}
                  1 &, F(x,y)>0 \\
                  0 &, \mbox{otherwise}.\\
                \end{array}
              \right.
  \]

\end{enumerate}
The final product is a datacube of time-ordered series of $N_{D}-1$ masks $M(x,y,t)$ that localize the bright wavefront of the EUV wave.  This is the AWARE stage 1 data product which is used to detect and characterize EUV wave propagation.  The mask datacube is clearly time-dependent.  However, the wave progress can be summarized for illustrative purposes as a static image by creating a wave progress map, defined as
\begin{equation}
\label{e:waveprogressmap}
    \mbox{wave progress map}(x,y)=\left\{
                \begin{array}{ll}
                  \max(t), & M(x,y,t)-M(x,y,t-1)>0, 1\le t \le N_{D}-1 \\
                  0, & \mbox{otherwise}.\\
                \end{array}
              \right.
 \end{equation}
The wave progress map summarizes the results of stage 1 of AWARE, indicating the time at which the leading edge of the wavefront is detected.  Figures \ref{fig:syntheticwave:results}a, \ref{fig:longetal4}a, \ref{fig:longetal7}a, \ref{fig:longetal6}a, \ref{fig:longetal8a}a and \ref{fig:longetal8e}a show examples of wave progress maps.  In the following section, the characterization of the EUV wave dynamics is described.

\subsection{Stage 2: Determining Wave Dynamics}
\label{sec:aware:dynamics}
The output of the first stage of AWARE is a time-ordered series of masks that indicate regions that were progressively brightening \rtr{(summarized by wave progress maps - see Equation \ref{e:waveprogressmap}, Section \ref{sec:aware:segment} and Figure \ref{fig:syntheticwave:results}a)}.  The second stage is to determine the dynamics of the wave using the following steps.

\begin{enumerate}
\item Isolate the increase in intensity due to the EUV wave. This is determined by calculating $S(x,y,t) = R(x,y,t)\times M(x,y,t)$ on a pixel-by-pixel basis, \IE\space the density increase caused by the wave is evaluated at the locations indicated by the mask.
\item It is known that EUV waves are associated with solar eruptive events and so this provides a candidate location and start time for the source of the wave. The candidate location is used as a source from which the wave is launched.  The equivalent pixel locations in the data $S$ of arcs emanating from the source location and propagating across the sphere of the Sun following the path of great circles are used, similar to \citet{2014SoPh..289.3279L}.
\item Each arc $A$ has a unique path $A(s)=(x_{s}, y_{s})$ $(0\le s \le 1)$ which is used to extract data from $S(x_{s}, y_{s}, t)$ at each time $t$.  This creates a two-dimensional segment of the data as a function of time and distance along the arc.  At each time, the  position $p(t)$ of the wavefront and an estimate of the error $e(t)$ in that position is calculated.  For the results shown here,
\begin{equation}
\label{e:position}
\mbox{position}(t)=\frac{\sum_{s=0}^{1}S(x_{s}, y_{s}, t)L(s)}{\sum_{s=0}^{1}S(x_{s}, y_{s}, t)},
\end{equation}
where $L(s)$ is the distance along the arc from the initiation point. The position of the wavefront is the weighted mean location of the wave, weighted according to the intensity of the wavefront. The error in the position is defined as the maximum width of the wavefront at time $t$, that is, 
\begin{equation}
\label{e:error}
e(t)=L(s_{max}) - L(s_{min}),
\end{equation}
where $s_{max} = \max_{s}S(x_{s}, y_{s}, t) \ne 0$ and $s_{min} = \min_{s}S(x_{s}, y_{s}, t) \ne 0$.  
Note that at some times, an estimate of $p(t)$ or $e(t)$ cannot be made. This can happen when there is no emission at a given time along the arc. These times are eliminated from further consideration.
\item It is common to find that noise in the data $S(x,y,t)$ yields
  estimates of $p(t)$ that are clearly outliers compared to the
  general trend of other nearby points. \citet{2014SoPh..289.3279L}
  also encounter this issue, and describe a novel iterative algorithm
  for deciding which points to consider.  Instead, AWARE uses the
  Random Sample Consensus (RANSAC) algorithm (\citet{RANSAC}, see also
  \citet{2014ApJ...793...70M}) to decide which $p(t)$ can be considered inliers and which can be discarded as outliers to a possible fit.  A point $p(t)$ is considered to be an inlier if the residual between a test fit and $p(t)$ is less than the median estimated error $\mbox{median}(e(t))$.  The \texttt{scikit-learn} \citep{scikit-learn} implementation of RANSAC is used\footnote{Version 0.19.1 of \texttt{scikit-learn} is used.}, with no other changes to the default selection of choices.
\item The points that remain after steps 3, 4 and 5 are indicated
    on fit participation maps (an example of the fit participation map is shown in Figure \ref{fig:syntheticwave:results}b). If more than three points $p(t)$ are found to be inliers via the RANSAC algorithm, the progress of the wave is fit with a quadratic function
\begin{equation}
\label{e:quad}
s(t) = s + \vinit t + 0.5\ainit t^{2}.
\end{equation}
\rtr{Increasing the number of samples fits and reducing noise ($e(t)$)
  increases the amount of information in the data, improving the fits \citep{2013A&A...557A..96B}.}
\end{enumerate}

\subsection{Adjusted CorPITA Score}\label{ss:adjusted}
\rtr{CorPITA \citep{2014SoPh..289.3279L} defines a quality score, referred to here as the \longscore\ to assess how well determined is each part of the EUV wavefront. The \longscore\ is defined as
\begin{equation}
\label{e:longscore}
\mbox{\longscore} = 100\times\left[E_{score} + D_{score}\right]/2
\end{equation}
The score weights the fit quality in two parts, an existence component
$E_{score}$ and a dynamic component $D_{score}$. The existence
component is the fraction of time that has a measurable extent.
CorPITA defines the denominator in this fraction as the number of
images processed.  However, it is not known {\it a priori} at what time the wave is no longer present or detectable.  In AWARE, we assume that the times of the first and last detections defines the start and end of when the wave was present and detectable.  The existence component measures how well AWARE captured the existence of the wave over the maximum extent of its estimated detectable range.}

\rtr{
The dynamic component of the score is defined as
\begin{equation}
\label{e:longscore:dynamic}
D_{score} = \frac{1}{3}\left(v_{score} + a_{score} + \sigma_{score}^{rel}\right),
\end{equation}
where each of the scores have value 1 if the measured value is within
pre-defined limits and zero otherwise. In \cite{2014SoPh..289.3279L},
if $1<\vfit< 2000 \speedunit$ then $v_{score}=1$ and if $-2<\afit< 2 \accunit$ then $a_{score}=1$.  If the mean value of the relative error in the fit position is less than 0.5, then $\sigma_{score}^{rel}=1$.  AWARE uses an adjusted version of the \longscore\ in which $v_{score}$ is defined as the integral of a normal probability distribution with mean value $\vfit$ and width equal to the error in $\vfit$, integrated between the same pre-defined limit range as \cite{2014SoPh..289.3279L}.  In comparison, this introduces more fit quality information to the existing \longscore.  The quantity $a_{score}$ is defined similarly.  AWARE calculates this adjusted \longscore\ for each arc.}

\section{Results}\label{sec:results}
To test the performance of the AWARE algorithm, it is useful to analyze signals where the underlying wave properties are already known. We therefore apply AWARE to a simulated wave with known  properties, and compare the detection and characterization of AWARE with the actual wave characteristics.  This allows us to identify systematic errors in the analysis process.

\subsection{Characterization of AWARE Using Simulated Data}
\label{ss:char}
The AWARE software suite contains a package that generates simulations of EUV waves propagating across the disk of the Sun.  Many different parameters of the wave kinematics can be altered, such as its kinematic profile, its width, dispersion, signal to noise ratio, and origin on the solar disk.  This simulation software is used to assess how well the AWARE detection and characterization software is performing (Section \ref{sec:aware}).  Section \ref{sss:offcenter} describes a simulated wave, the application of AWARE to these simulated data, and compares the AWARE results to the known properties of the wave.  Section \ref{sss:bic} tests the ability of AWARE to determine if a wave is accelerating or not, and Section \ref{sss:fitbias} demonstrates the existence of a systematic bias in the values of $\vinit$ and $\ainit$ that must be present in any algorithm (including AWARE) that allows for the wavefront to accelerate.

\subsubsection{Detection and Characterization of an Example Simulated Wave}
\label{sss:offcenter}
A circularly propagating wave with an initial velocity $\vtrue=466.5 \speedunit$ and acceleration $\atrue=1.5 \accunit$ is launched from heliographic position $(-22\Deg, -33\Deg)$.  The wave has a Gaussian profile in the direction of propagation, with a width of $0.49\Deg$.  The wave has an amplitude of 1 (in arbitrary units).  The position of the wave is calculated once every 12 seconds for 60 consecutive images. Images are 1024 by 1024 pixels, with diameter of the disk the same width as the image.  Noise in each image is Poisson-distributed with a mean value of 1.  

\begin{figure*}
\begin{center}
\includegraphics[width=6.0cm]{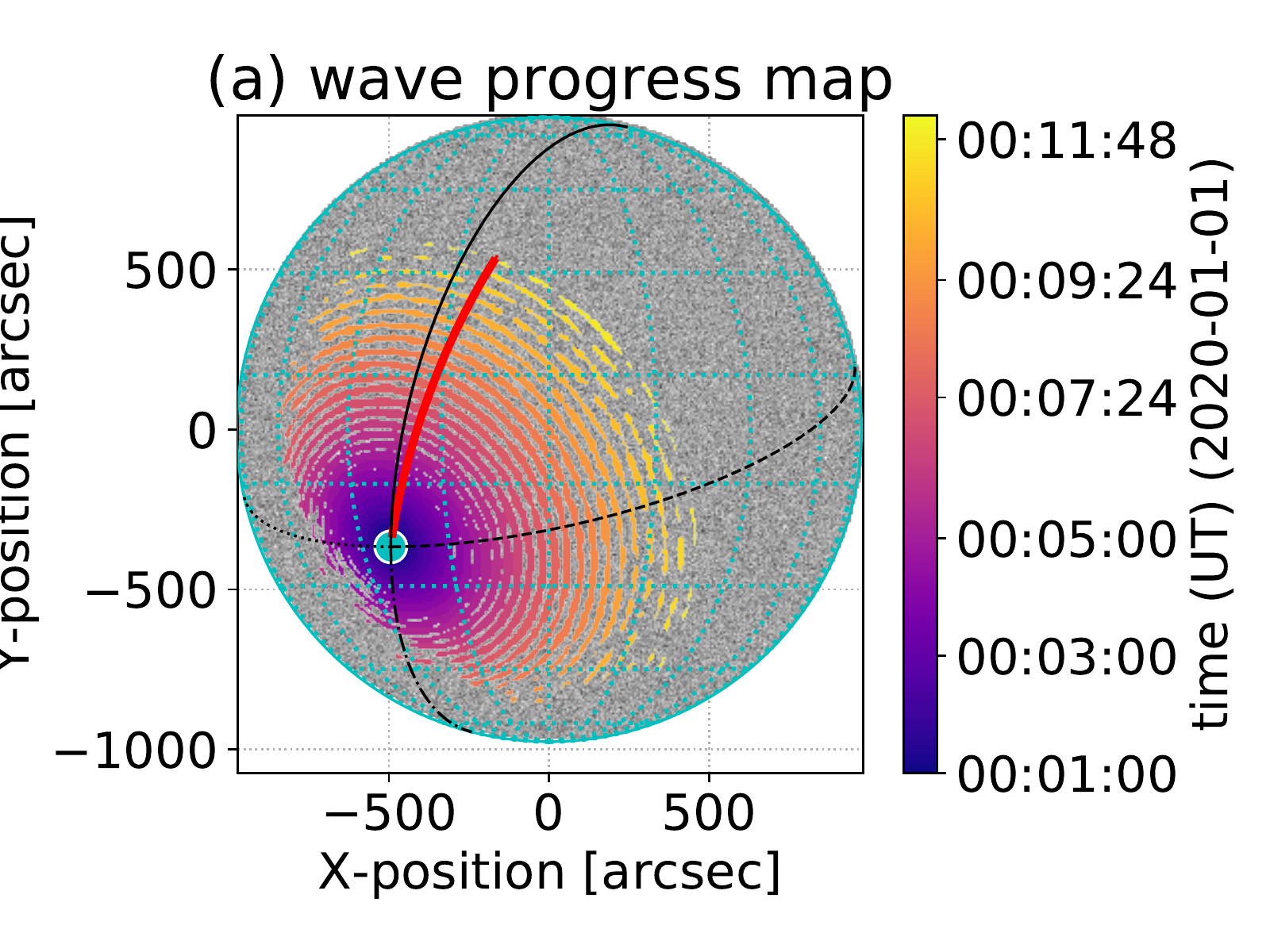}
\includegraphics[width=6.0cm]{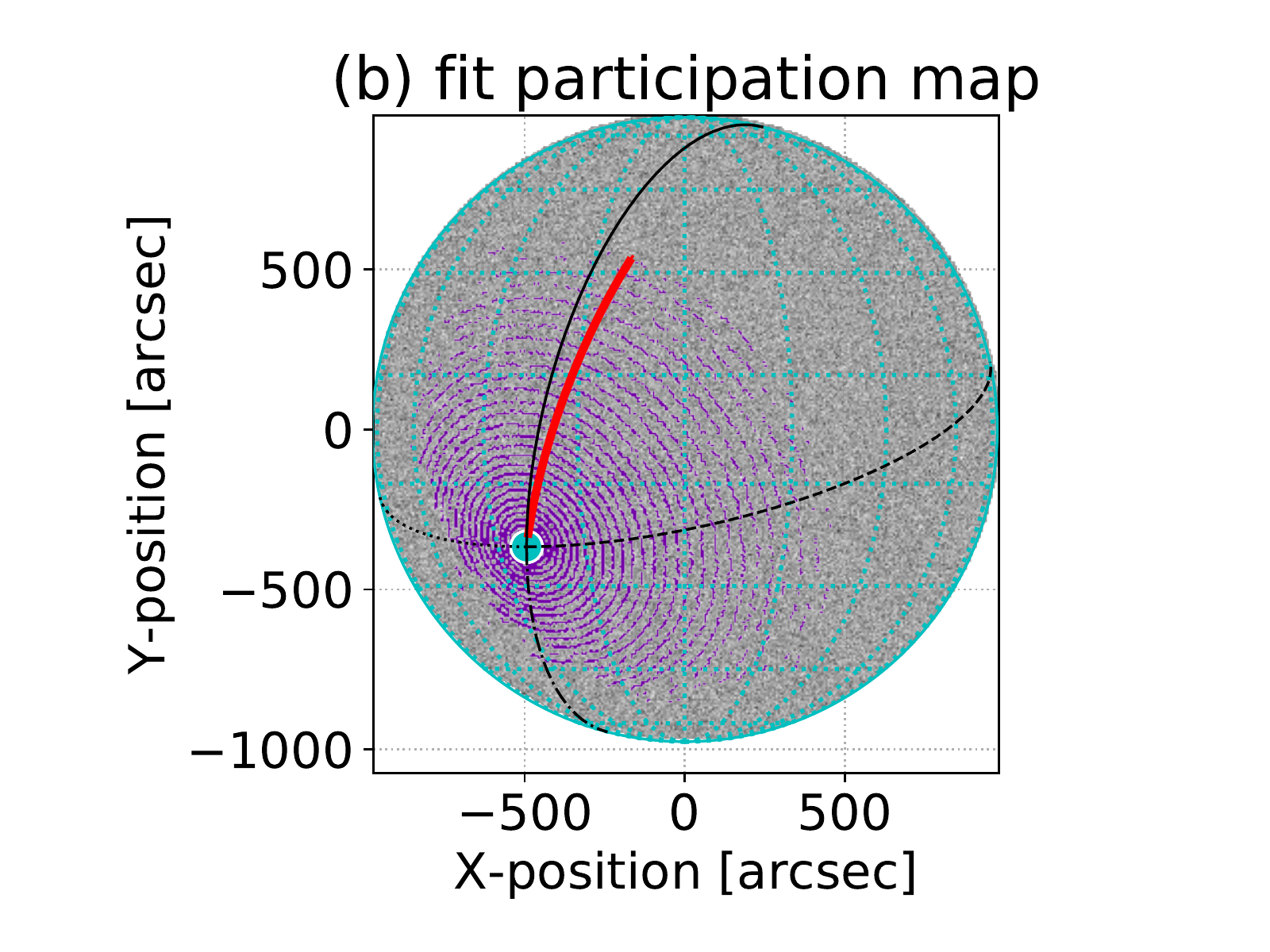}
\includegraphics[width=6.0cm]{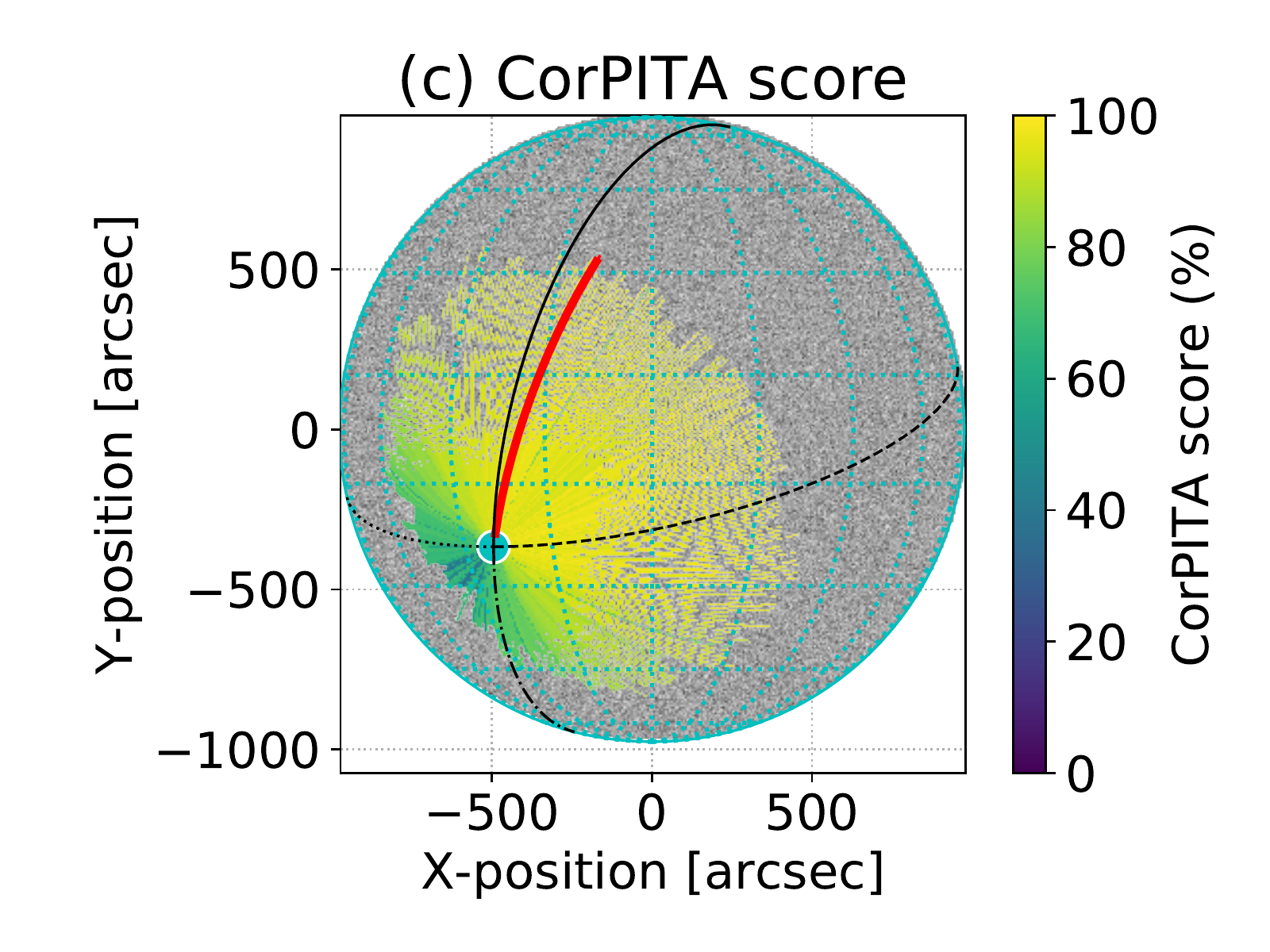}
\includegraphics[width=6.0cm]{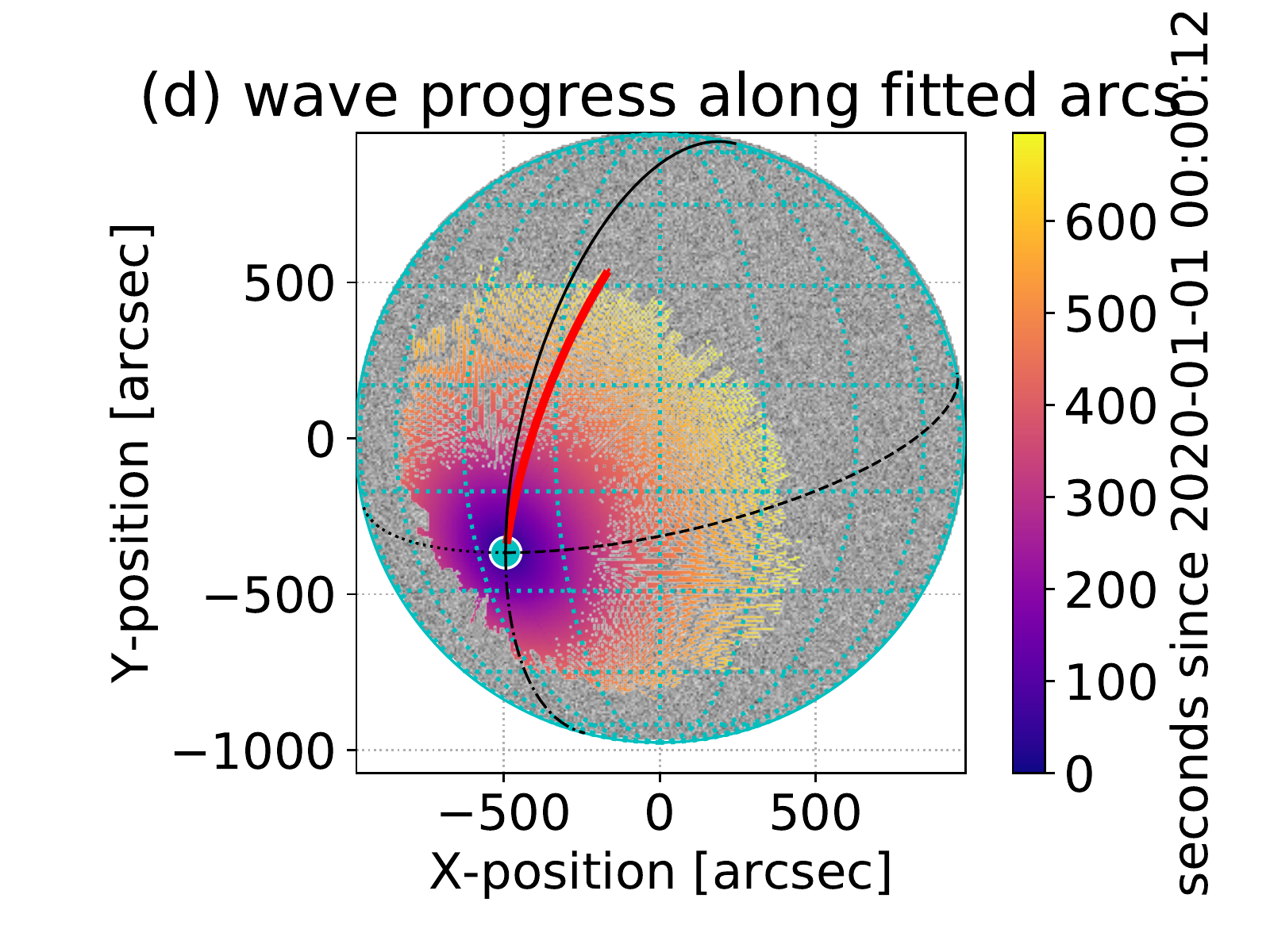}
\includegraphics[width=6.0cm]{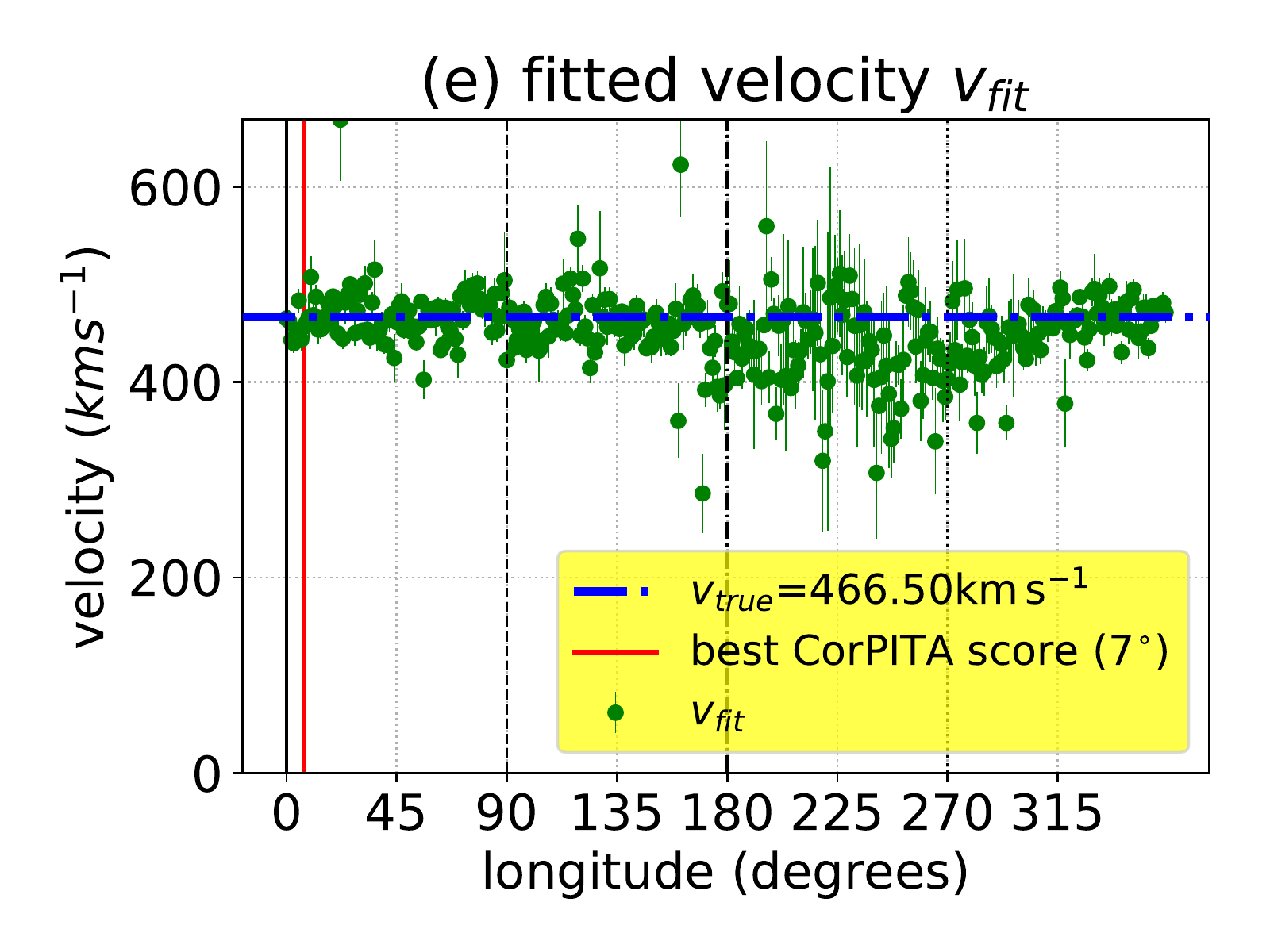}
\includegraphics[width=6.0cm]{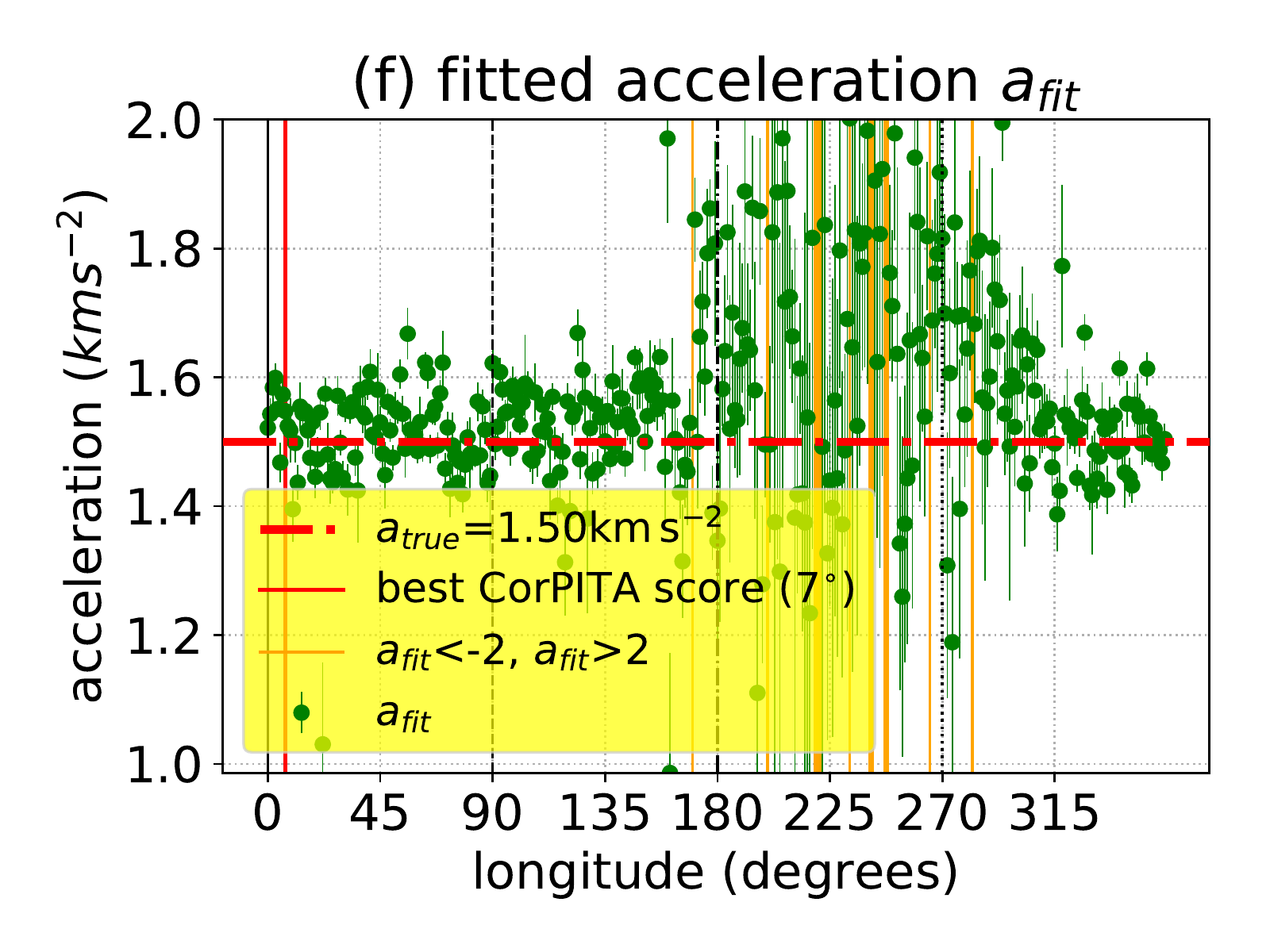}
\includegraphics[width=6.0cm]{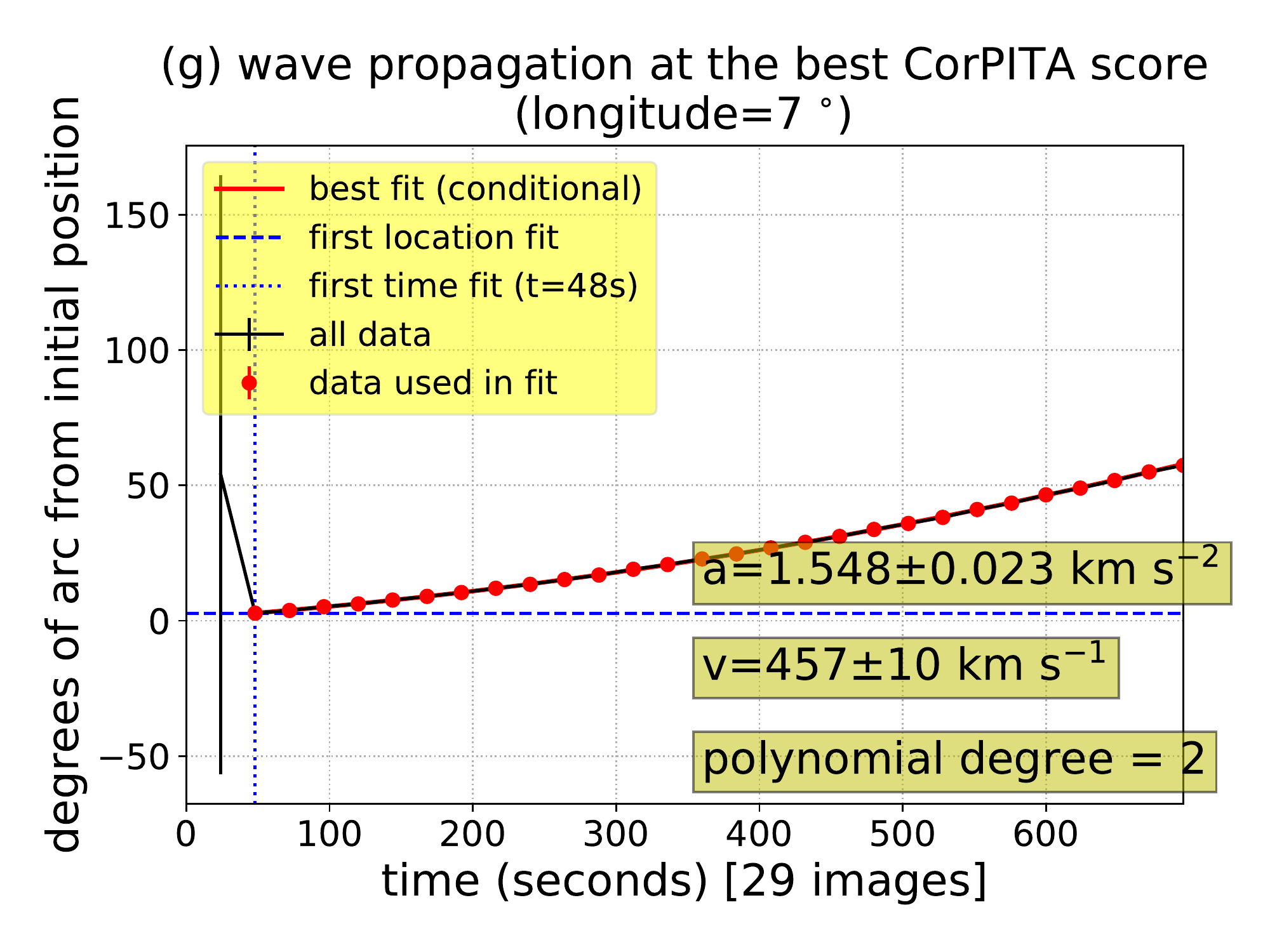}
\caption{AWARE applied to simulated data, a wavefront propagating
  evenly out from an off-disk center initiation point with an initial
  velocity of $466.5 \speedunit$ and an acceleration of $1.5
  \accunit$. (a) Is the result of stage 1 (Section
  \ref{sec:aware:segment}), the wave progress map (Equation
  \ref{e:waveprogressmap}). (b) Is the fit participation map showing
  the locations at which the wavefront is fit, and is the result of
  AWARE stage 2 (Section \ref{sec:aware:dynamics}). (c) Is the
  \longscore for each of the arcs. (d) Shows the fitted arcs and the fitted wave propagation.  In (a--d) the red line indicates the position and extent of the arc with the highest \longscore.  The solid, dashed, dot-dashed and dotted black lines indicate the $0\Deg, 90\Deg, 180\Deg$, and $270\Deg$ angular location around the wave source. (e, f) Show the mean and standard deviation of the wavefront velocity and accelerations respectively, as a function of longitudinal angle around the source. (g) Shows the wave progress along the angle with the best \longscore.}
\label{fig:syntheticwave:results}
\end{center}
\end{figure*}

\begin{figure*}
\begin{center}
\includegraphics[width=12.0cm]{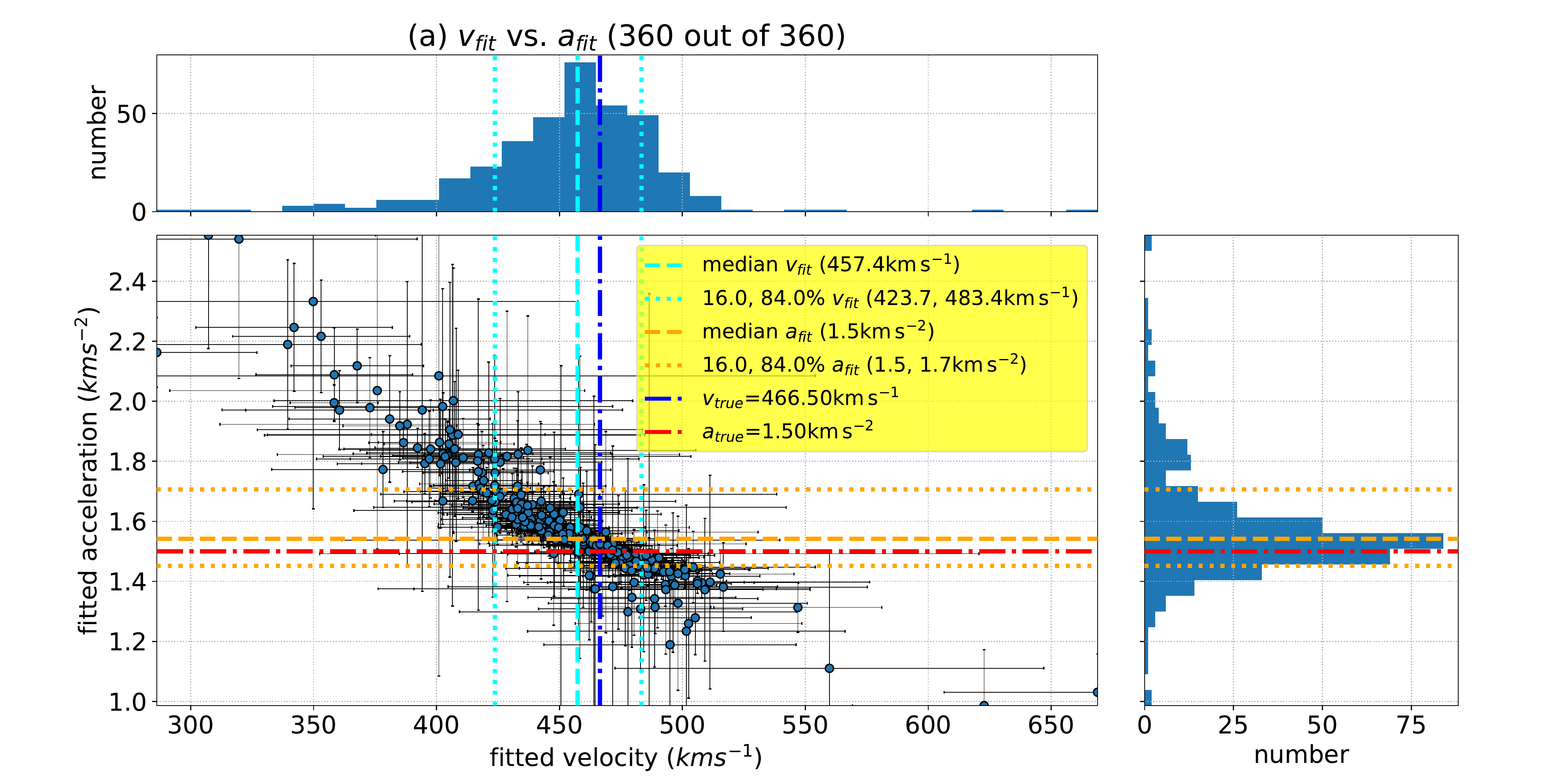}
\includegraphics[width=12.0cm]{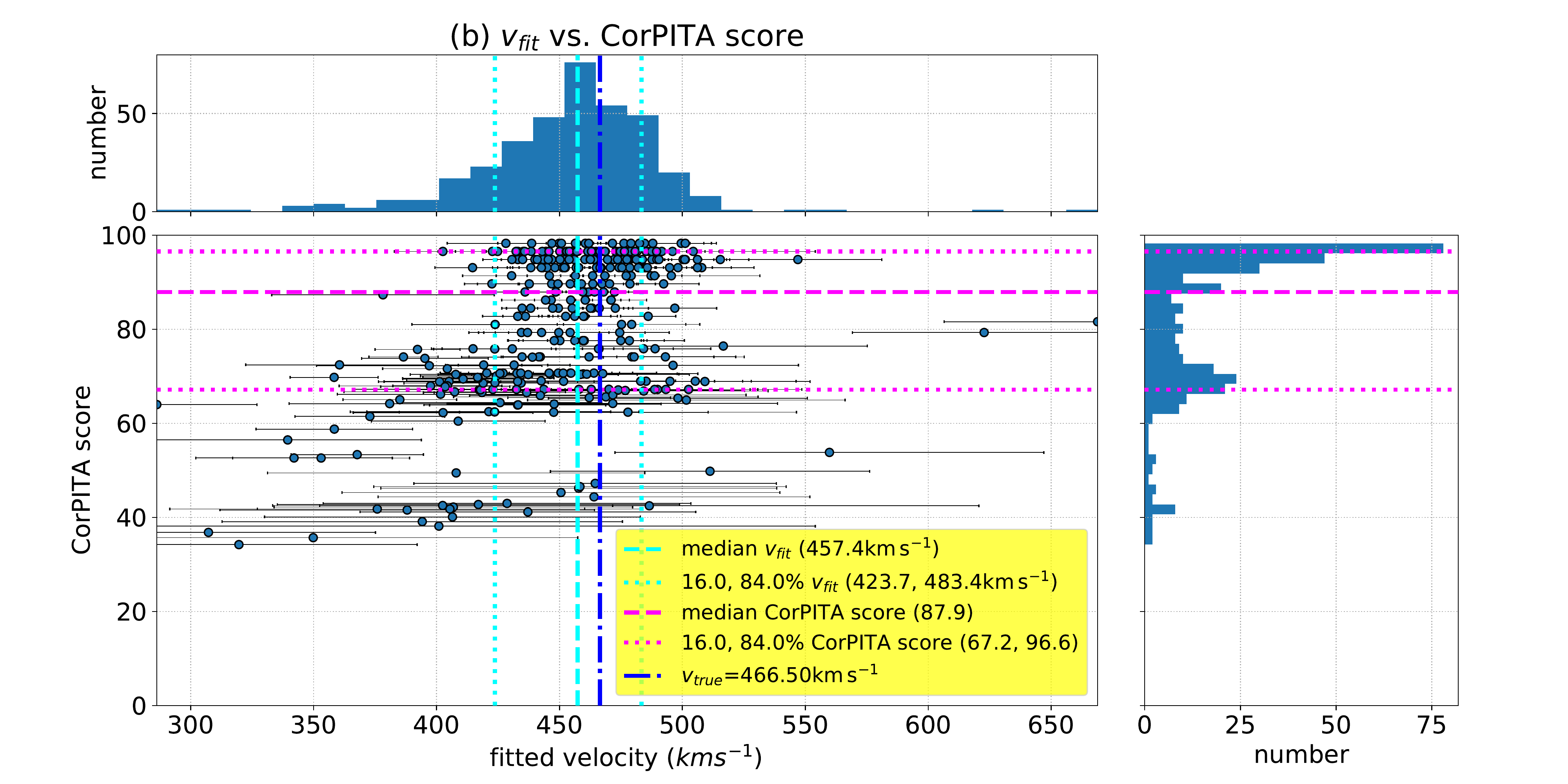}
\includegraphics[width=12.0cm]{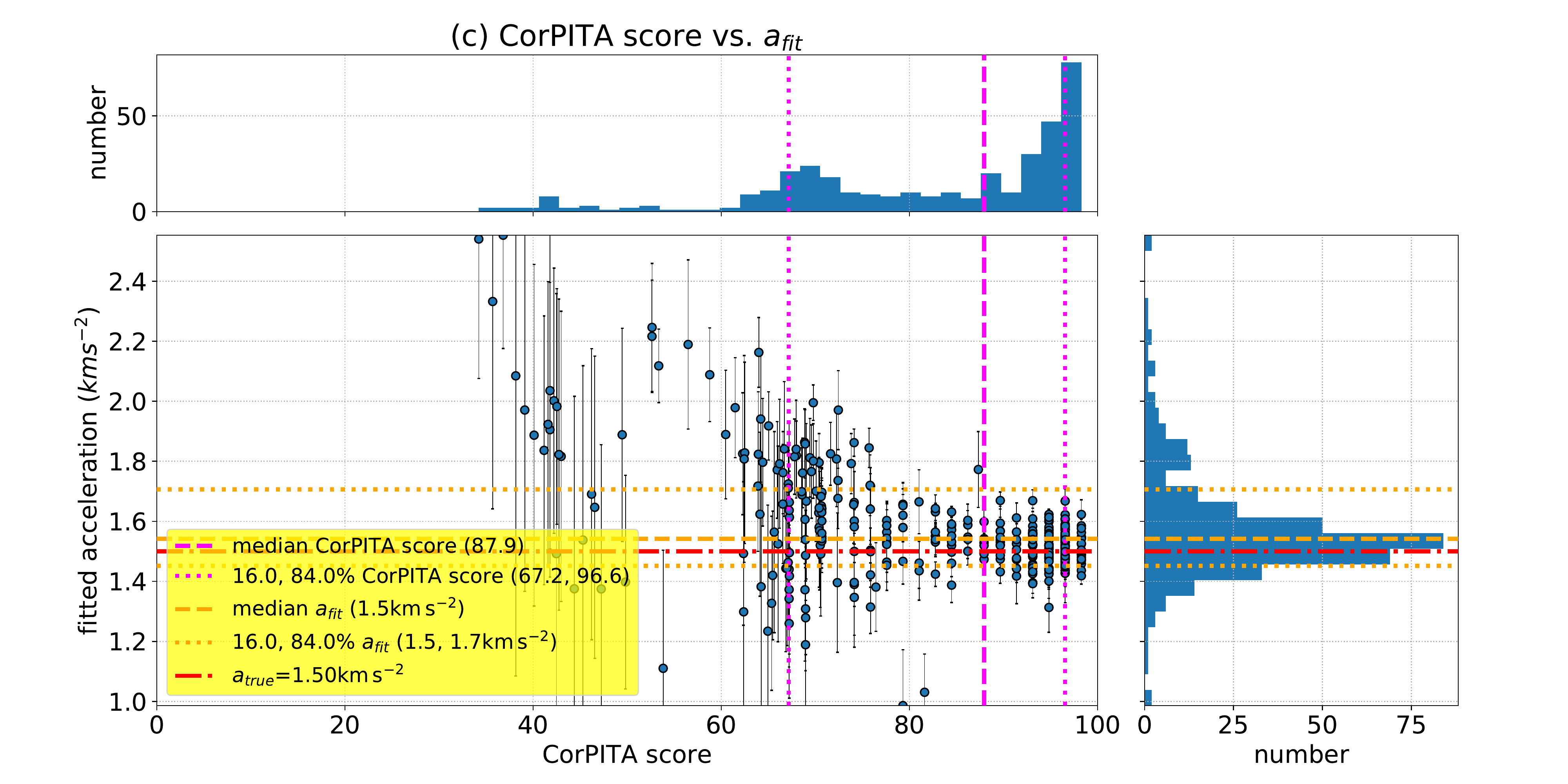}
\caption{Distributions of $\vfit$, $\afit$ and the \longscore\ for the
  fitted arcs of the simulated data. (a) Shows $\vfit$ \VERSUS\
  $\afit$; median values and 16\% and 84\% percentile values of each
  quantity are shown (for normal distributions the 16\% and 84\%
  percentile values indicate the mean value plus/minus one sigma).
  The title of (a) shows the number of arcs fitted out of a possible
  360).  Error bars are the one-sigma errors reported by the fit
  algorithm (stage 2, step 7).  (b) shows $\vfit$ \VERSUS\ the
  \longscore, and plot (c) shows the \longscore\ \VERSUS\ $\afit$.}
\label{fig:syntheticwave:dynamics}
\end{center}
\end{figure*}

AWARE is implemented on the resulting simulated dataset as follows. To increase the signal-to-noise ratio (see Section \ref{sec:aware:segment}), the images are binned in to $2\times2$ super-pixels and are pair-summed in time. Finally, the noise reduction and morphological closing operations are applied (see Section \ref{sec:aware:segment}) using disks of radius $22\arcsec$.  The results are shown in Figure \ref{fig:syntheticwave:results} and Figure \ref{fig:syntheticwave:dynamics}. 

Figures \ref{fig:syntheticwave:results} and
\ref{fig:syntheticwave:dynamics} illustrate various data products
produced by AWARE. \rtr{Figure \ref{fig:syntheticwave:results}a is a
  wave progress map (Equation \ref{e:waveprogressmap}) and shows the
  pixels at which wave progress is detected by stage one of AWARE
  (Section \ref{sec:aware:segment}).  Figure \ref{fig:syntheticwave:results}b shows the fit
  participation map,  the locations that  determine the wave dynamics
  (found through AWARE stage 2 - see Section
  \ref{sec:aware:dynamics}).  Figure \ref{fig:syntheticwave:results}c shows the fitted arcs and
  their \longscore.  Figure \ref{fig:syntheticwave:results}d shows the wave progress along each arc
  as a function of time since the initiation time. Figures \ref{fig:syntheticwave:results}e and f
  show the values $\vfit$ and $\afit$ respectively and their estimated
  error as a function of angle around the initiation point.   Figure
  \ref{fig:syntheticwave:results}g shows the wave location and its
  error as a function of time along the arc which has the highest
  \longscore.   Red lines in each of the plots (a-f) indicate the location of the arc with the highest \longscore. The remaining lines in Figures \ref{fig:syntheticwave:results}a-f indicate the $0\Deg, 90\Deg, 180\Deg$, and $270\Deg$ longitudinal extent around the wave source.}

Figure \ref{fig:syntheticwave:dynamics} shows the distributions of the
dynamics of the wave detection.  Figure
\ref{fig:syntheticwave:dynamics}a shows a clear dependence of $\vfit$
and $\afit$, even although the simulated wave does not include this
dependence.  The reasons for this dependence are explored more fully
in Section \ref{sss:fitbias}. Figures
\ref{fig:syntheticwave:dynamics}b and c show the distribution of
$\vfit$ and $\afit$ with the \longscore.  All plots demonstrate that
although there is a spread in the values recovered compared to the
true values, the mean value is correctly recovered (see also Figure
\ref{fig:syntheticwave:results}e and f).

\begin{figure*}
\begin{center}
\includegraphics[width=0.48\textwidth]{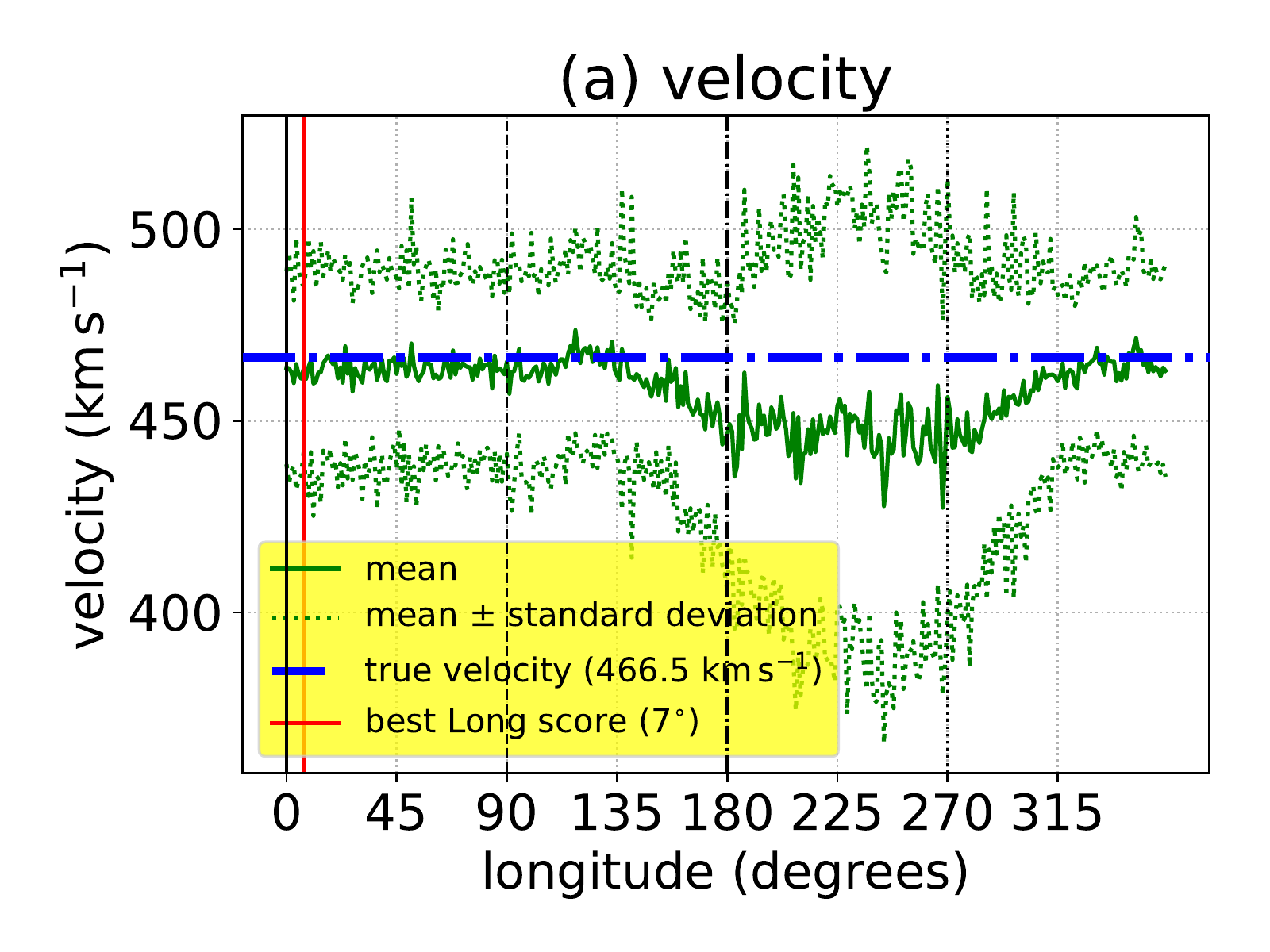}
\includegraphics[width=0.48\textwidth]{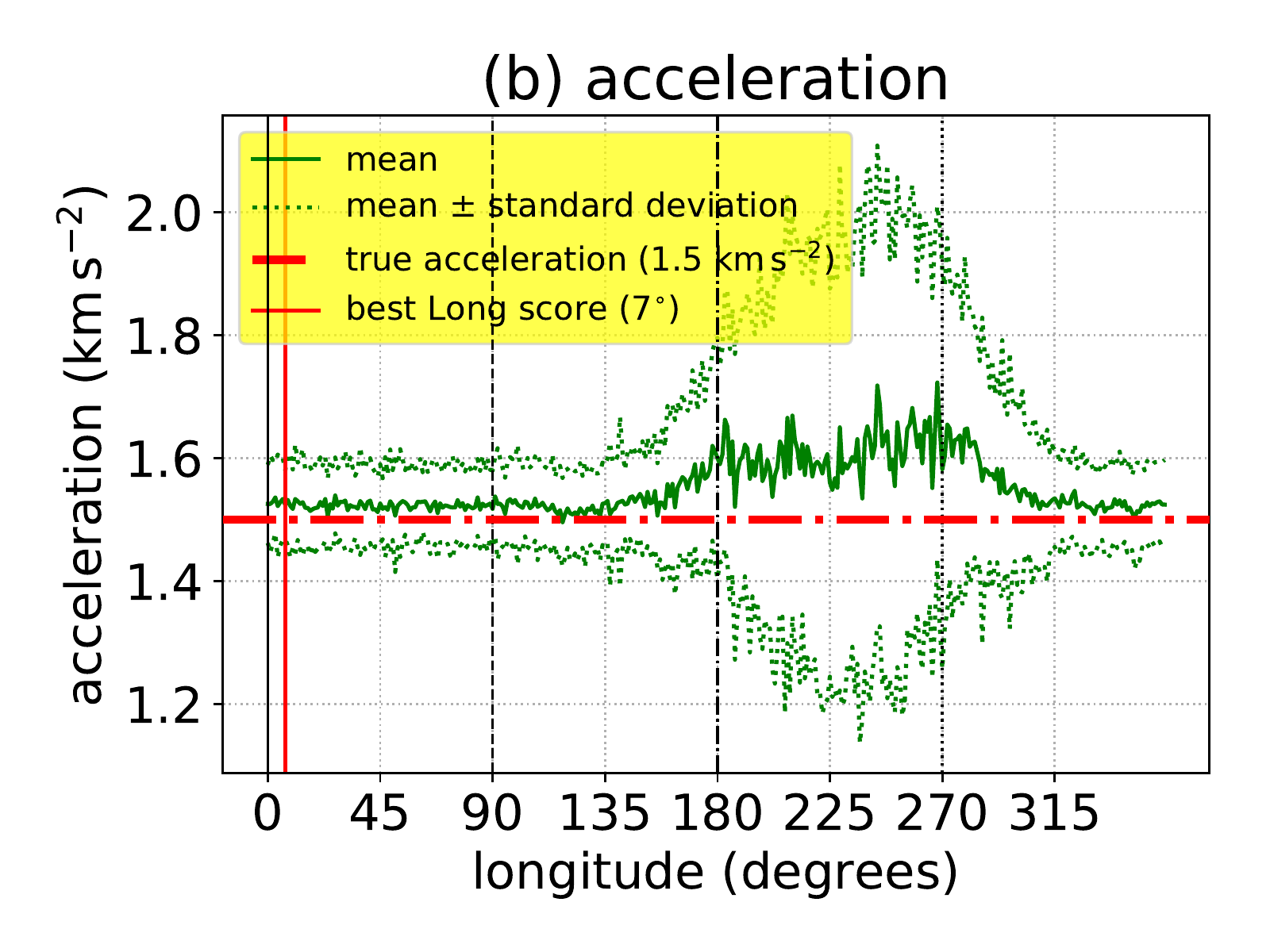}
\caption{(as) Scatter of recovered values of $\vfit$ and  $\afit$ (b) for 100 runs of AWARE on different noisy versions of the simulated wave.}
\label{fig:syntheticwave:statistics}
\end{center}
\end{figure*}

Figure \ref{fig:syntheticwave:statistics} shows the average behavior of AWARE as a function of longitude around the source in recovering the true initial velocity $\vtrue$ and acceleration $\atrue$.  The mean value and standard deviation error bars are shown.  These results are found by running AWARE on 100 noisy realizations of the wave setup described in Section \ref{sss:offcenter}.  Even with the very low signal-to-noise ratio of the simulated wave, the algorithm recovers a range of values that include the $\vinit$ and $\ainit$ values along all arcs, but with a substantial error.  There is a small but systematic offset in the fit value of the acceleration in the range $0-180\Deg$.  These systematic effect is caused by projection effects in creating the simulated data, and in determining which pixels lie on arcs from the initial point.  These systematic effects are more notable in the angular range $180-270\Deg$ and are due to foreshortening making fewer pixels available (Figure \ref{fig:syntheticwave:results}).  However, the range of values indicated by the error bar suggests that the systematic offset is much less than the expected error due to noise.  The error in both $\vfit$ and $\afit$ increases in the angular range $180-270\Deg$, due to the lower number of positions at which AWARE finds a propagating wave.  The average value of $\vfit$ decreases here, and the average value of $\afit$ increases, suggesting that the two quantities are not independent of each other, as was already seen in Figure \ref{fig:syntheticwave:dynamics}a.  This dependence is explored further in Section \ref{sss:fitbias}.

\subsubsection{Velocity-Acceleration Correlation as a Source of Bias}\label{sss:fitbias}
\begin{figure*}
\begin{center}
\includegraphics[width=6.0cm]{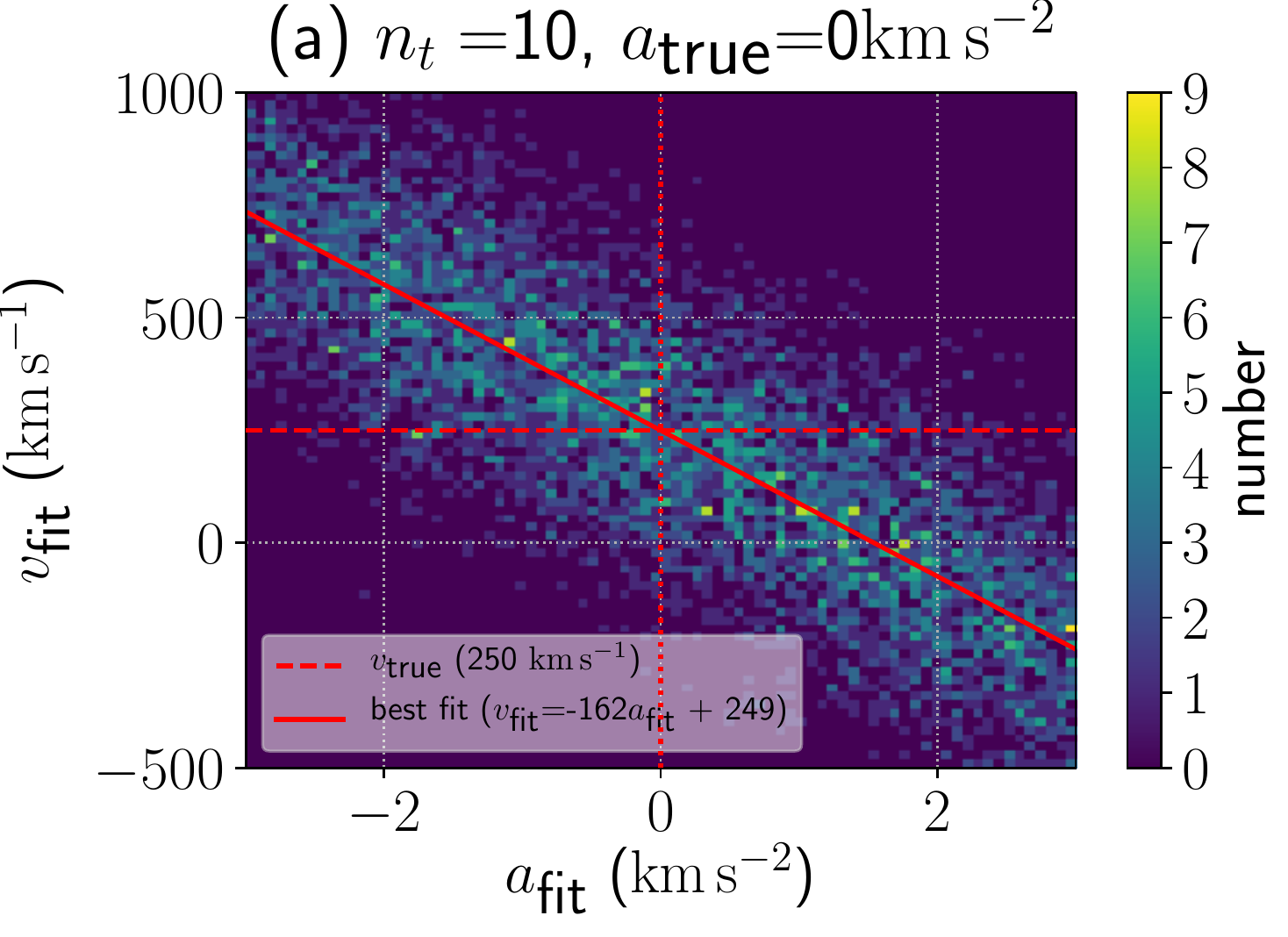}
\includegraphics[width=6.0cm]{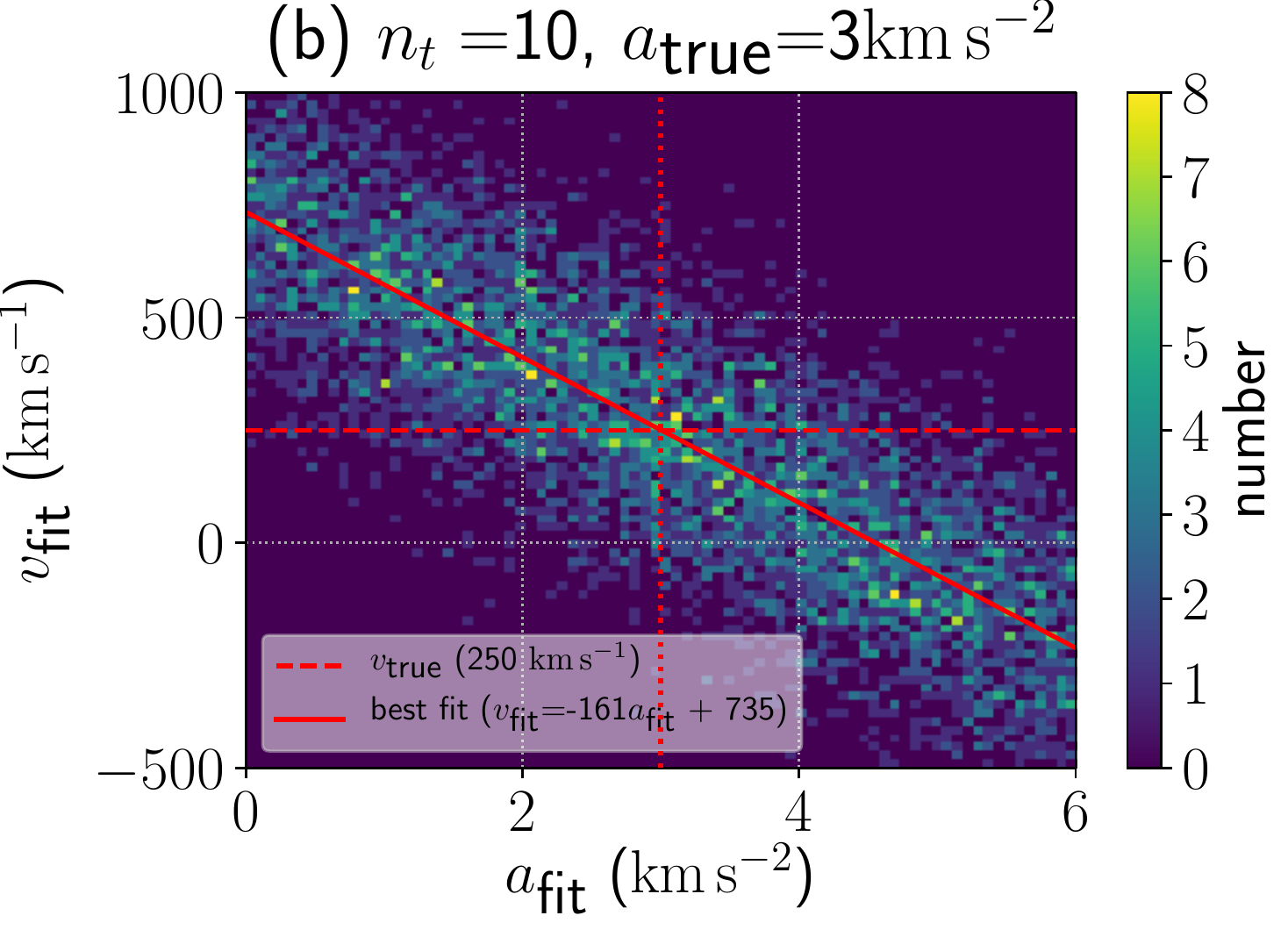}
\includegraphics[width=6.0cm]{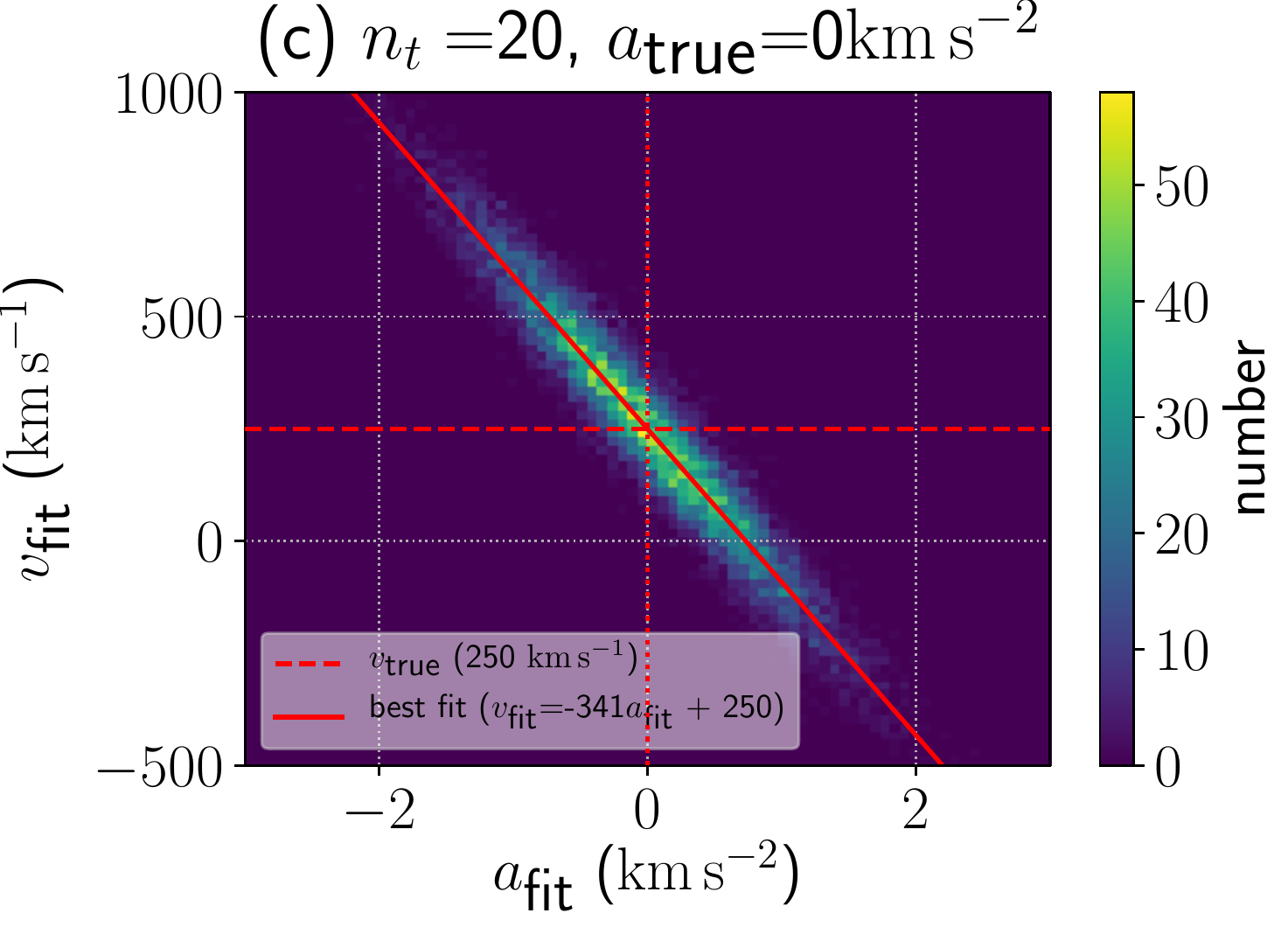}
\includegraphics[width=6.0cm]{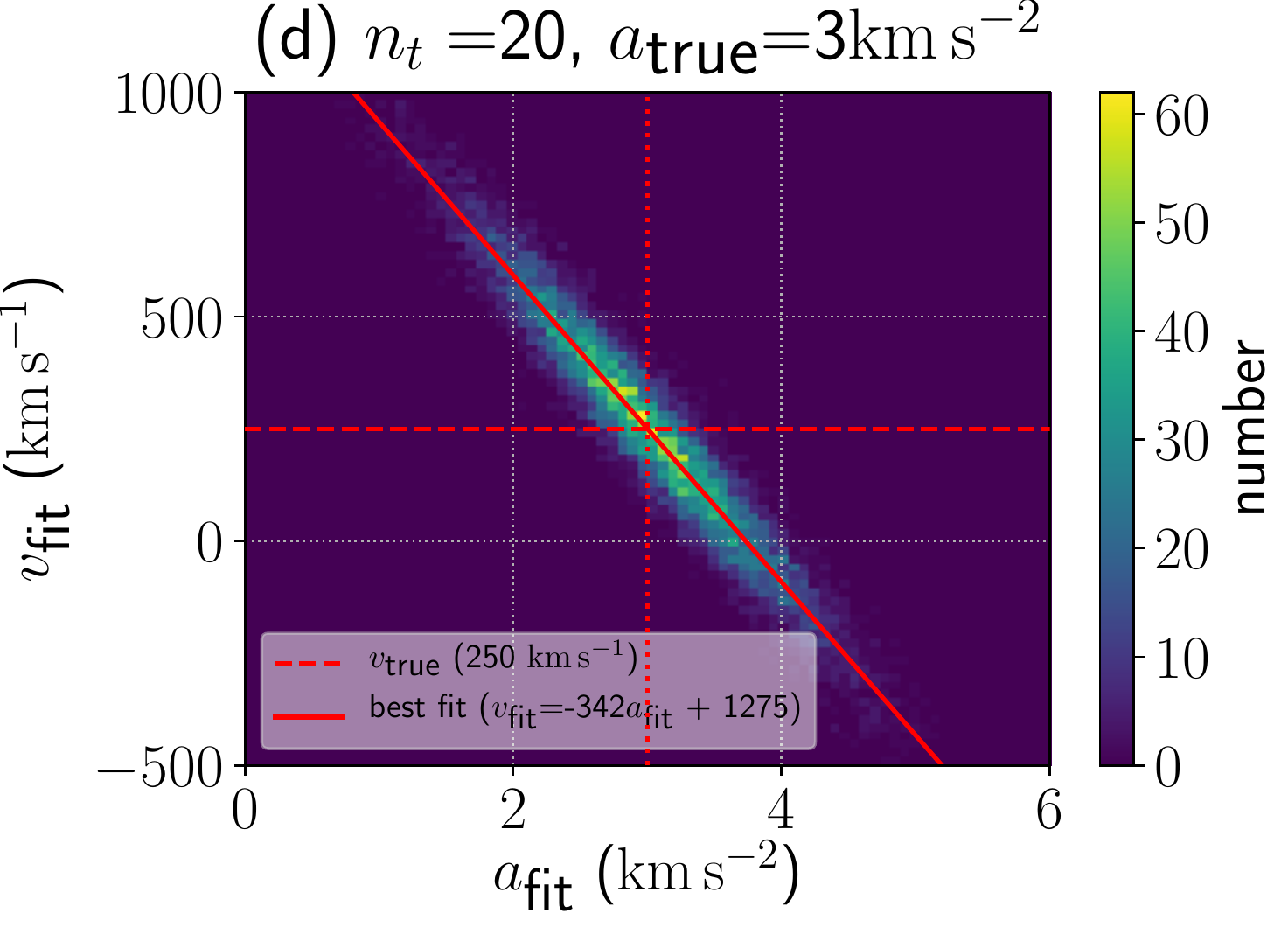}
\includegraphics[width=6.0cm]{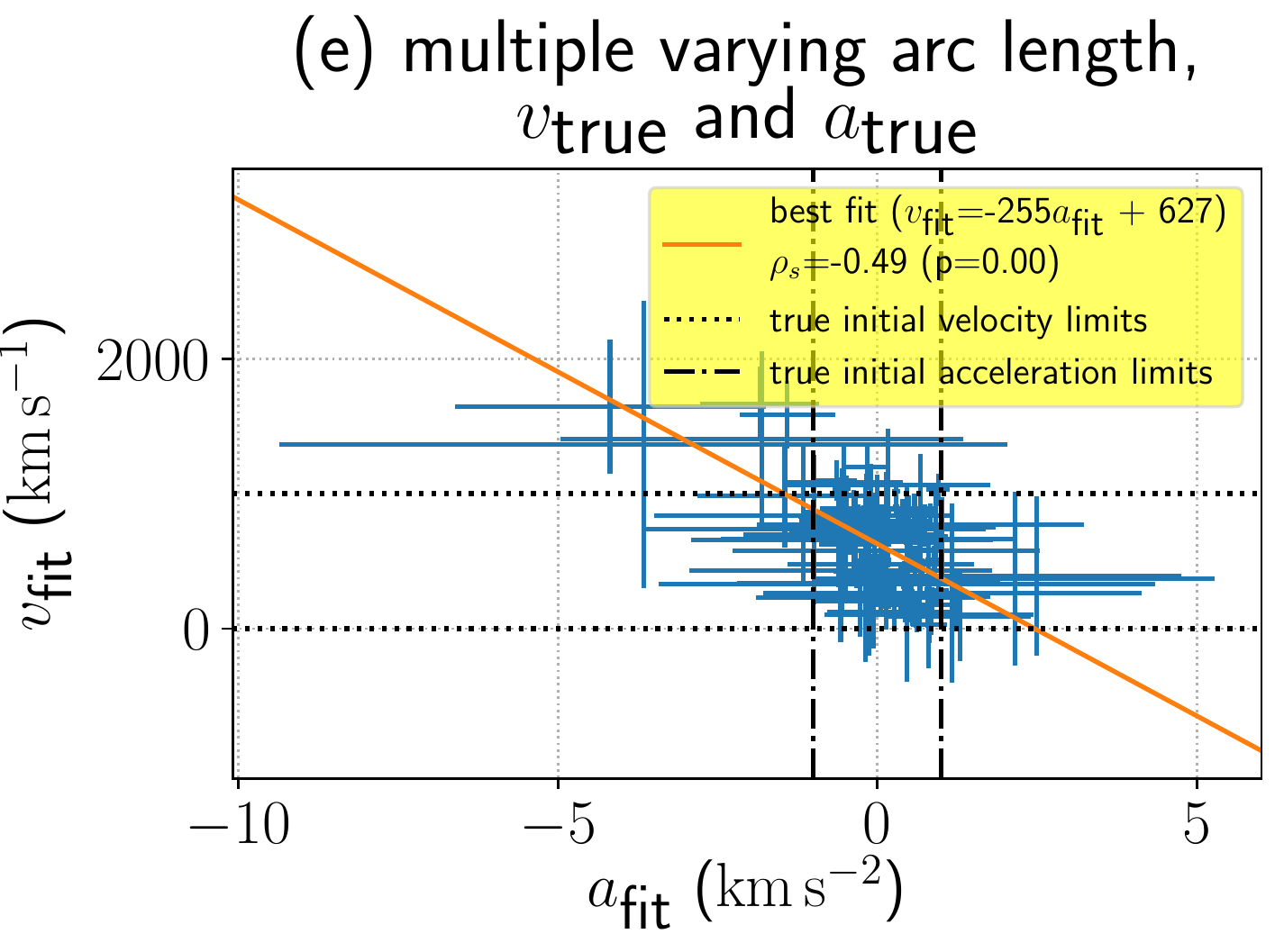}

\caption{(a--d) Shows two dimensional histograms of the $\vfit$ and
  $\afit$ correlation.  The histograms of $\vfit$ and $\afit$ are
  found by fitting a quadratic polynomial to 10000 noisy realizations
  of one arc in a wavefront.  All plots were generated assuming a 5
  degree error in locating the wavefront. (a, c) Shows
  results with $\atrue=0 \accunit$.  (b, d) shows results with
  $\atrue=3 \accunit$.  Comparing (a, b) to (c, d) shows that the
  $\afit, \vfit$ dependency is a function of the number of samples in
  the time-series $n_{t}$ (sample cadence is 36 seconds for all
  plots). It is clear that there is a dependency between $\afit$ and
  $\vfit$ and that the number of samples in the time series varies the
  dependence.  The solid red line in the plot indicates the line of best fit assuming a linear dependence of $\vfit$ on $\afit$. (e) Shows how it is possible to create a correlation even when each arc has a different initial velocity, initial acceleration and number of samples $n_{t}$. See Section \ref{sss:fitbias} for more details.}
\label{f:biased_fitting}
\end{center}
\end{figure*}

The error in determining the position of the wave is a significant source of error in determining the errors in $\vfit$ and $\afit$.  Figure \ref{f:biased_fitting}a--d shows the scatter in the fit initial velocity and acceleration found in 10000 noisy realizations of a test arc.  The plots show that fitting any particular realization of the true wave progress can lead to velocities and accelerations that are very distant from $\vtrue$ and $\atrue$.  It is easy to see how the correlation between the fit velocity and acceleration arises by considering Equation \ref{e:quad} and the fitting process.  The fitting process calculates
\begin{equation}
D=\min
\sum_{i=1}^{n_{t}}
\left[
\frac{s(t_{i}) - p(t_{i})}{e(t_{i})}
\right]^{2}
\label{e:min}
\end{equation}
by varying the values of $a_{0}$ and $v_{0}$.  If the fit finds a value of $a_{0}$ that is a higher than the true value then a lower value of $v_{0}$ is preferred for the minimization.  Similarly, if the fit finds a value of $a_{0}$ that is lower than the true value then a higher value of $v_{0}$ is sought by the minimization routine.  This leads to correlated values of $\afit$ and $\vfit$.

It is also notable that the histograms shown in the top and bottom rows of Figure \ref{f:biased_fitting} illustrate different scattering behavior around the true values.  This is due to the number of samples in each time series.  For smaller numbers of samples, there is less information to constrain fit values, and so the scatter about the true value increases.  Increasing the number of samples not only reduces the scatter but changes the $\afit$, $\vfit$ correlation.  With more information, a small change in the value of $a_{0}$ means that the fit requires a bigger change in the value of $v_{0}$ in order to reach a minimum.  
This can be understood by considering Equation \ref{e:min} for two different fits.  Assume there is a fit $s(t) = S + Vt + At^{2}/2$ giving a value $D$; now perturb that fit and consider $s'(t) = (S + s) + (V + v)t + (A + a)t^{2}/2$ (for simplicity we also assume that $e(t_{i})$ is 1 for all $t_{i}$) giving a value $D'$.  If we assume that $D\approx D'$ (both fits give approximately the same value), and consider each summand in Equation \ref{e:min} then
\begin{eqnarray}\label{e:bias}
v
&\approx
&
\frac{ -at^{4} \left(2A + a\right)/4 - Vat^{3}  + t^{2} \left(As + Sa + as + v^{2} - ay\right) - 2Vst - 2Ss - s^{2} + 2sy}{t^{3} \left(A + a\right) + 2 V t^{2} + 2t \left(S + s - y\right)}
\nonumber\\
& \rightarrow & -\frac{a}{4}\left(\frac{2A + a}{A + a}\right)t
\end{eqnarray}
as $t\rightarrow\infty$.  Hence as larger times are considered, small changes in the acceleration inevitably lead to larger changes in the velocity to achieve the same quality of fit.  Equation \ref{e:bias} also explains the negative correlation between the fit velocity and acceleration.

Figure \ref{f:biased_fitting}e is generated by creating 200 simulated wavefront propagation profiles with random properties and using the fit process described in Section \ref{sec:aware:dynamics} to derive values of $\vfit$ and $\afit$.  The number of samples in each profile is drawn from $U_{integer}(10, 60)$\footnote{$U_{integer}(x, y)$ is the uniform probability distribution for integers between $x, y$ and $U(x, y)$ is the uniform probability distribution for real numbers between $x, y$.} , and the true initial velocity is drawn from $U(0, 1000)$and the true initial acceleration is drawn from $U(-1, 1)$.  Plotting $\vfit$ against $\afit$ shows an apparent correlation, even although the true values are uncorrelated since they are drawn from independent uniform distributions.  This complicates the discussion of \cite{2017SoPh..292..185L} as to if there is true, physical correlation between the initial velocity and the acceleration.  One way of estimating if there is a true physical correlation is to fit multiple noisy realizations of the original data to generate a probability distribution $P(\rho)$ of correlation coefficients and then compare that distribution to the correlation coefficient $\rho_{o}$ derived from the original velocity \VERSUS\ acceleration scatter.  If the complementary cumulative probability value $1-\int_{-\infty}^{\rho_{o}}P(\rho')d\rho'$ is close to zero then it is very unlikely that the correlation value $\rho_{o}$ arose by chance, lending support that the initial velocity and acceleration are indeed correlated.

\subsubsection{Detection of Accelerating Versus Non-Accelerating Wavefronts}\label{sss:bic}
\begin{figure*}
\begin{center}
\includegraphics[width=6.0cm]{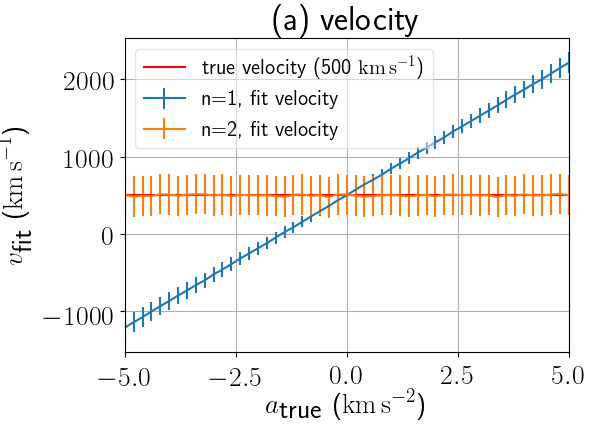}
\includegraphics[width=6.0cm]{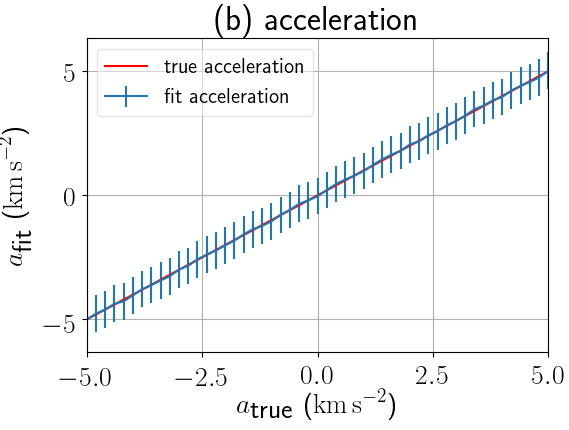}
\includegraphics[width=6.0cm]{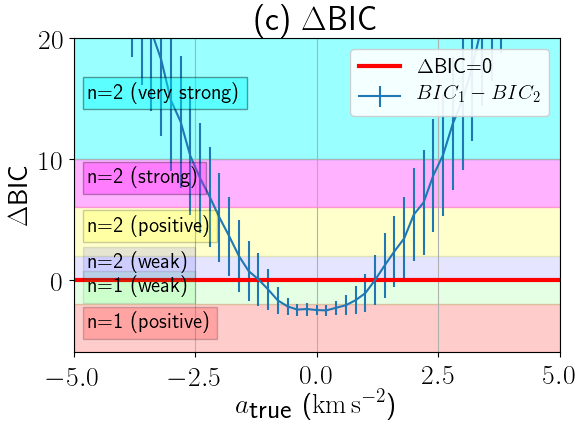}
\caption{The effect of the degree of the polynomial used to fit a
  simulated wave arc as a function of the wave acceleration.  The
  x-axis indicates the true value of the acceleration in the simulated
  wavefront. (a) Shows the median and median absolute deviation
  velocity derived from fitting either a $n=1$ or an $n=2$ polynomial.
  (b) Shows the  median and median absolute deviation acceleration
  derived from fitting a $n=2$ polynomial (the red true acceleration
  line is fully covered by the blue line indicating the value derived
  from fitting).  (c) Shows the median and median absolute deviation
  difference in the Bayesian information criterion (BIC) calculated from each polynomial fit.  Positive values indicate that the BIC correctly prefers the $n=2$ polynomial fit containing an acceleration term.  (note that BIC$_{1}$ refers to the BIC value for the $n=1$ model, and BIC$_{2}$ refers to the BIC $n=2$ value model). The degree of preference for each model ($n=1$ or $n=2$) is indicated on the left hand side of the plot \citep{kassr95}.  The simulation assumed that the position of the wavefront was measured at 20 times, with a cadence of 36 seconds between each measurement.  The wave position had a normally distributed error of 5 degrees of arc. The propagation of the wavefront along the arc was simulated 1000 times to generate the results.}
\label{fig:acc_versus_non_acc}
\end{center}
\end{figure*}
One test for the existence of accelerating wavefronts is to fit linear ($n=1$) or quadratic ($n=2$) functions to the progress of the wave along each arc, and then decide which fit is a reasonable explanation for the data.  For illustrative purposes, \rtr{we use the Bayesian information criterion (BIC) \citep{schwarz1978} to perform model selection, in this case, to decide between the linear and quadratic fits to the arc progress}.  The BIC is defined as
\begin{equation}
BIC = -2\ln(L) + k\ln(N),
\label{eqn:bic}
\end{equation}
where $L$ is the likelihood function evaluated at its maximum, $k$ is the number of parameters in the fitting function, and $N$ is the number of values fit.  Since we are assuming that the location of the wavefront is Gaussian-noisily distributed, maximizing the likelihood function is equivalent to the least squares fit used in AWARE to fit either linear or quadratic functions to the data.  The model with the lowest BIC is the preferred model.

Figure \ref{fig:acc_versus_non_acc}a shows the fit initial velocity
$\vfit$ derived from fitting the simulated wavefront using linear and
quadratic functions.  As expected, the linear function performs poorly
except close to $\atrue=0$.  Away from $\atrue=0$, the linear function
is attempting to compensate for the accelerating progress of the wave
by changing the initial velocity, leading to initial velocities with
magnitudes much larger than the true value of $\vinit$.  Figure
\ref{fig:acc_versus_non_acc}b shows that the quadratic fit recovers
the true acceleration with a relatively small error.  Figure
\ref{fig:acc_versus_non_acc} c shows the difference in the BIC for
each model.  Positive values indicate a preference for the correct
model $n=2$ which includes an acceleration.  Negative values indicate
a preference for the linear model.  The degree of preference is also
indicated by the different colored regions, following the
classification of  \cite{kassr95}.  The plot shows that the
accelerating model is not strongly preferred until the magnitude of
the acceleration is around $2 \accunit$.  Therefore at lower magnitude accelerations, fits with and without accelerations are not easily distinguishable.

\subsection{Application to Observational Data}
\label{sec:obs}
\begin{figure*}
\begin{center}
\includegraphics[width=6.0cm]{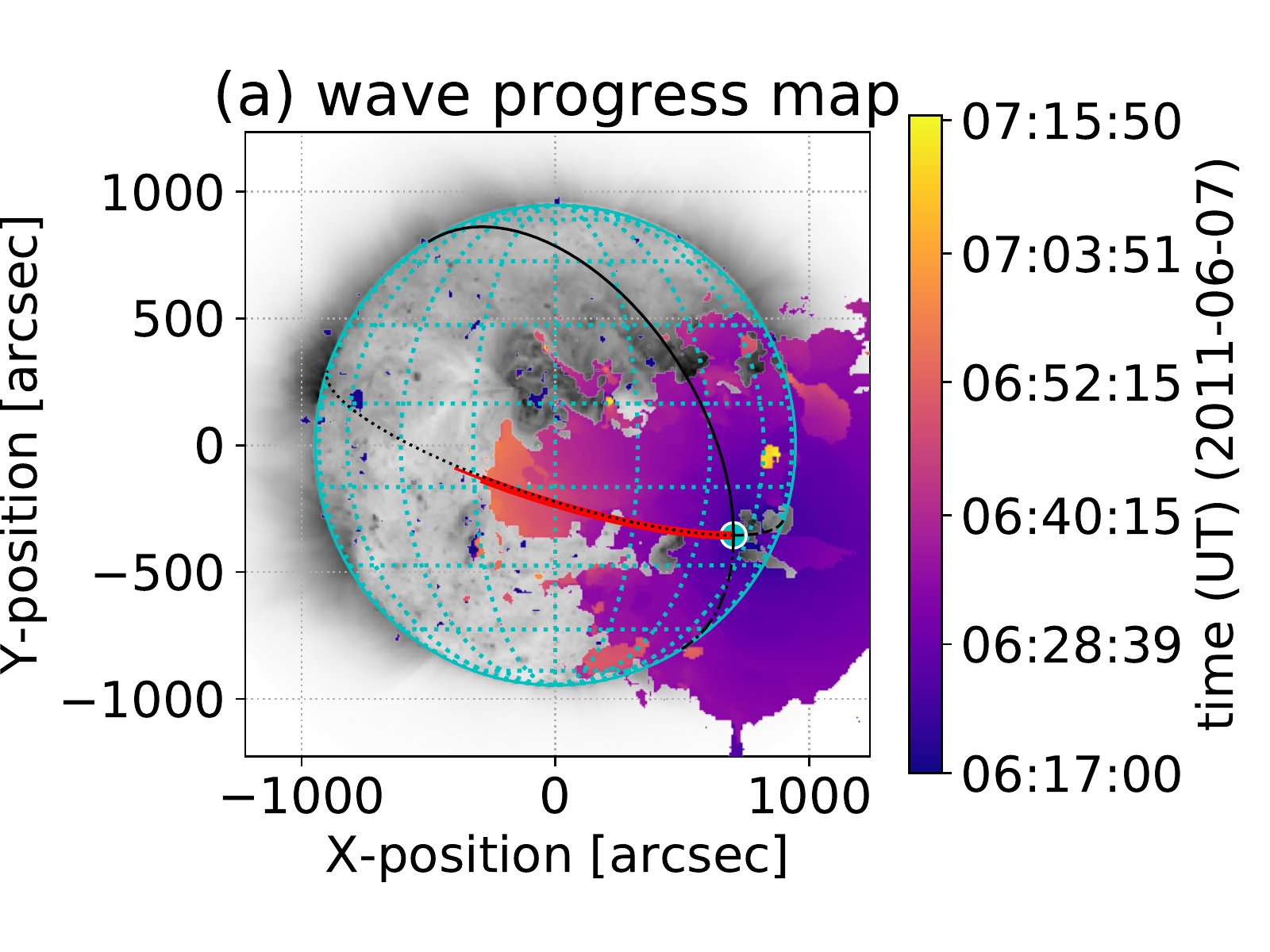}
\includegraphics[width=6.0cm]{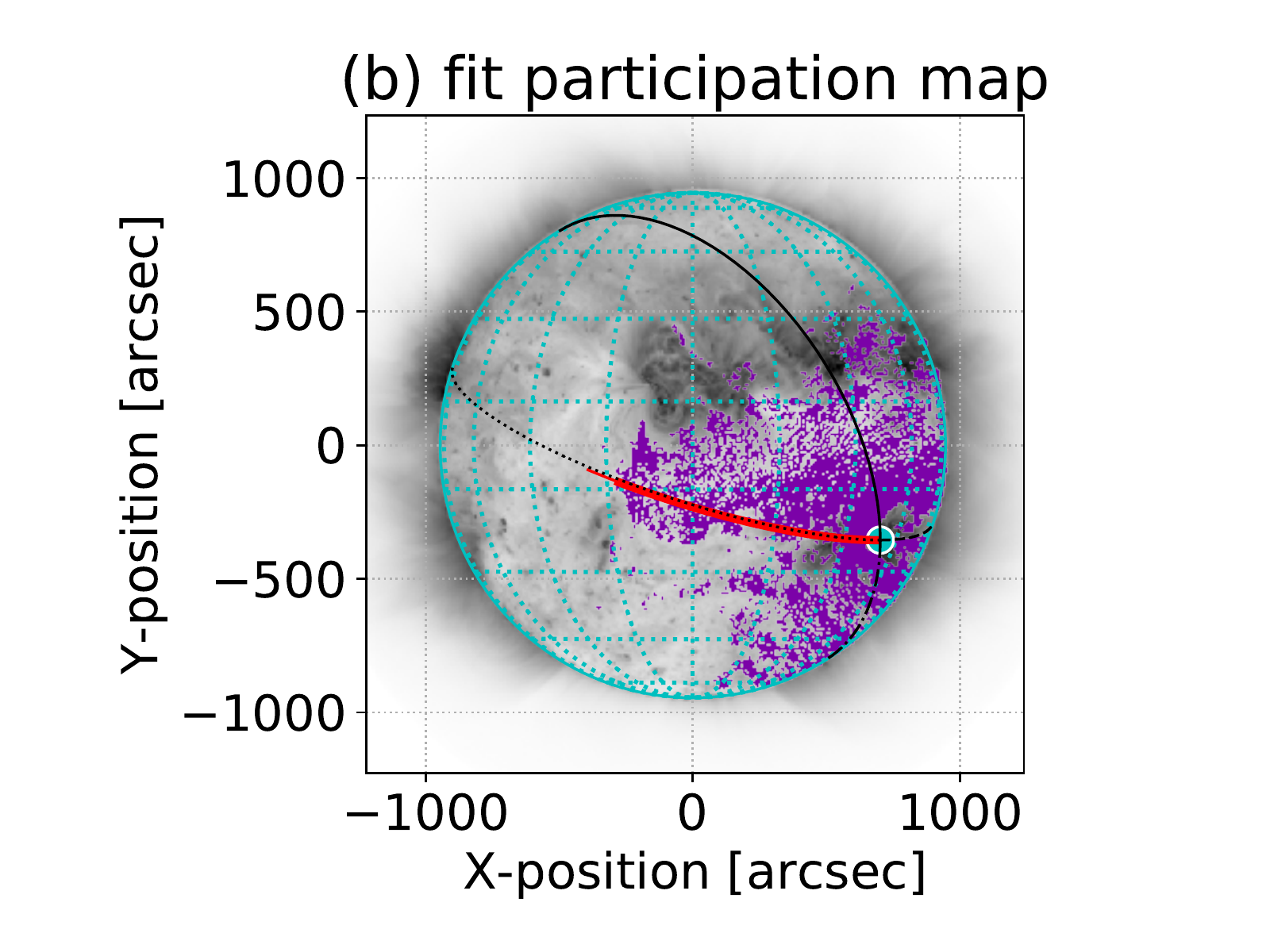}
\includegraphics[width=6.0cm]{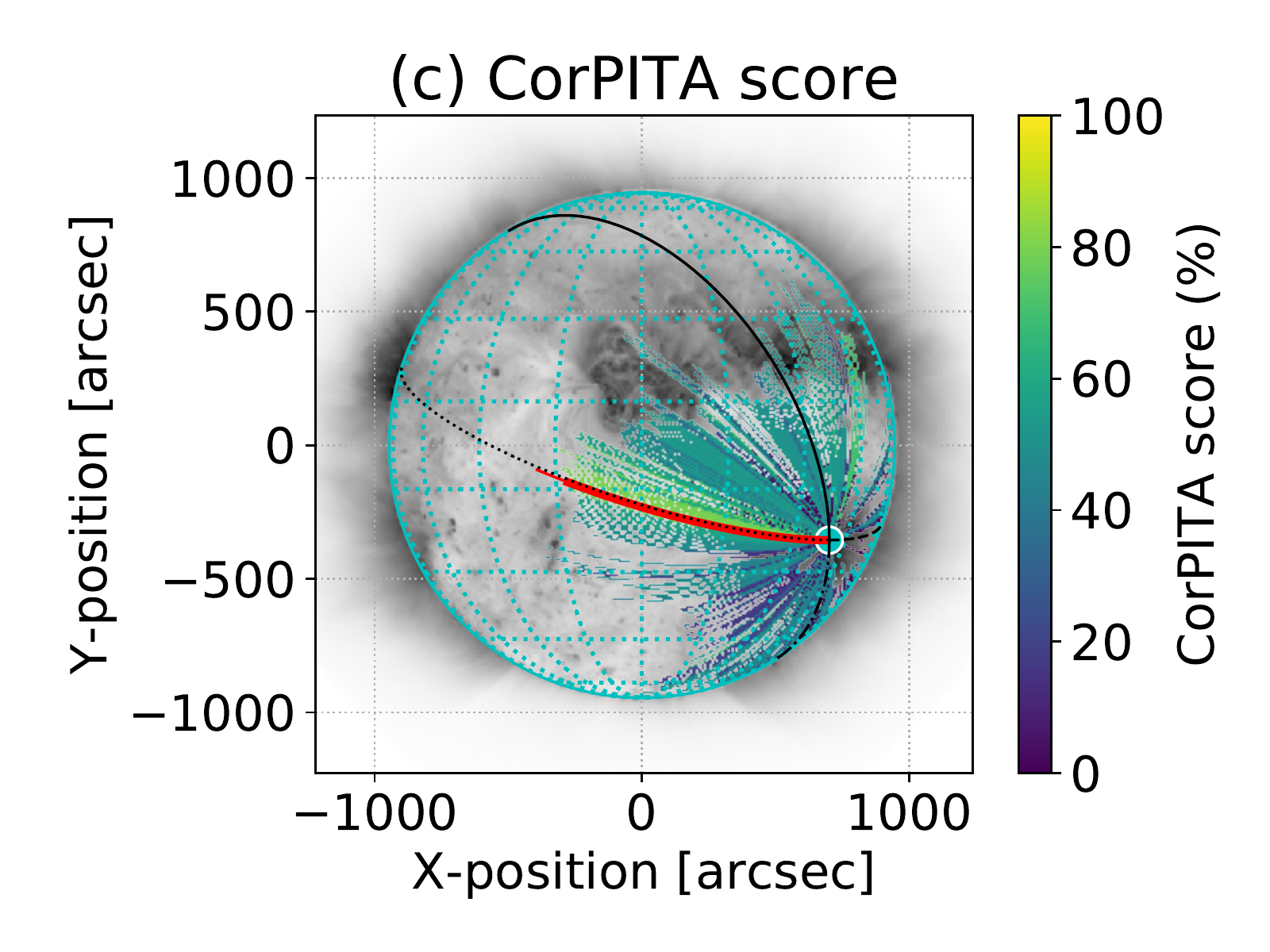}
\includegraphics[width=6.0cm]{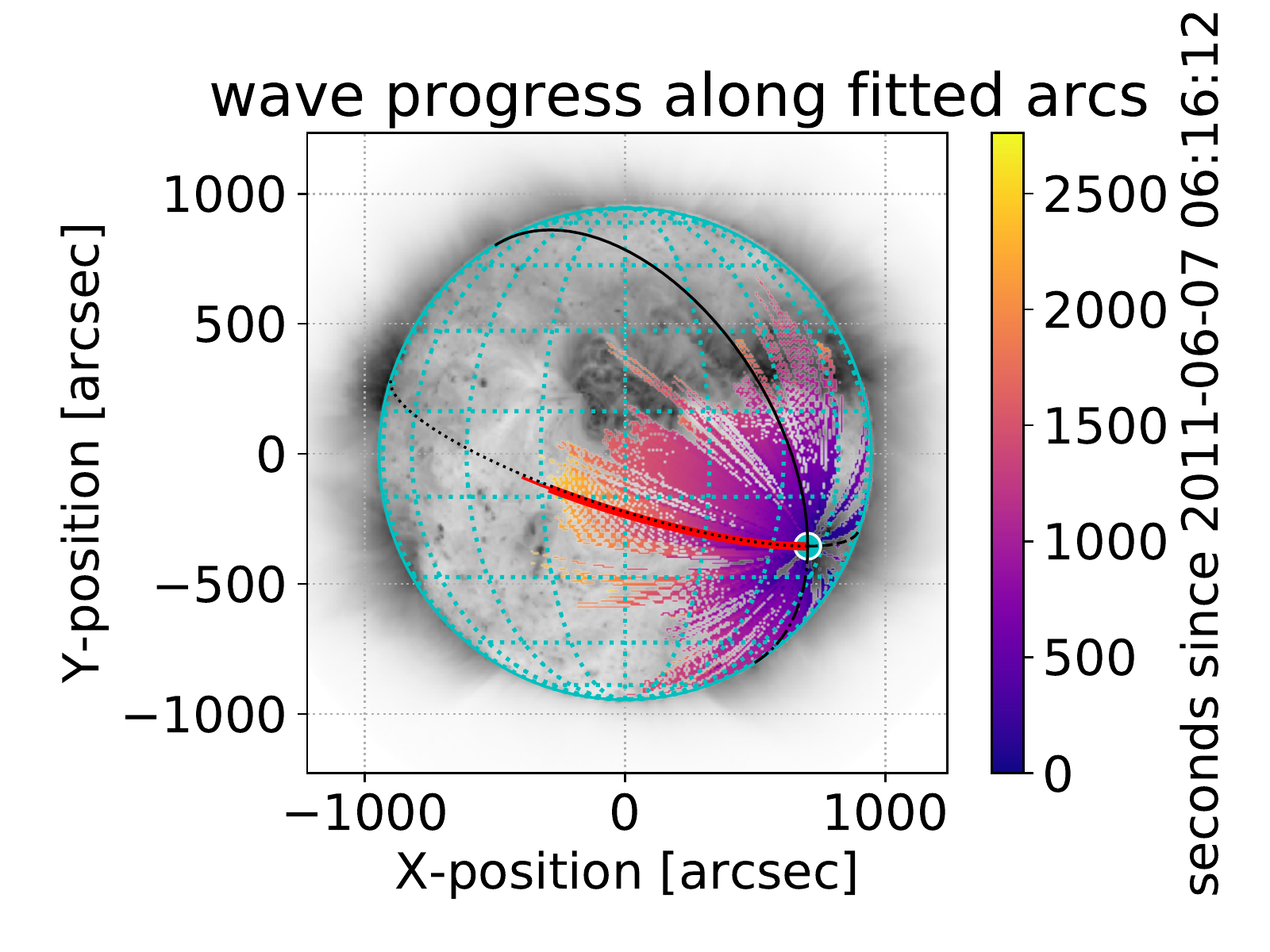}
\includegraphics[width=6.0cm]{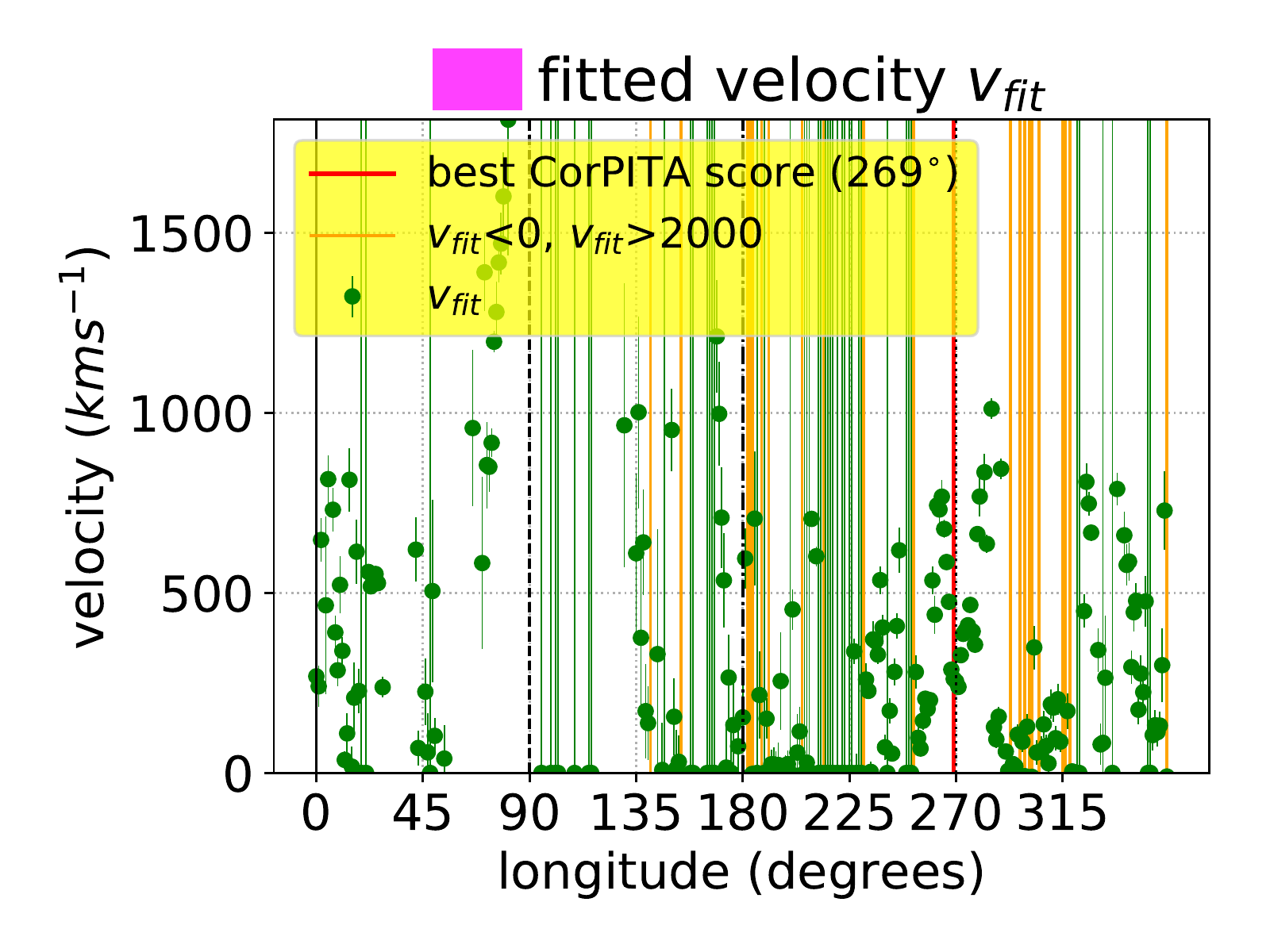}
\includegraphics[width=6.0cm]{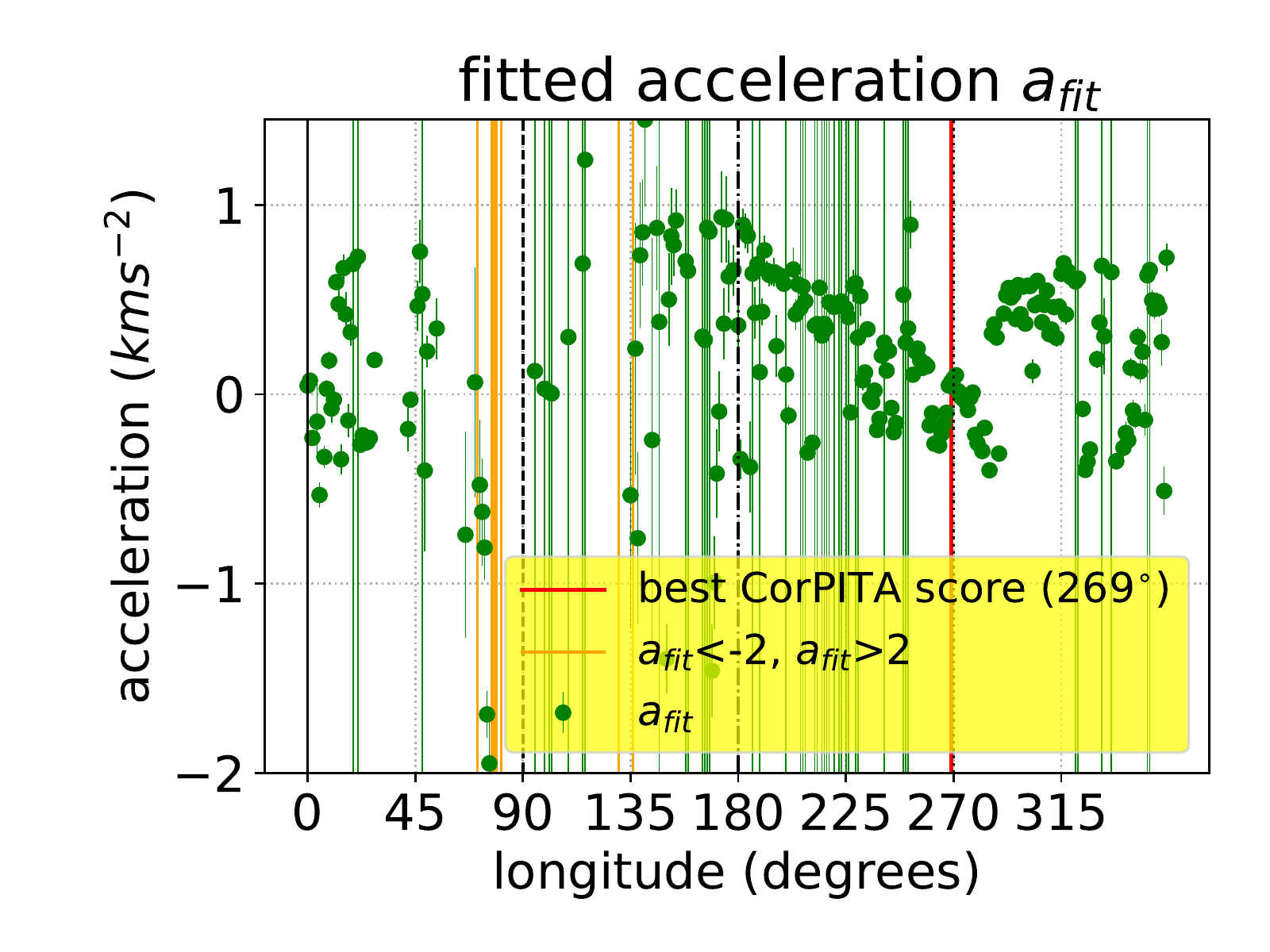}
\includegraphics[width=6.0cm]{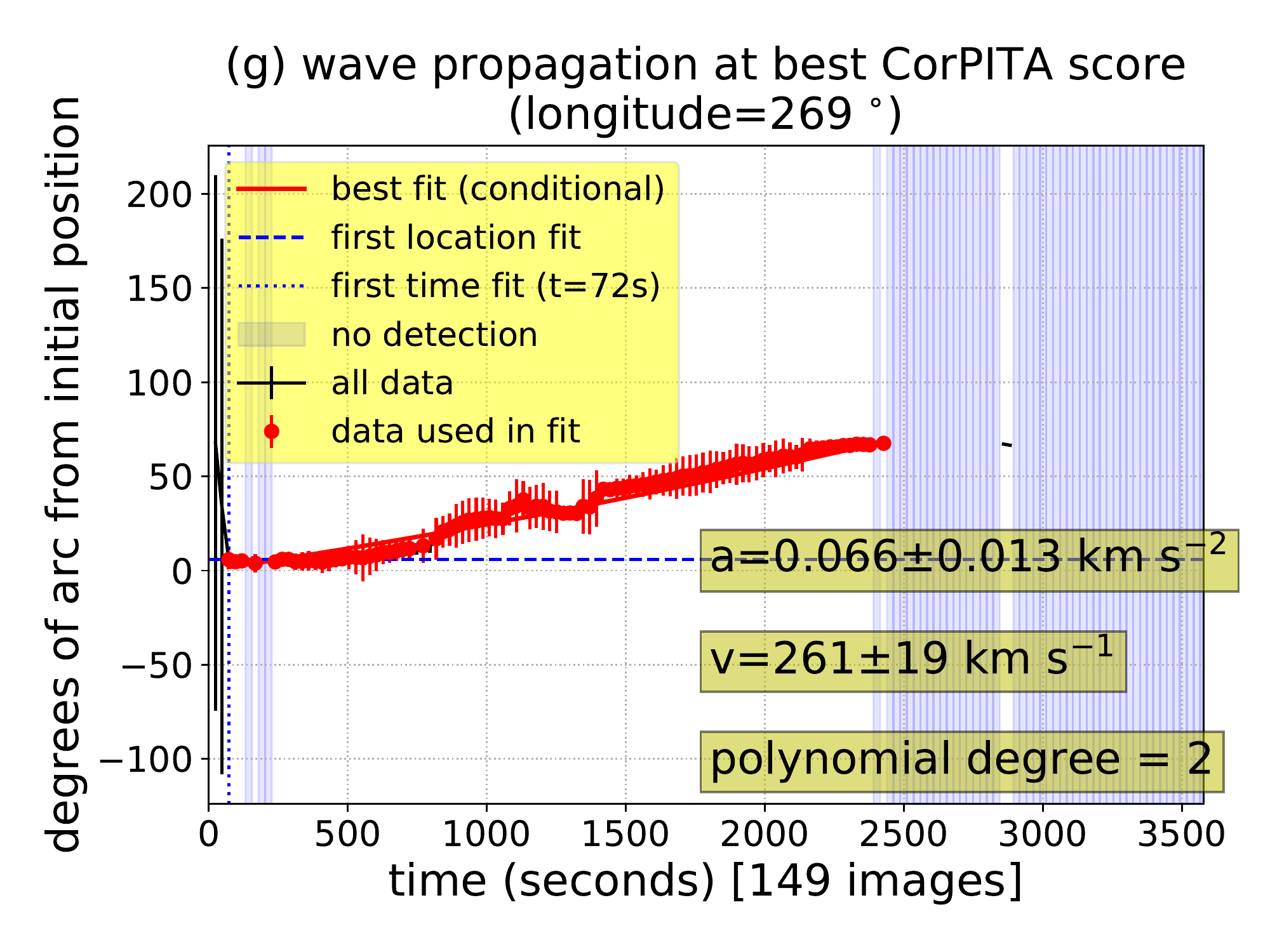}
\caption{AWARE performance for the EUV wave of 7 June 2011. See the caption of Figure \ref{fig:syntheticwave:results} for more details.}
\label{fig:longetal4}
\end{center}
\end{figure*}

The AWARE algorithm is applied to datasets derived from the results
shown in the CorPITA article \citep{2014SoPh..289.3279L}, the EUV waves
of 7 June 2011 (Figure \ref{fig:longetal4}, 13 February 2011 (Figure
\ref{fig:longetal7}), 15 February 2011 (Figure \ref{fig:longetal8a}), and
16 February 2011 (Figure \ref{fig:longetal8e}).  The algorithm is also
applied to the same time range analyzed by \citep{2014SoPh..289.3279L}
from 8 February 2011 (Figure \ref{fig:longetal6}) to illustrate a null
result.  In all cases, full spatial resolution FITS level 1.0 AIA 211 \AA\ image data at 12 second cadence \rtr{in the hour following the initiating eruptive event are downloaded and analyzed}.

The wave progress maps (Figures \ref{fig:longetal4}a, \ref{fig:longetal6}a, \ref{fig:longetal7}a, \ref{fig:longetal8a}a, and \ref{fig:longetal8e}a) illustrate the initial identification of the wave location.  It is notable that the algorithm captures changes that are quite distant from the wave source at earlier times.  These are clearly unphysical, as the wave cannot have reached those locations so soon after its initiation.  These initial wave progress locations are obtained from stage 1 of the algorithm.  At these earlier times, small changes at distant locations are captured by the RDP image processing and expanded into contiguous areas by the morphological processing.  The effect of this is mitigated somewhat by stage 2 of the algorithm (fitting a profile to the wave propagation along radial arcs beginning at the initiation point) where the mean location of the wavefront and its error often lead to the elimination of earlier times from consideration in the final fit.

\begin{figure*}
\begin{center}
\includegraphics[width=12.0cm]{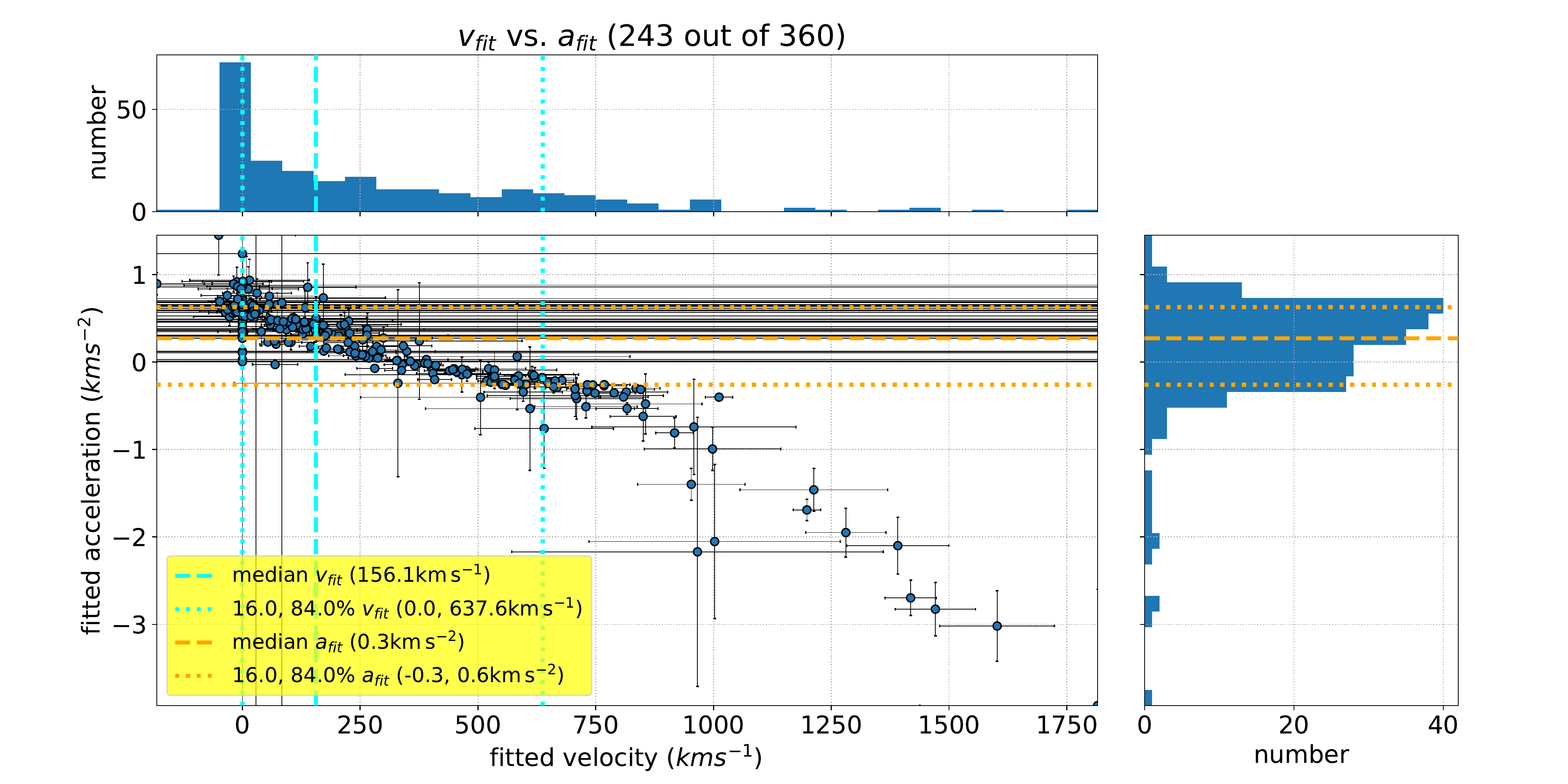}
\includegraphics[width=12.0cm]{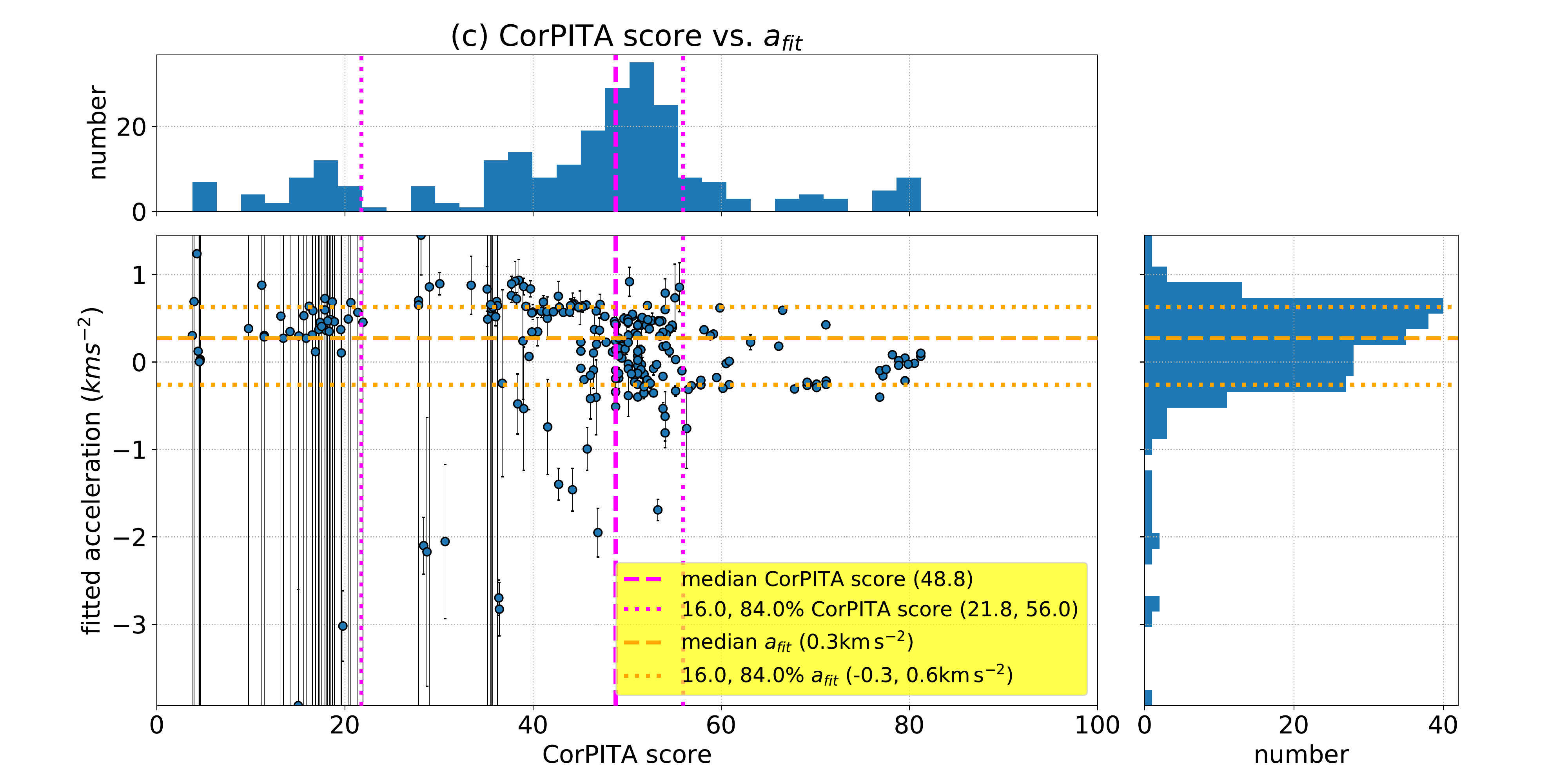}
\includegraphics[width=12.0cm]{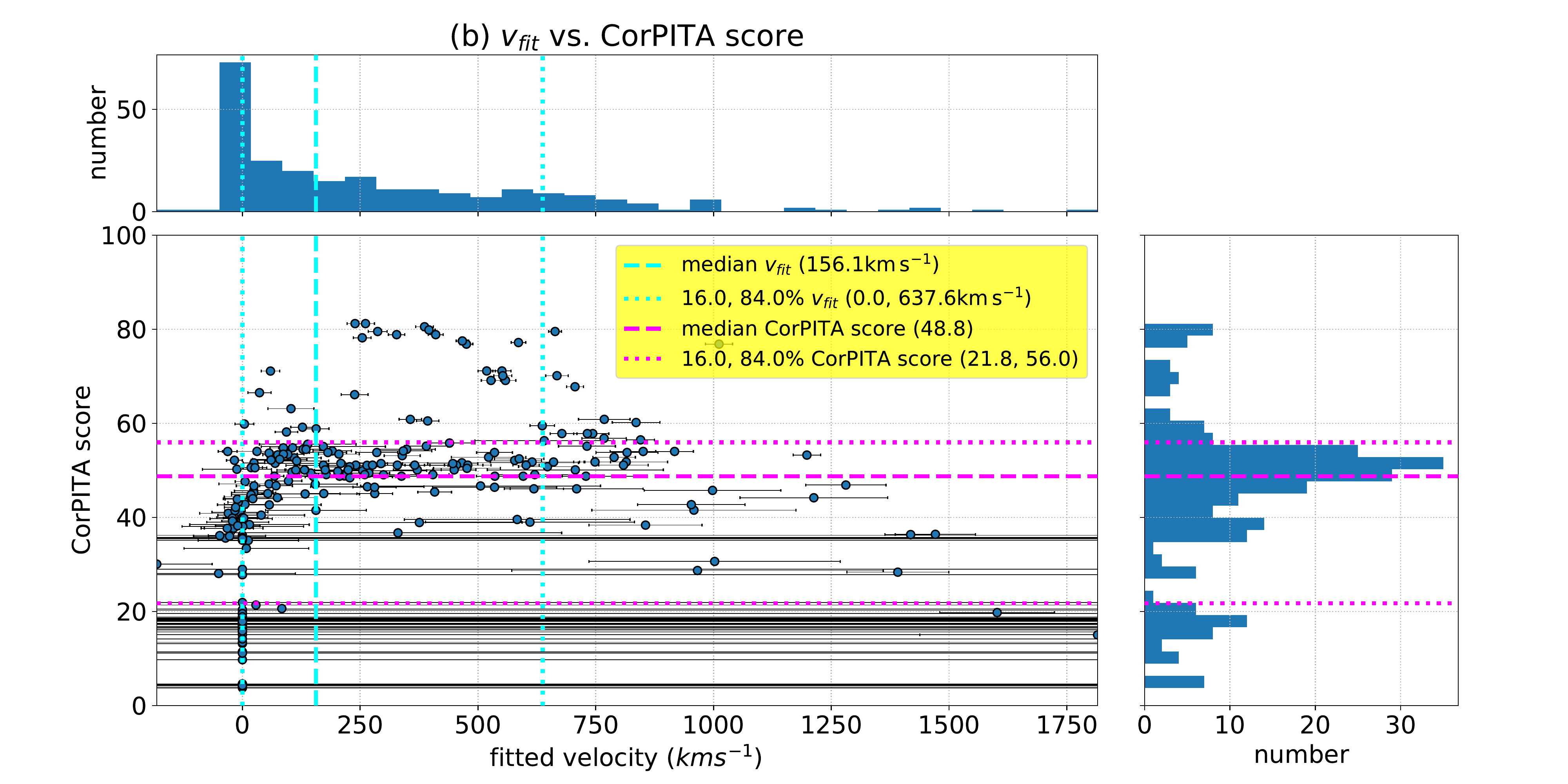}
\caption{Distributions of $\vfit$, $\afit$, and the \longscore\ for the fitted arcs of the 7 June 2011 EUV wave. See the caption of Figure \ref{fig:syntheticwave:dynamics} for more details.}
\label{fig:longetal4dynamics}
\end{center}
\end{figure*}

\begin{figure*}
\begin{center}
\includegraphics[width=6.0cm]{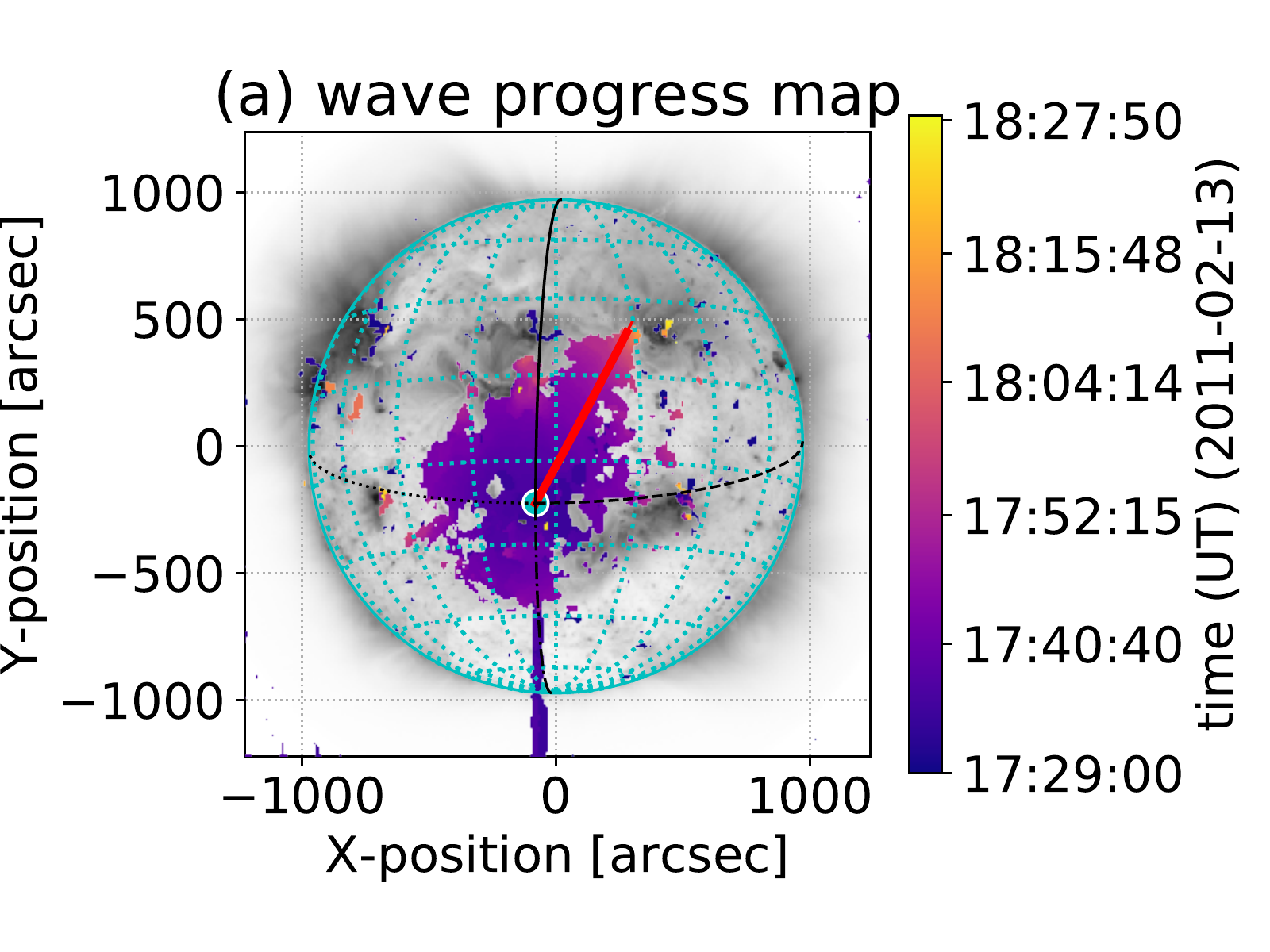}
\includegraphics[width=6.0cm]{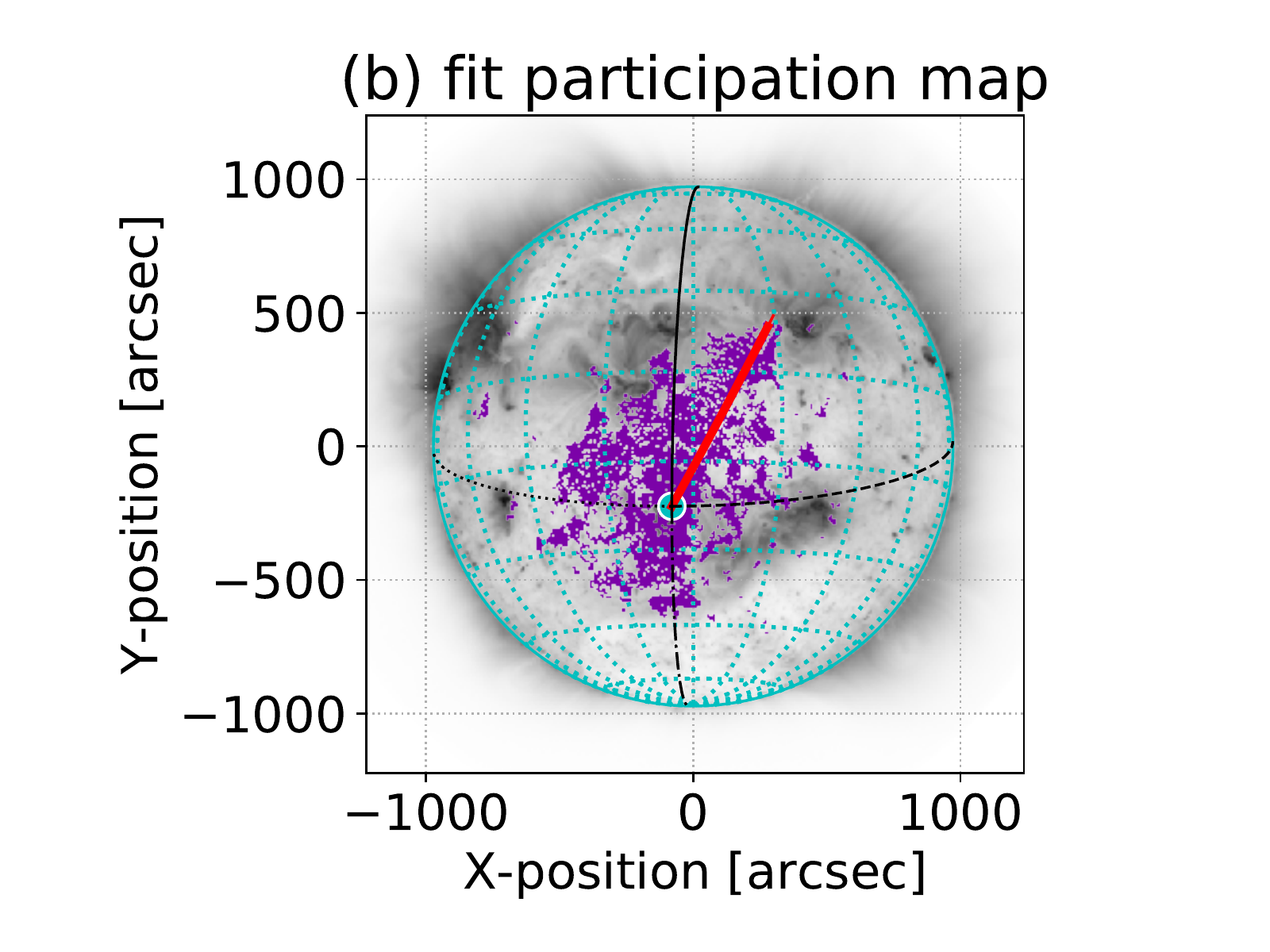}
\includegraphics[width=6.0cm]{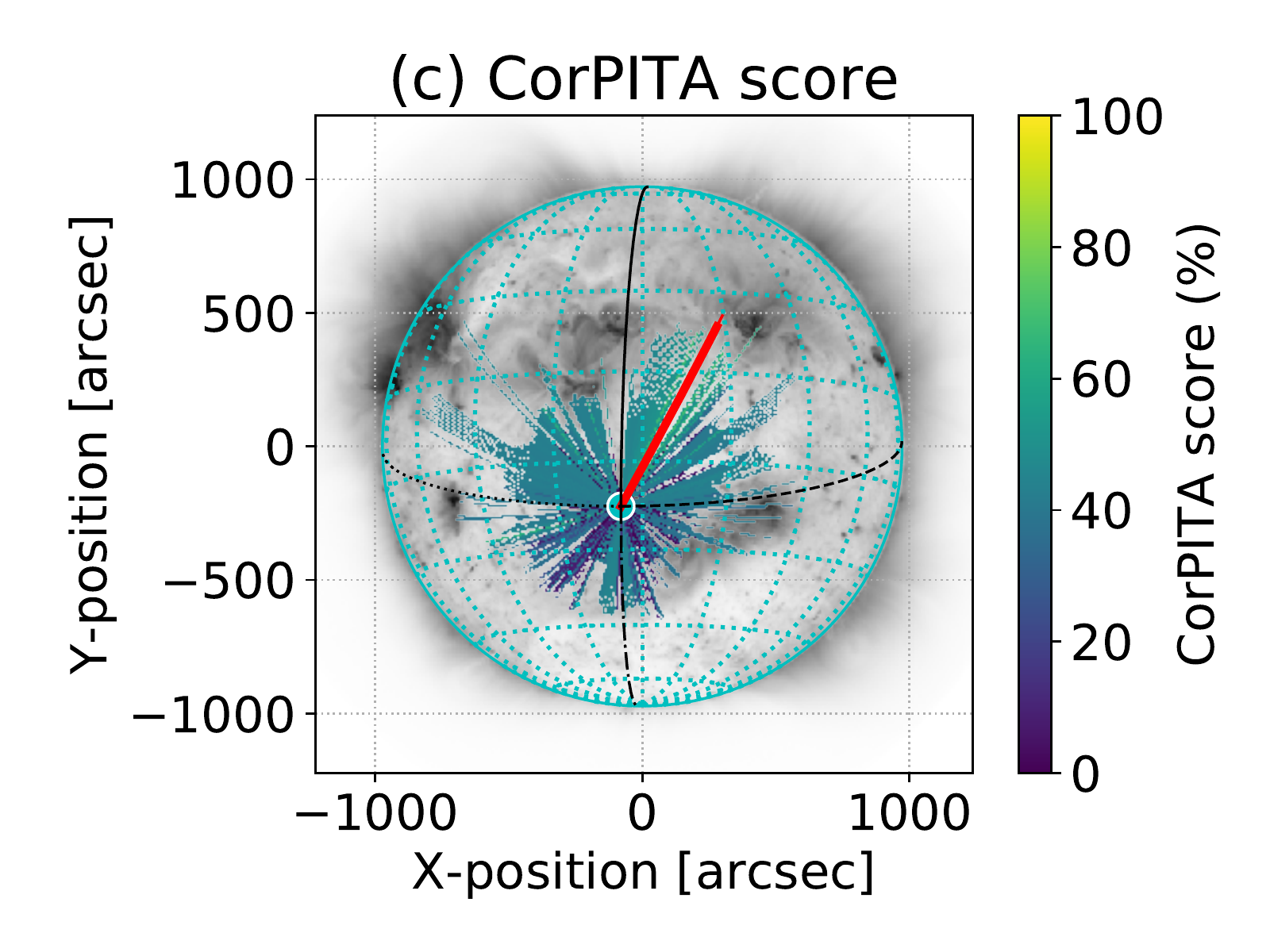}
\includegraphics[width=6.0cm]{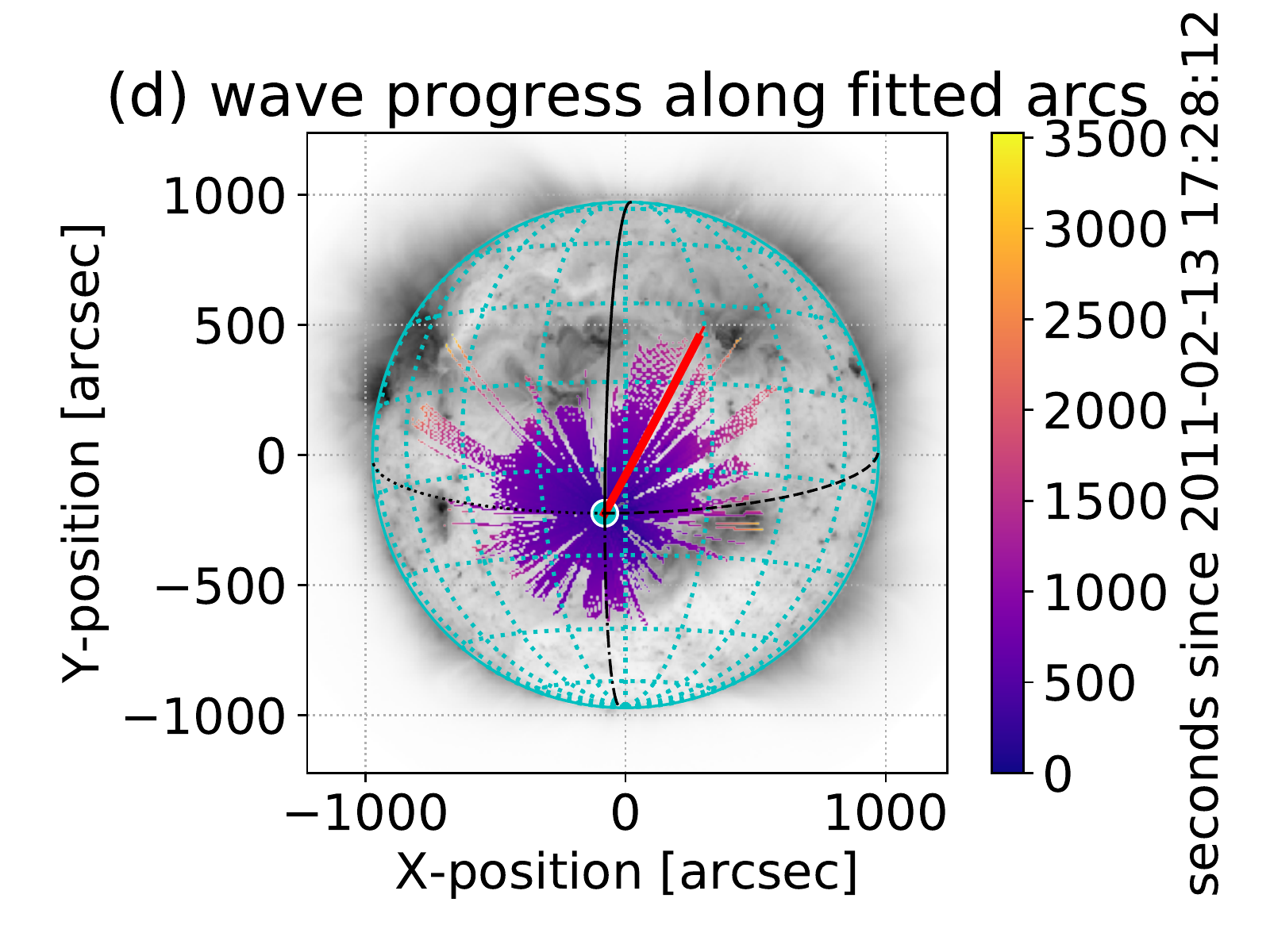}
\includegraphics[width=6.0cm]{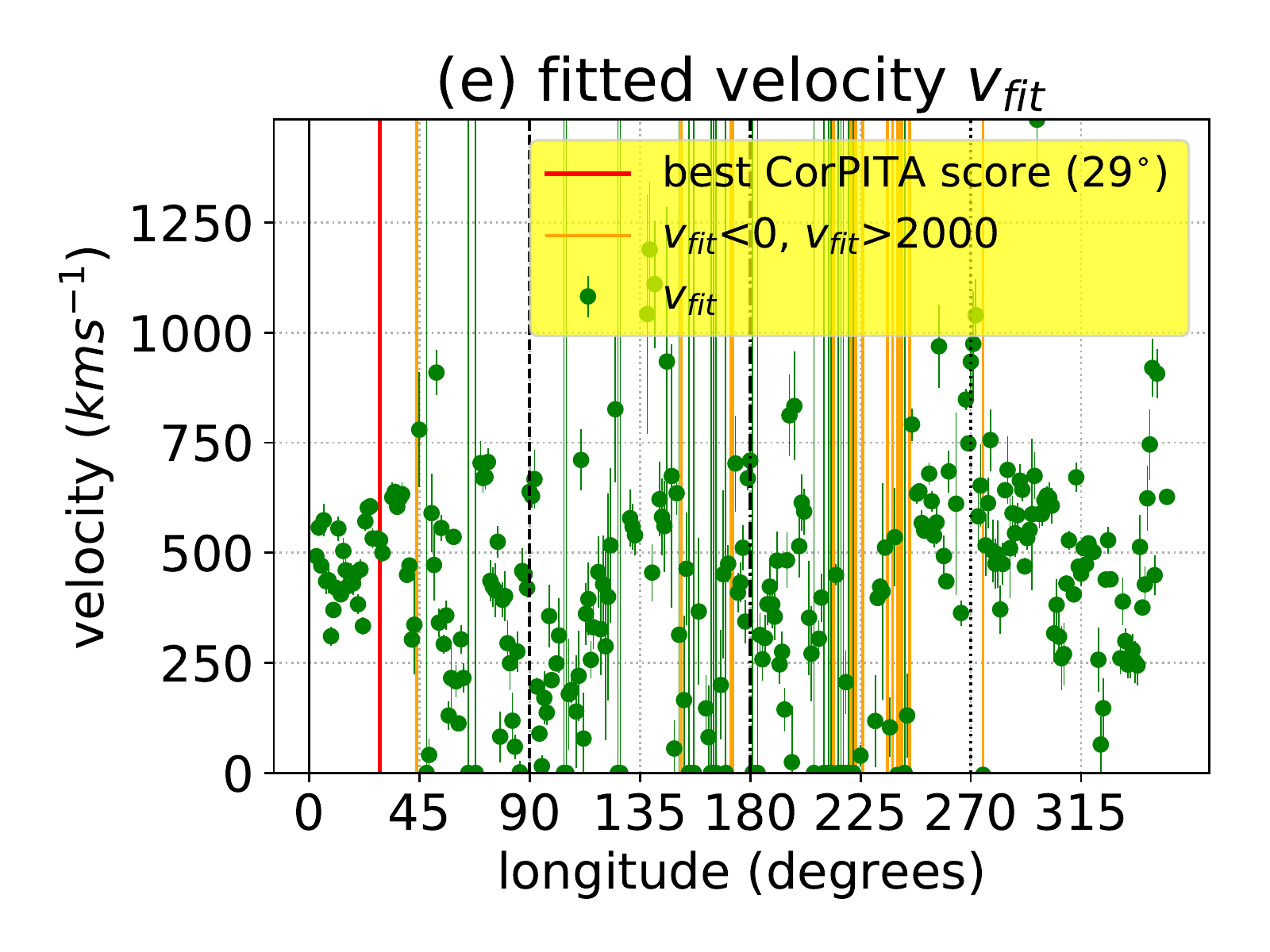}
\includegraphics[width=6.0cm]{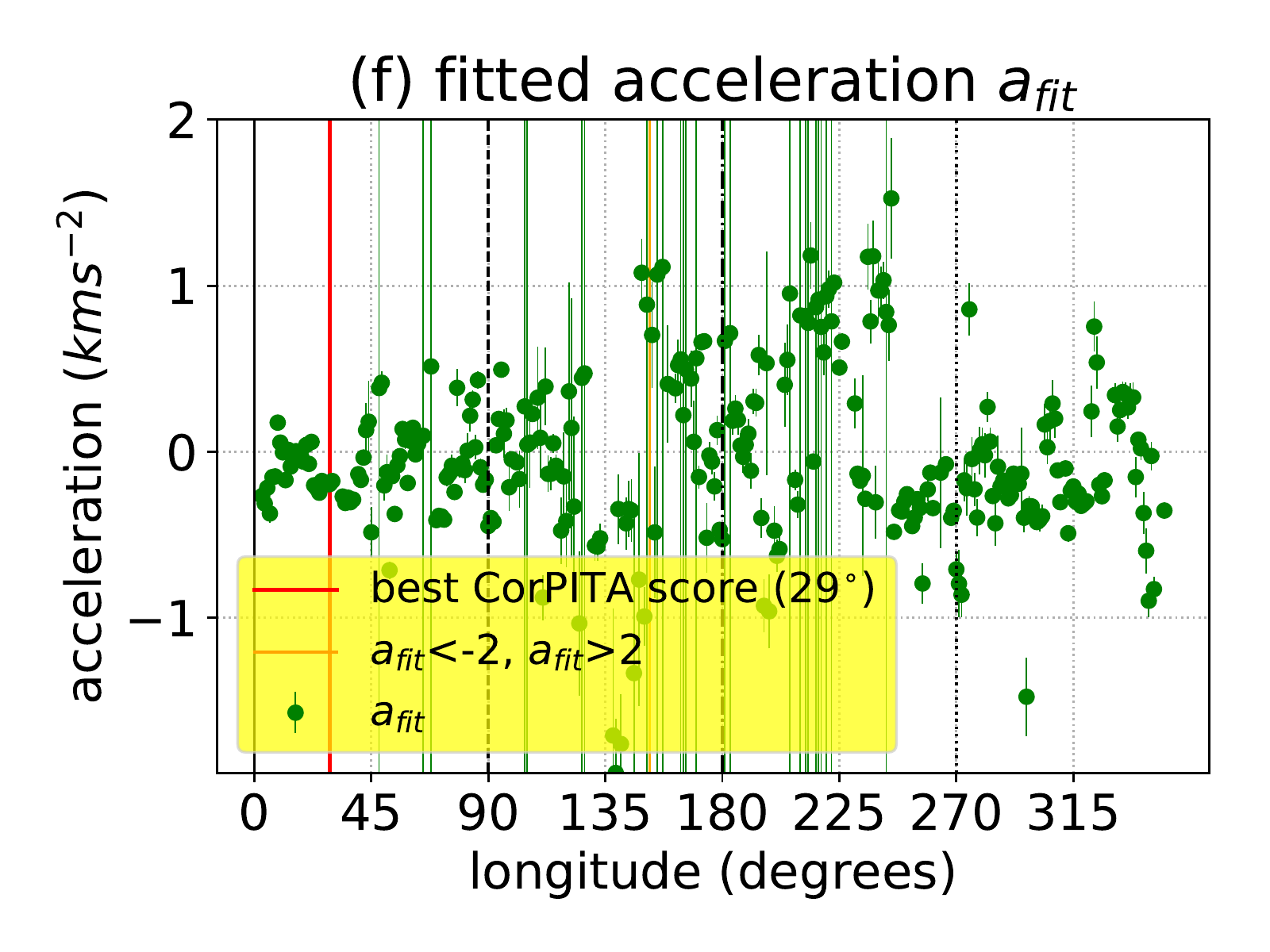}
\includegraphics[width=6.0cm]{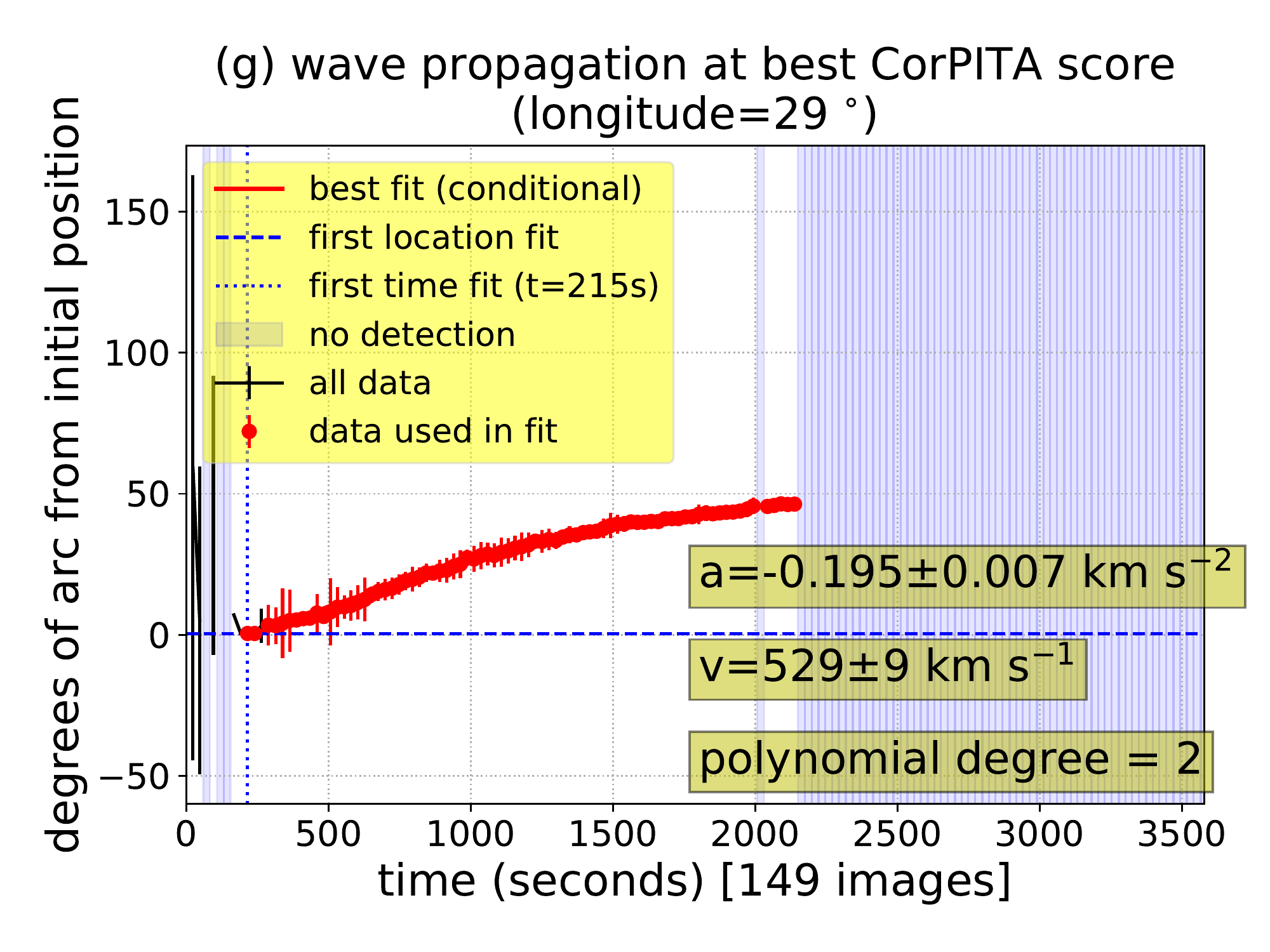}
\caption{AWARE performance for the EUV wave of 13 February 2011. see
  the caption of Figure \ref{fig:syntheticwave:results} for more details.}
\label{fig:longetal7}
\end{center}
\end{figure*}

\begin{figure*}
\begin{center}
\includegraphics[width=12.0cm]{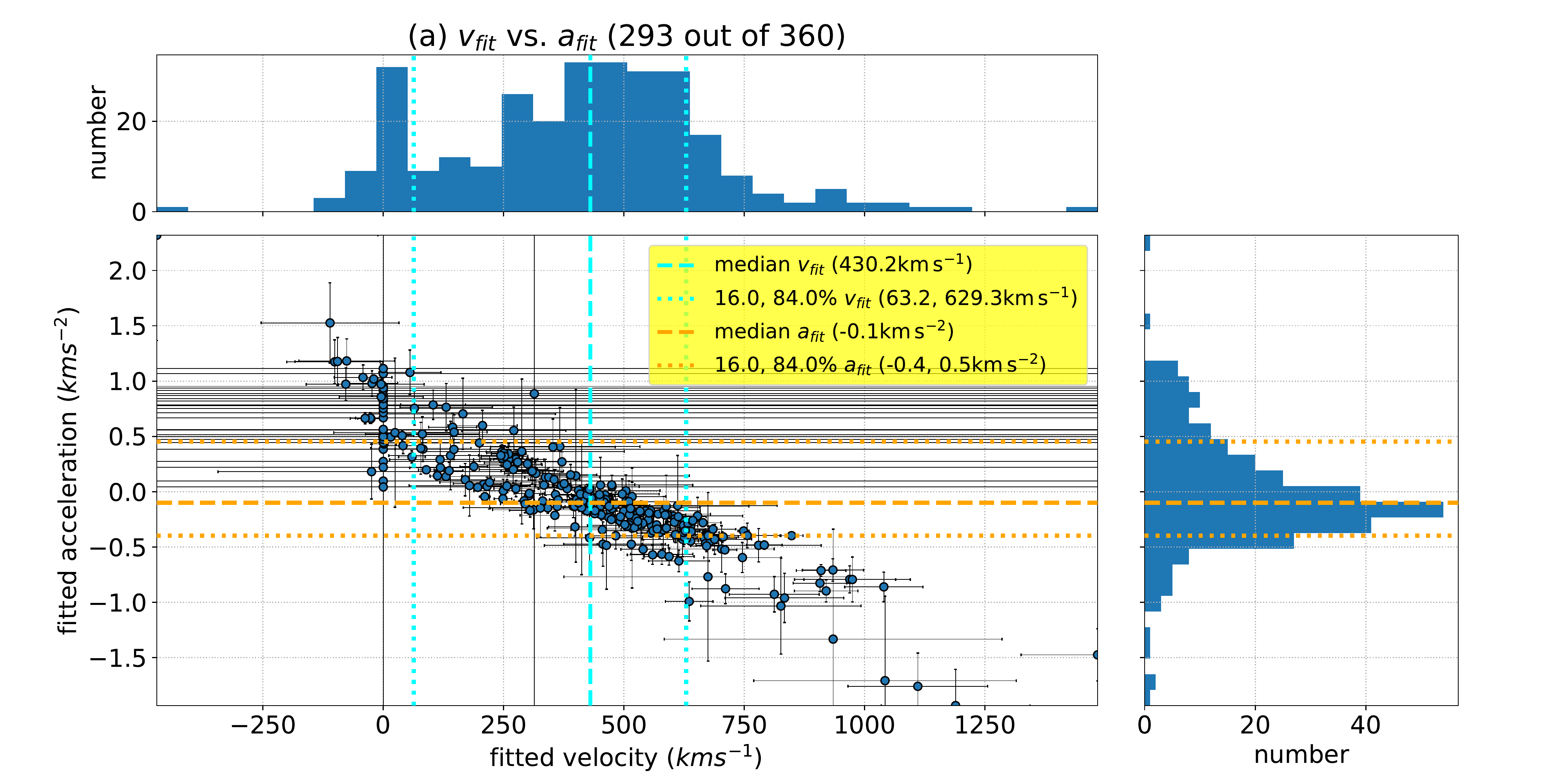}
\includegraphics[width=12.0cm]{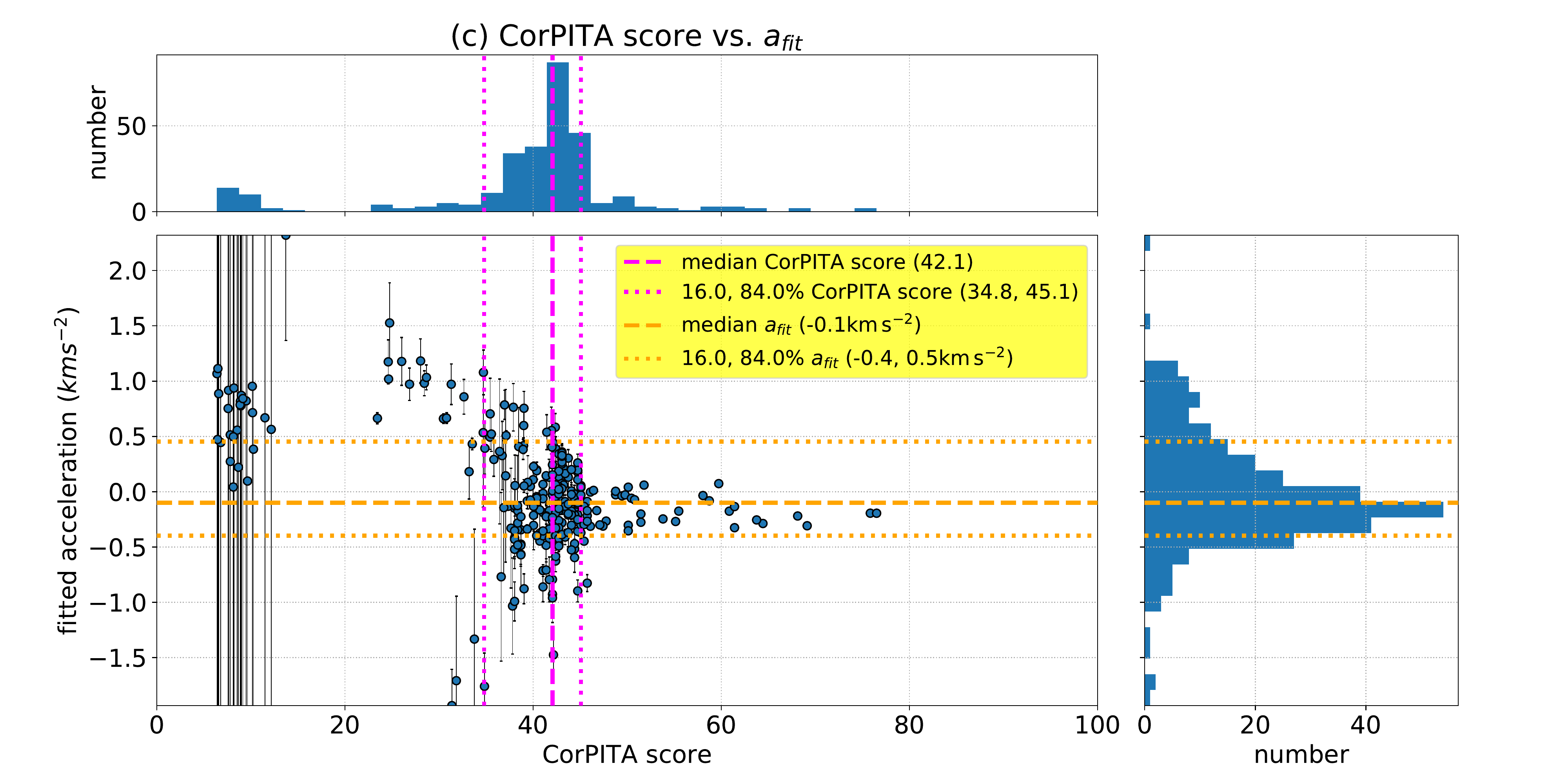}
\includegraphics[width=12.0cm]{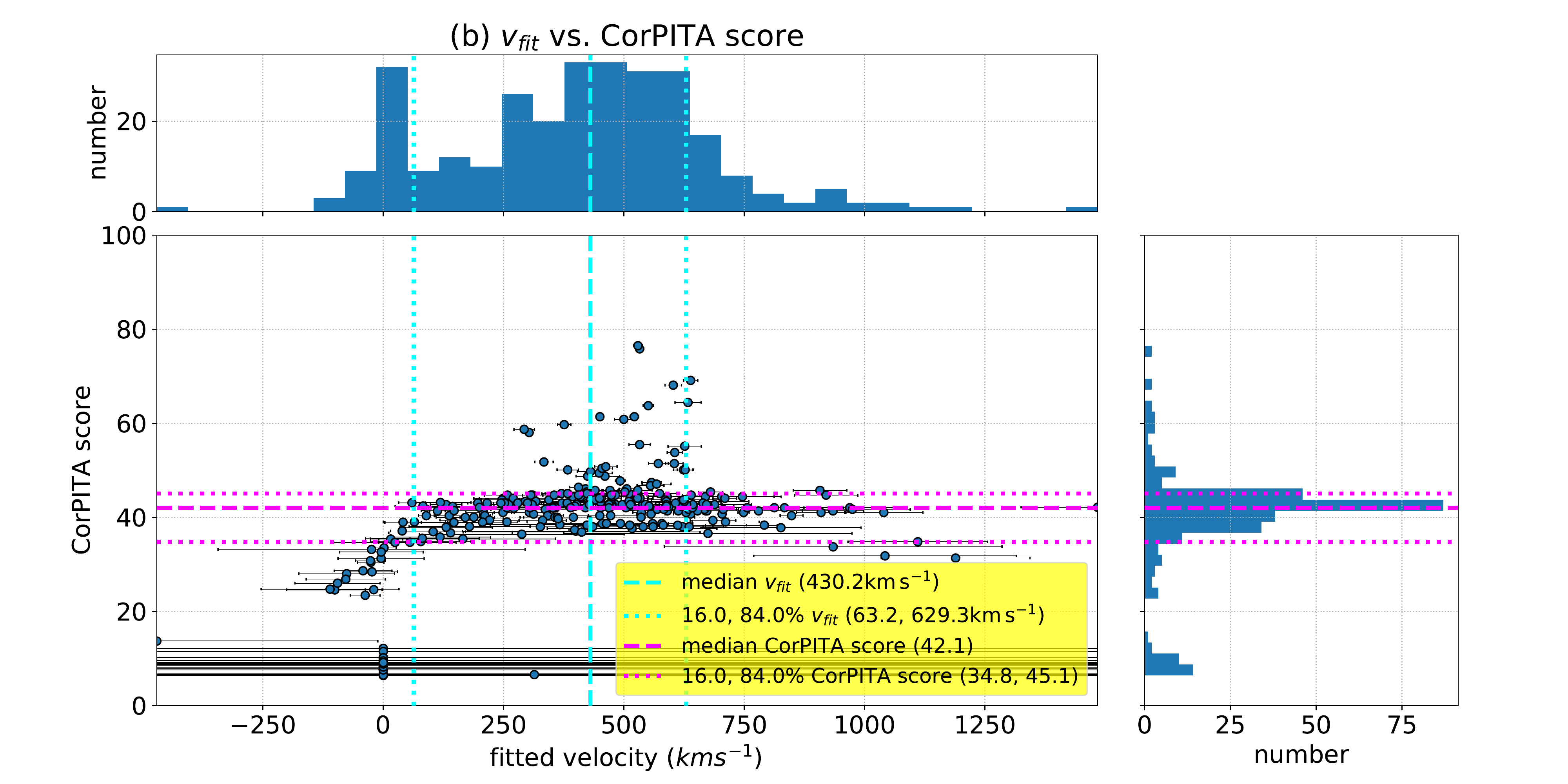}
\caption{Distributions of $\vfit$, $\afit$ and the \longscore\ for the fitted arcs of the 13 February 2011 EUV wave (see Figure \ref{fig:longetal7}).  See the caption of Figure \ref{fig:syntheticwave:dynamics} for more details.}
\label{fig:longetal7dynamics}
\end{center}
\end{figure*}

\begin{figure*}
\begin{center}
\includegraphics[width=6.0cm]{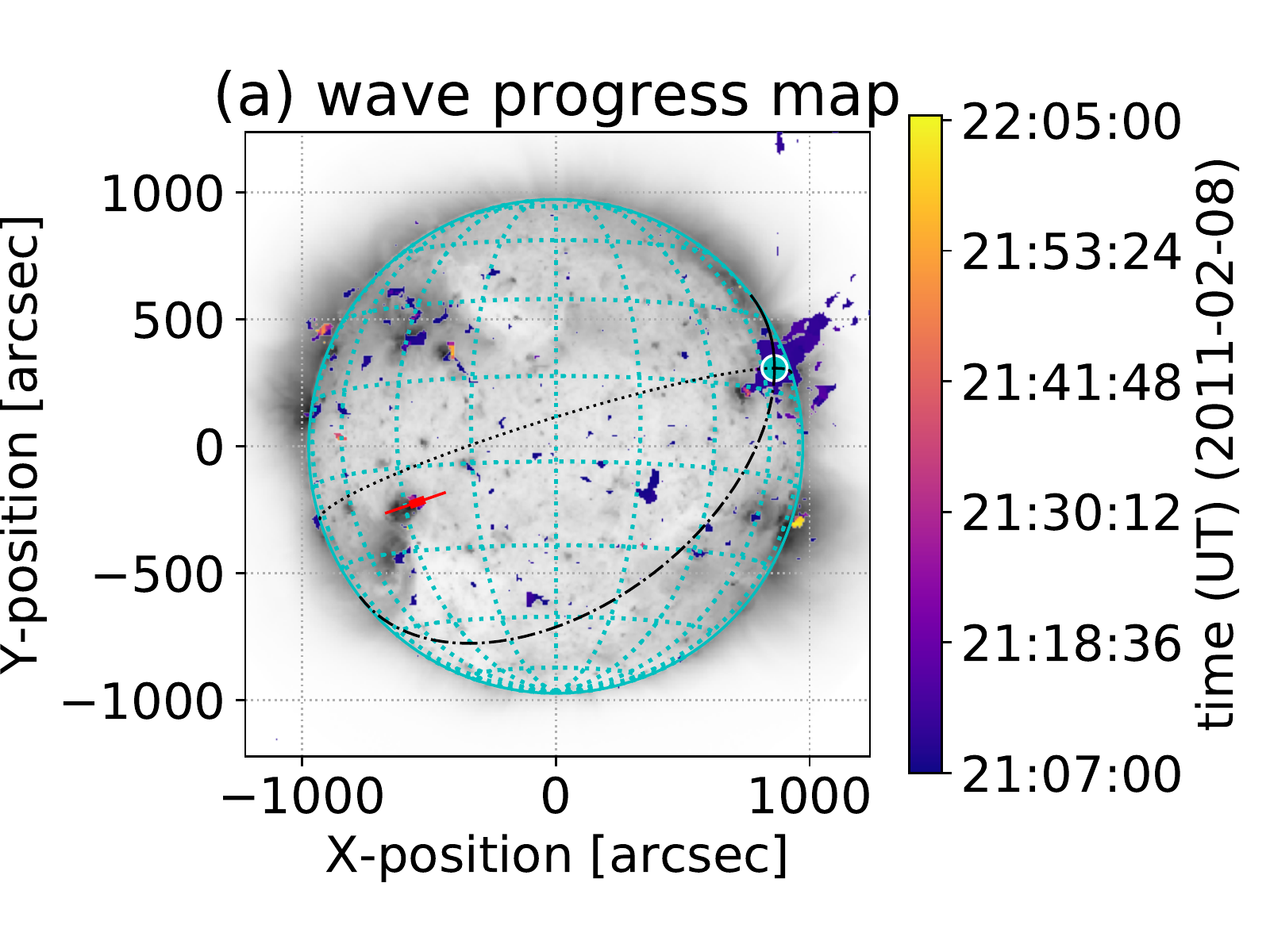}
\includegraphics[width=6.0cm]{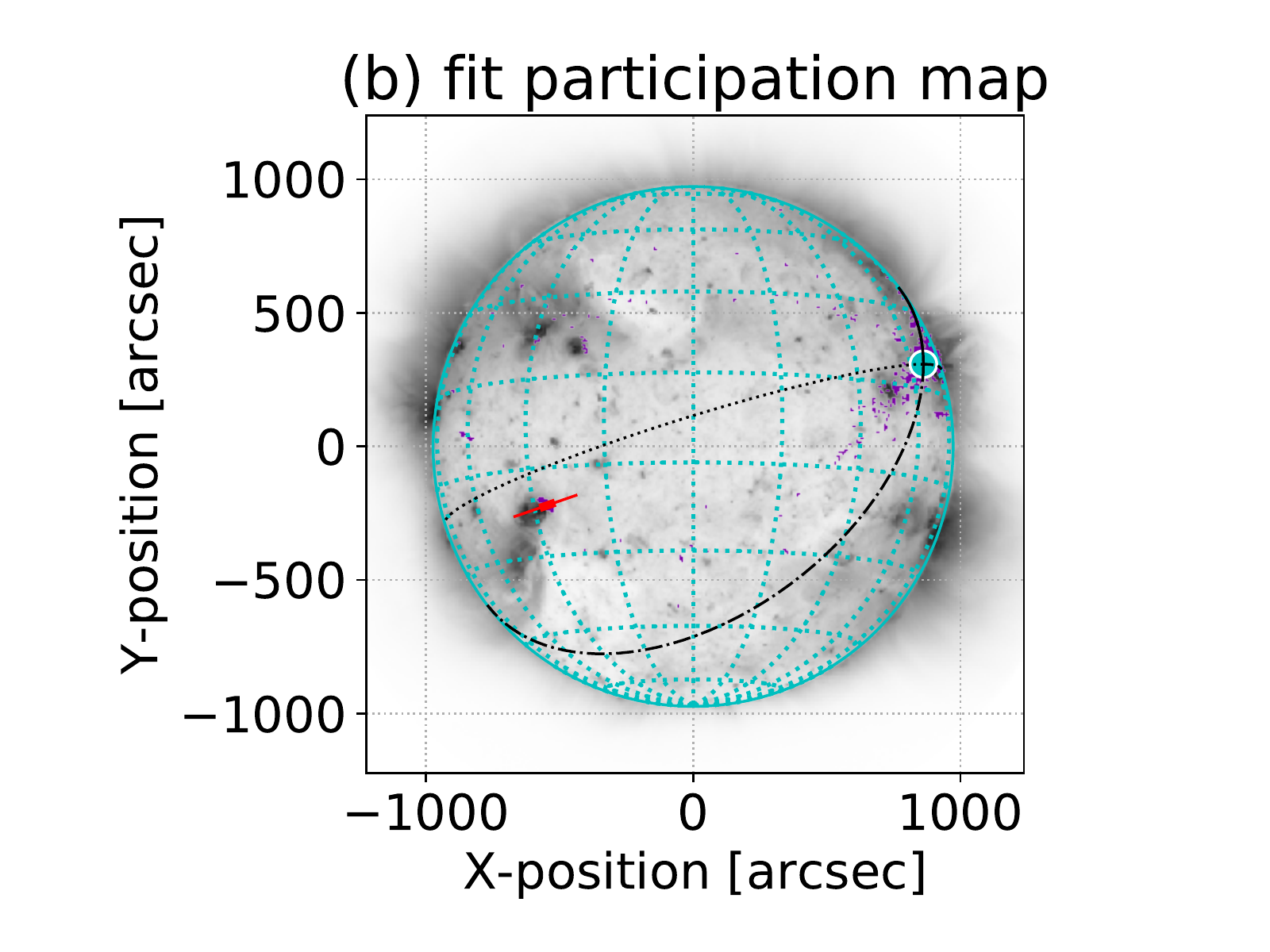}
\includegraphics[width=6.0cm]{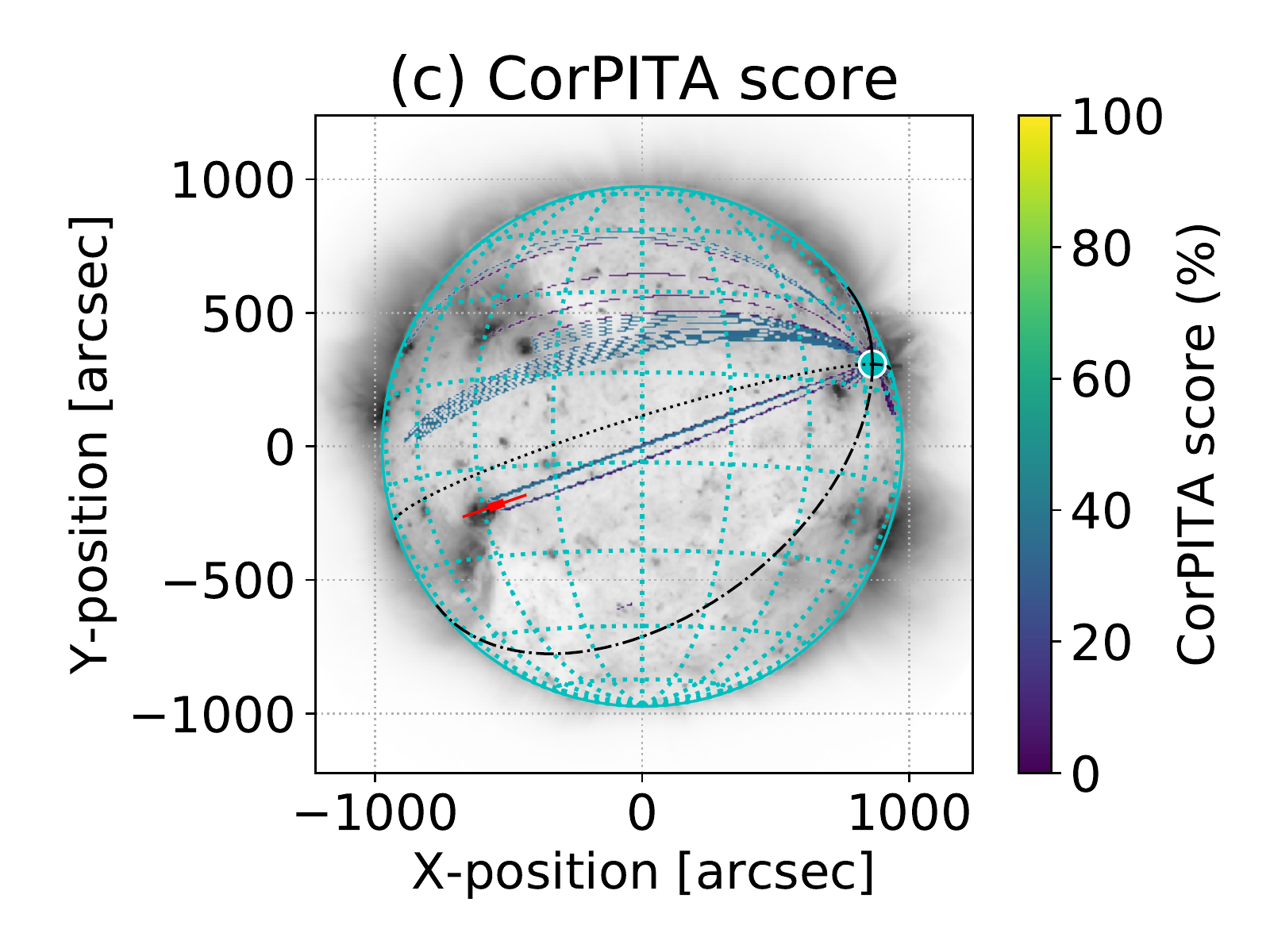}
\includegraphics[width=6.0cm]{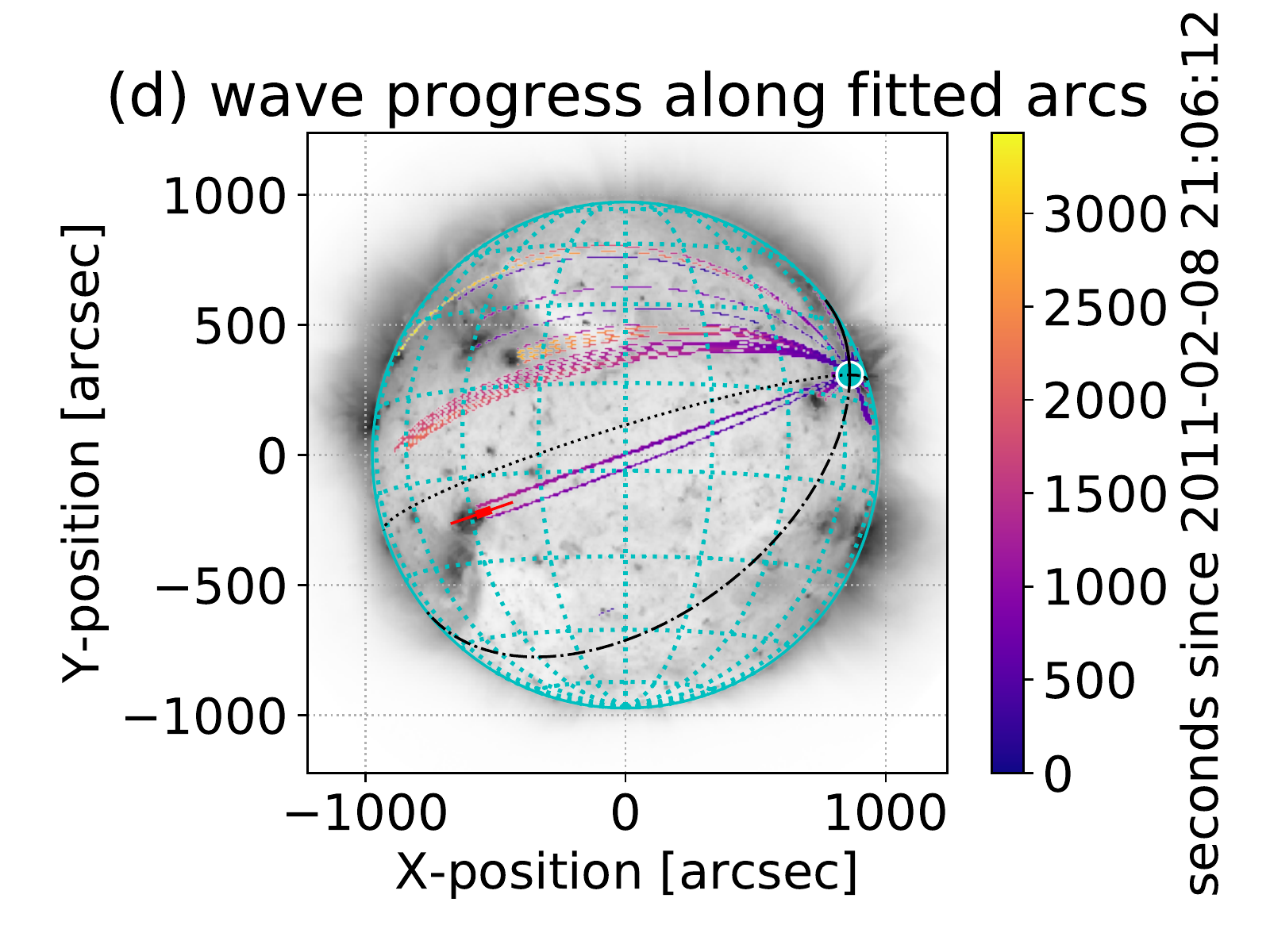}
\includegraphics[width=6.0cm]{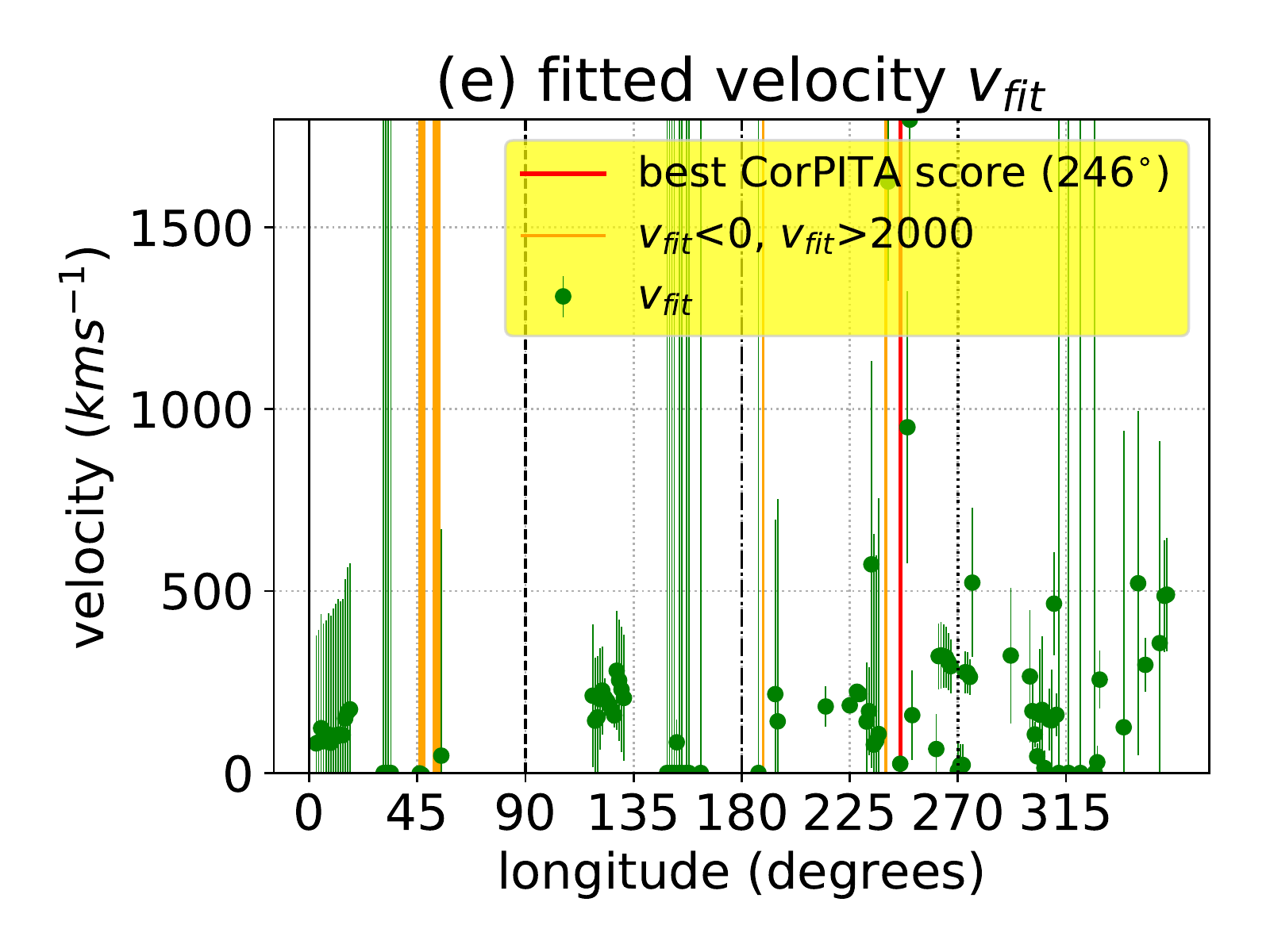}
\includegraphics[width=6.0cm]{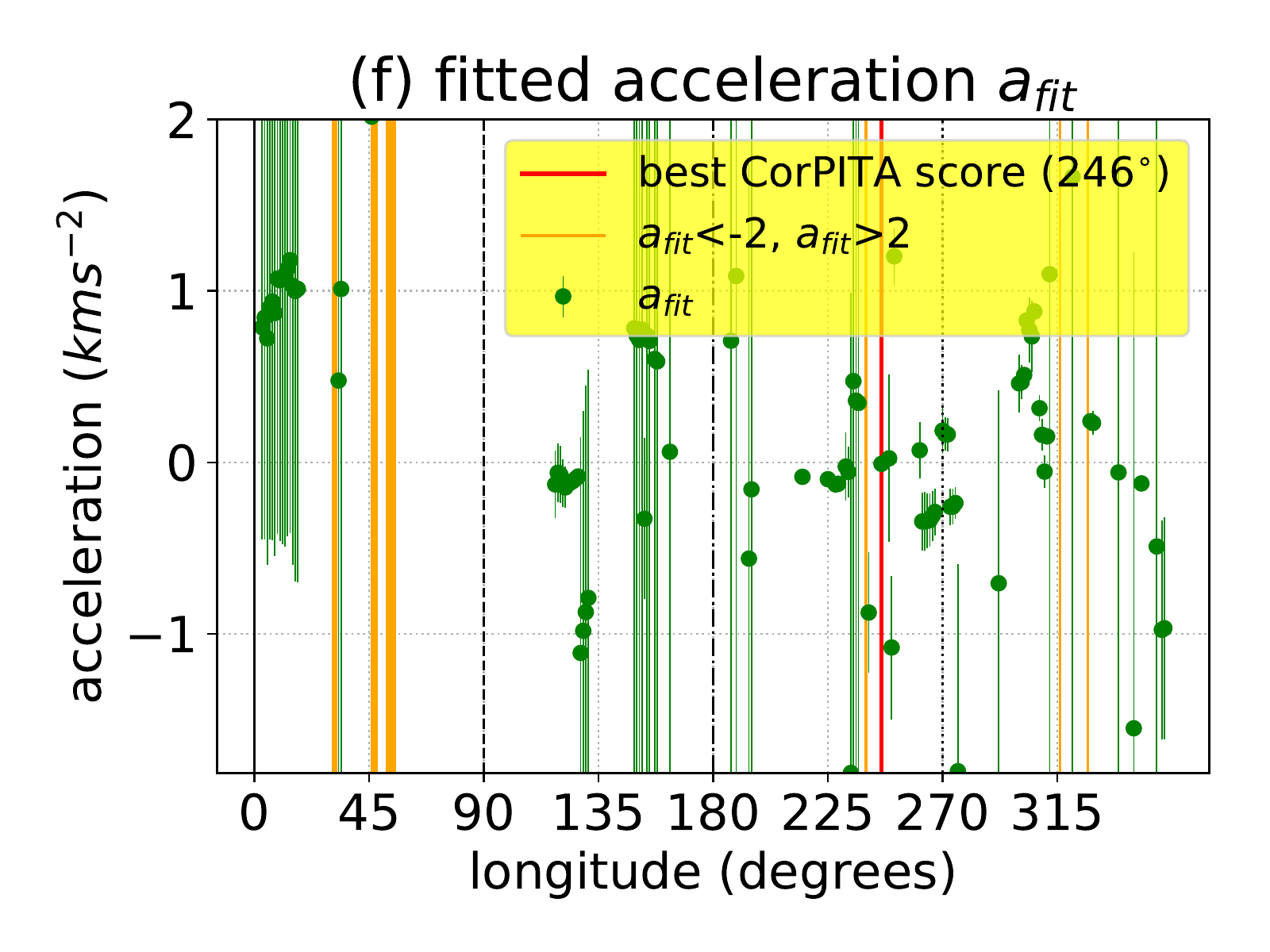}
\includegraphics[width=6.0cm]{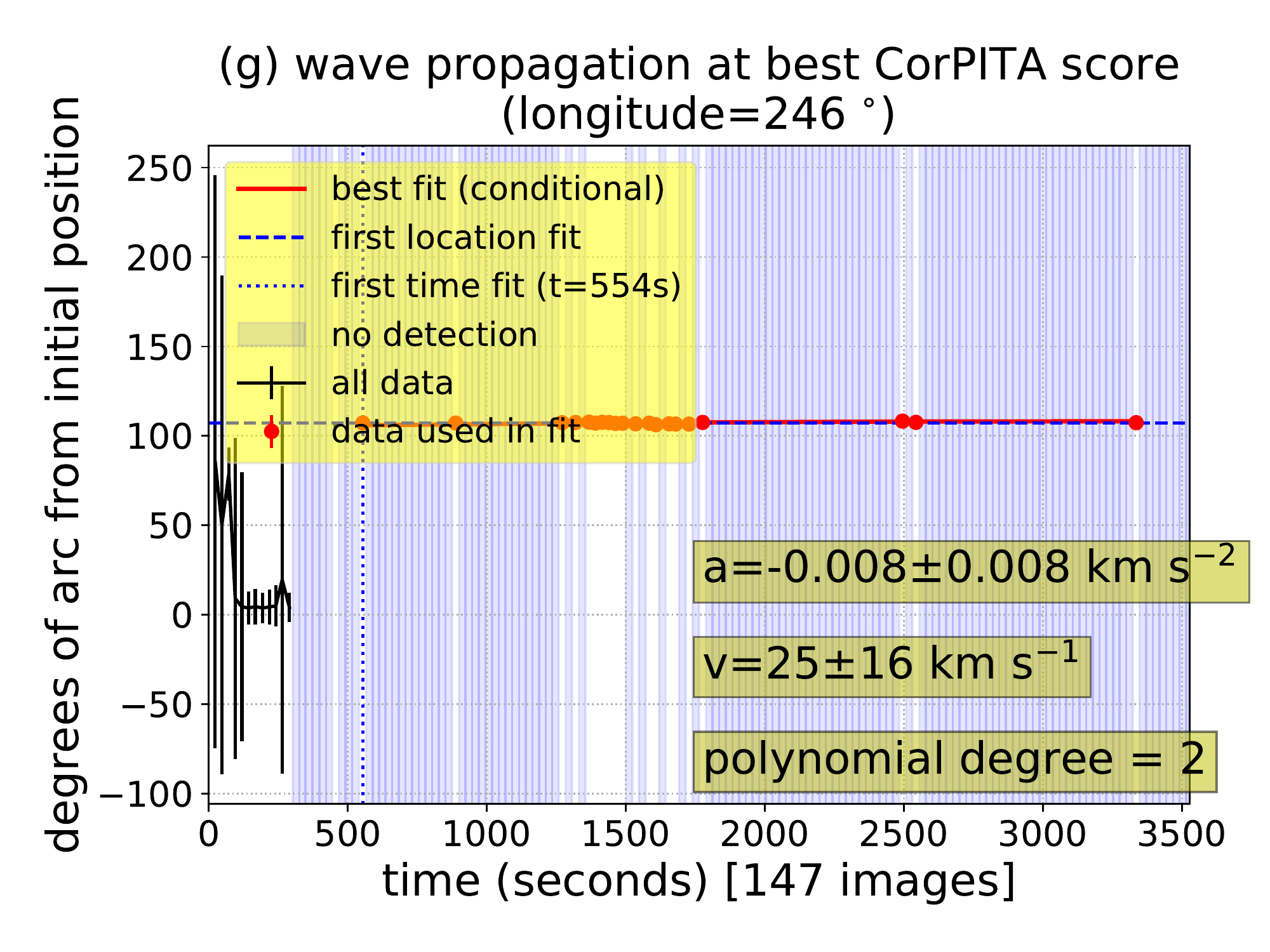}
\caption{AWARE performance when there is no EUV wave identifiable by eye, using the same data analyzed by \cite{2014SoPh..289.3279L} and shown in their Figure 6. See the caption of Figure \ref{fig:syntheticwave:results} for more details.}
\label{fig:longetal6}
\end{center}
\end{figure*}

\begin{figure*}
\begin{center}
\includegraphics[width=6.0cm]{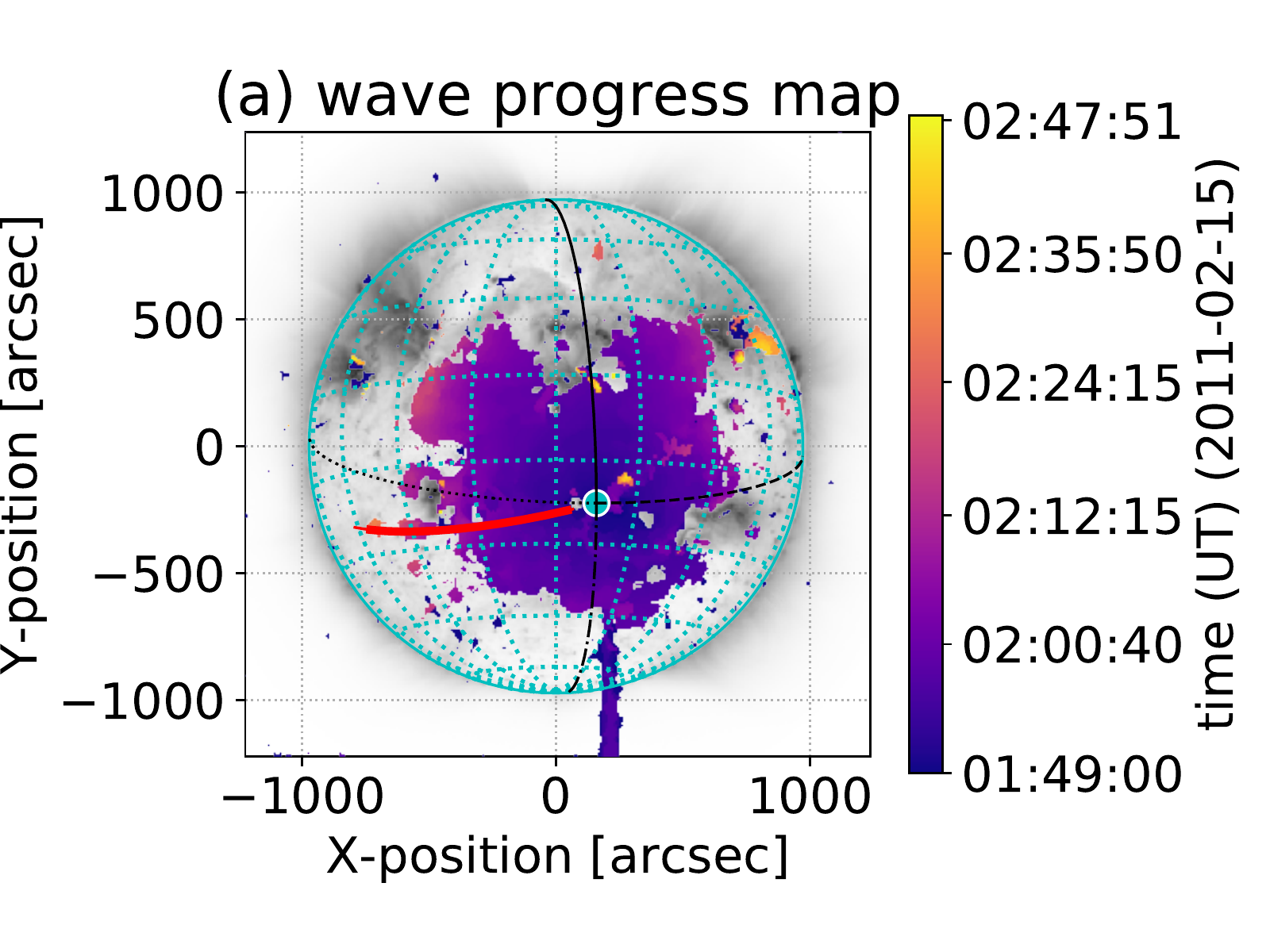}
\includegraphics[width=6.0cm]{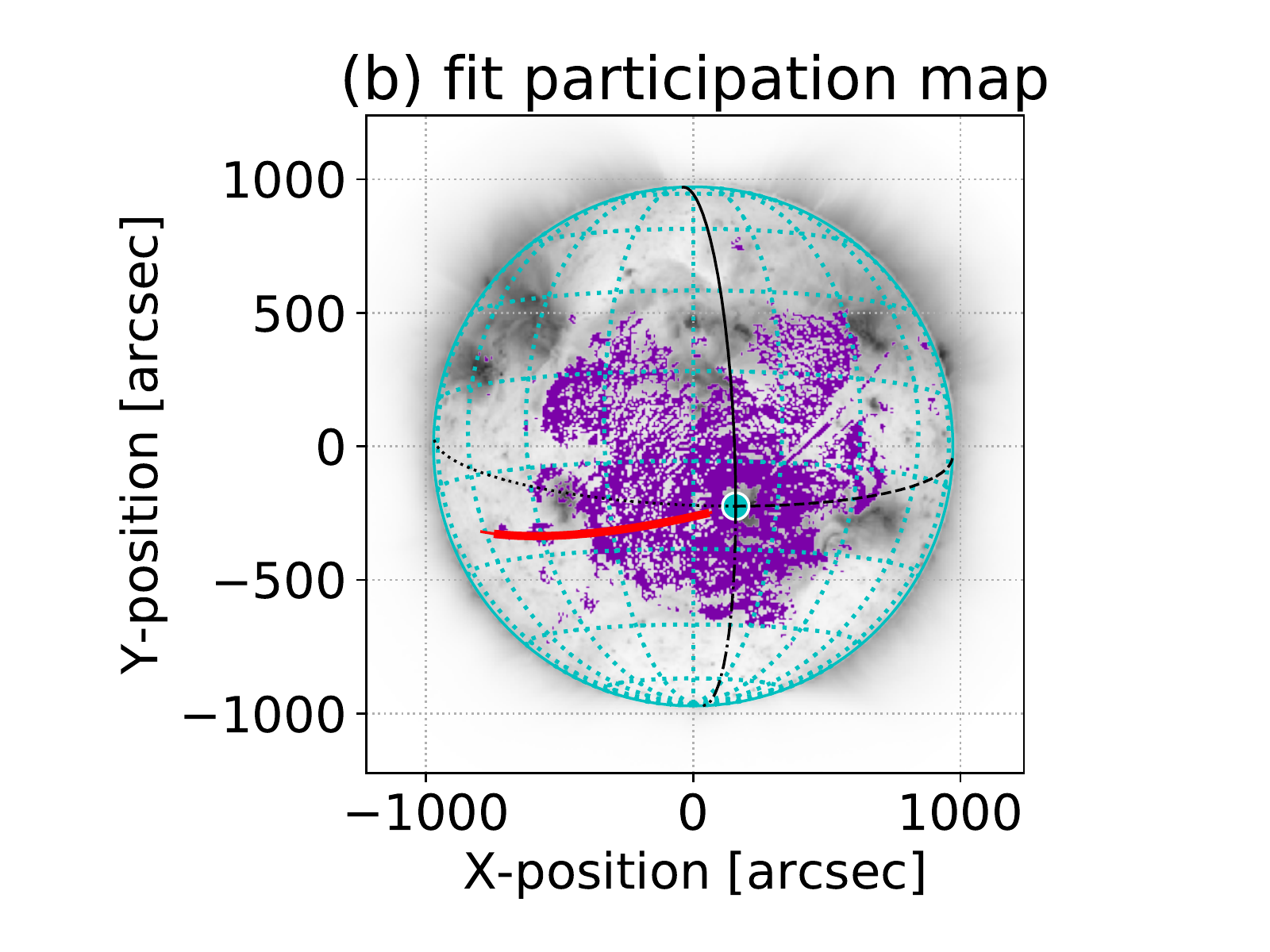}
\includegraphics[width=6.0cm]{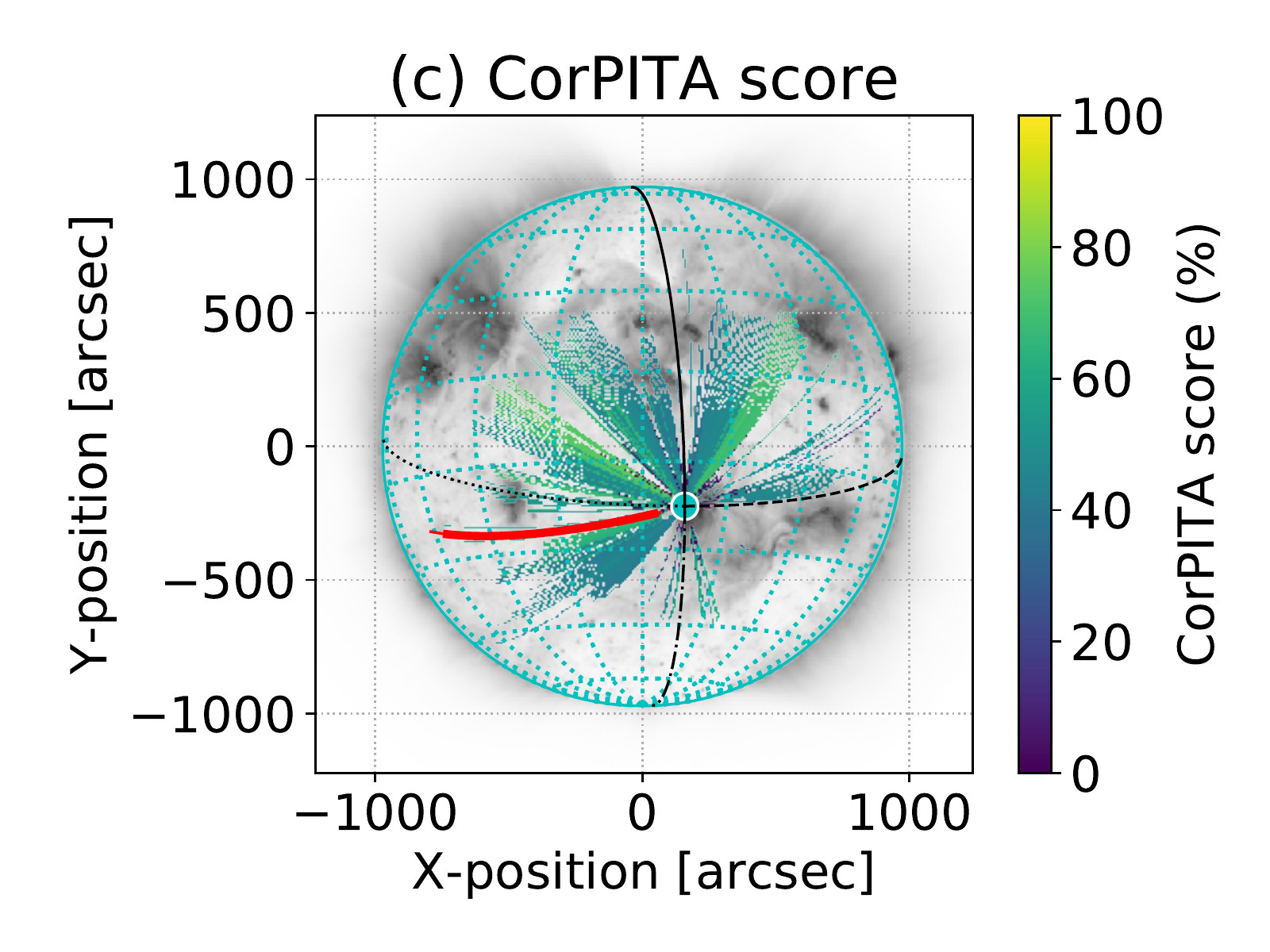}
\includegraphics[width=6.0cm]{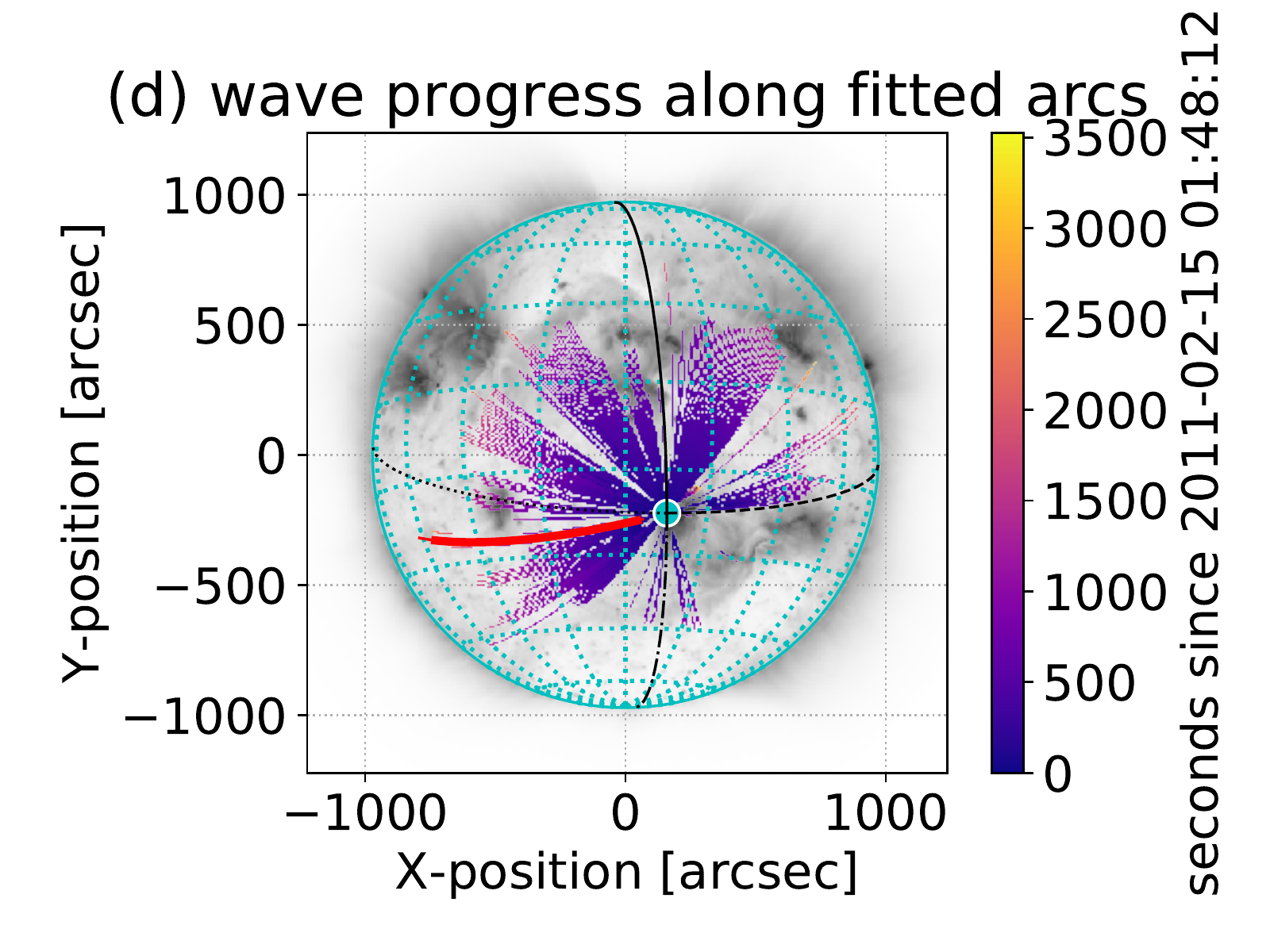}
\includegraphics[width=6.0cm]{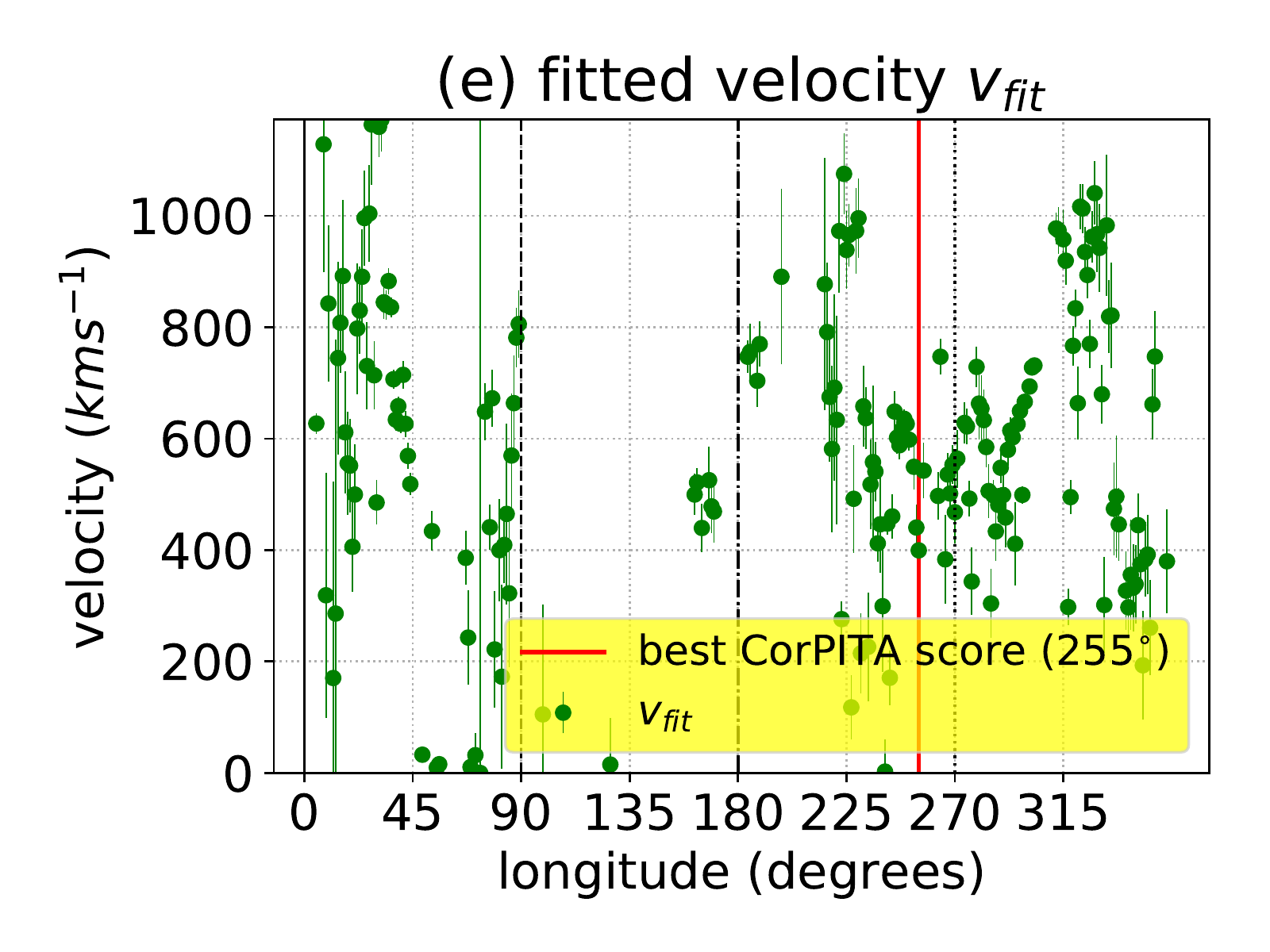}
\includegraphics[width=6.0cm]{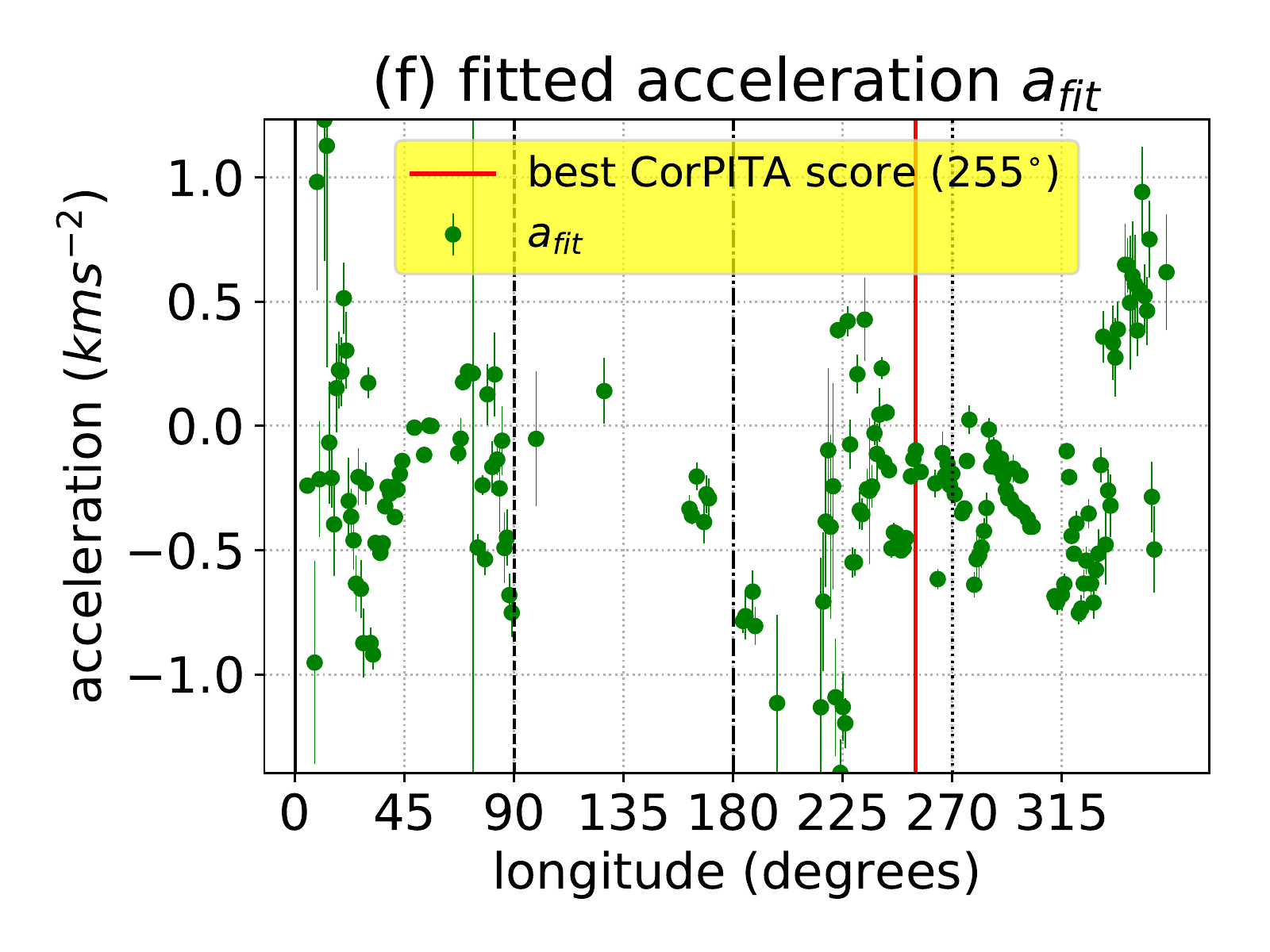}
\includegraphics[width=6.0cm]{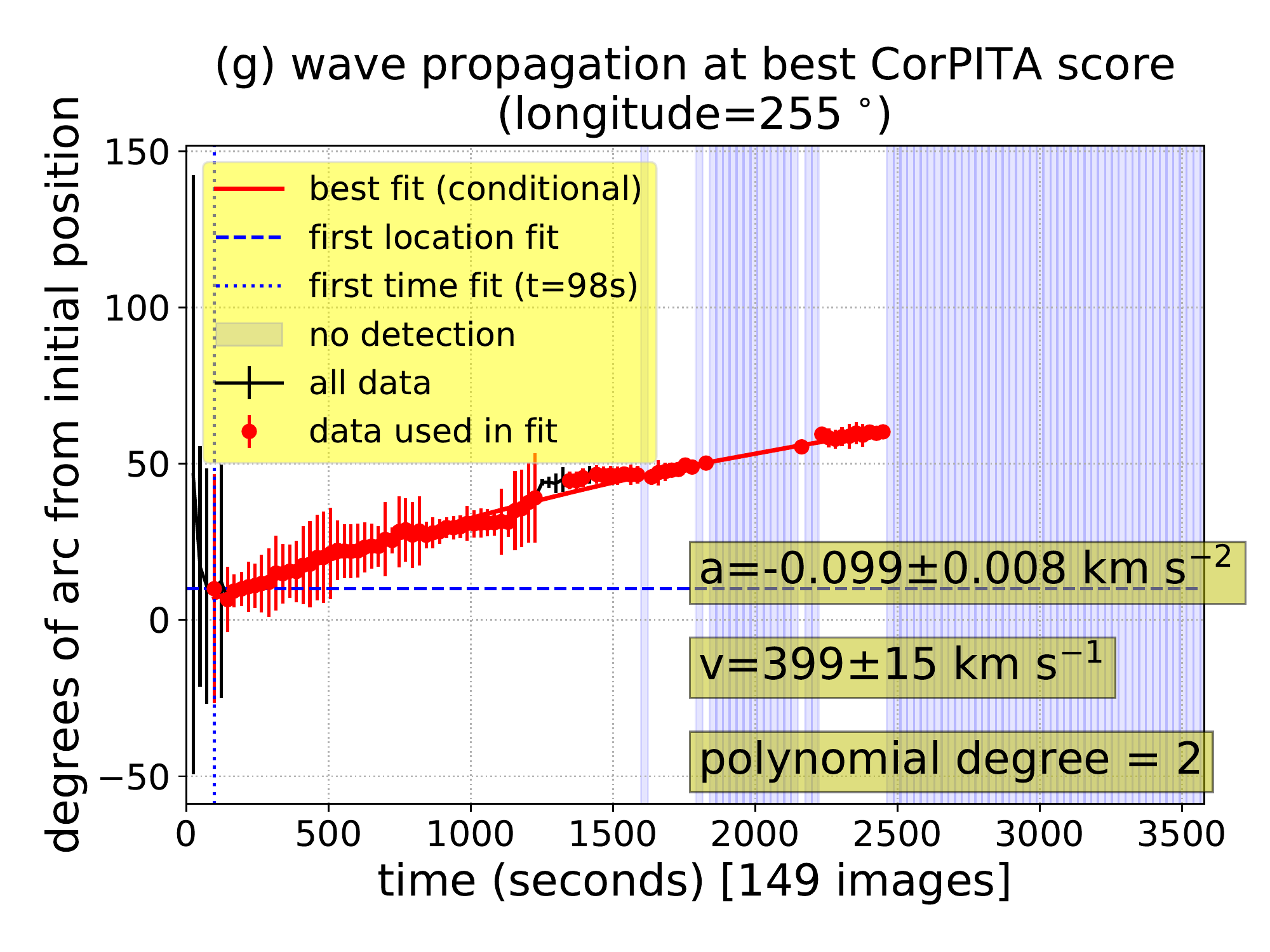}
\caption{AWARE performance for the EUV wave of 15 February 2011. See the caption of Figure \ref{fig:syntheticwave:results} for more details.}
\label{fig:longetal8a}
\end{center}
\end{figure*}

\begin{figure*}
\begin{center}
\includegraphics[width=12.0cm]{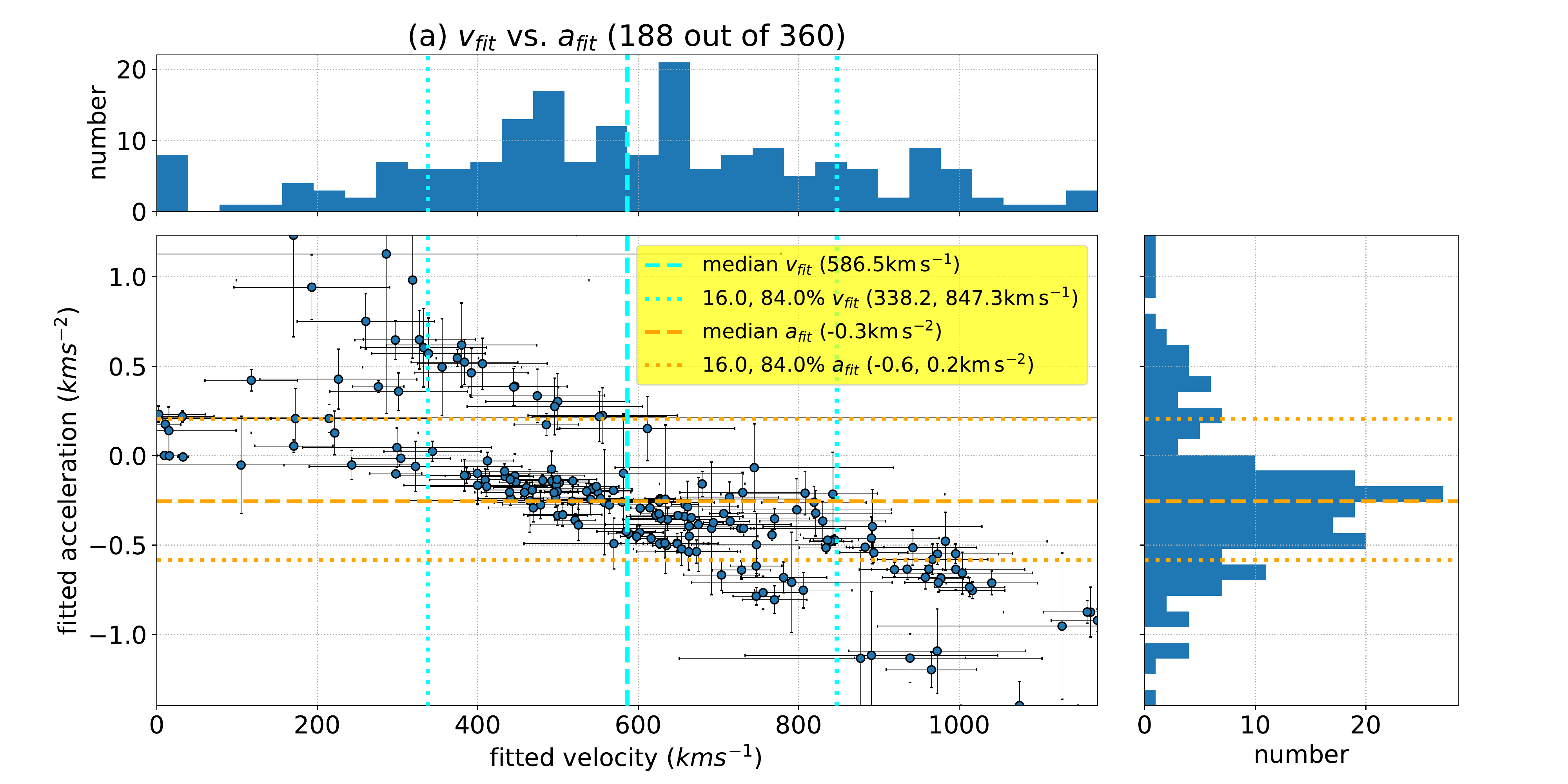}
\includegraphics[width=12.0cm]{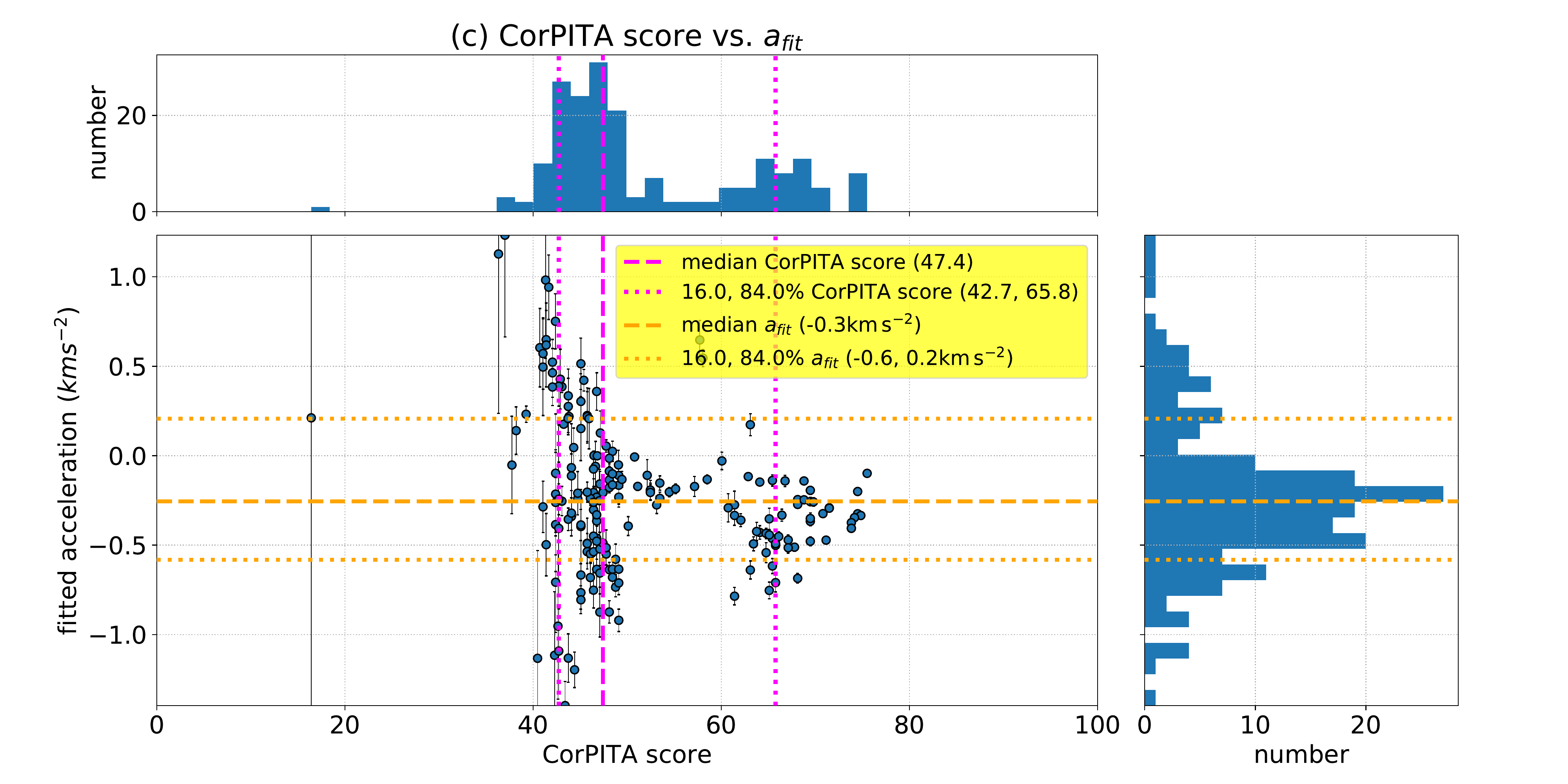}
\includegraphics[width=12.0cm]{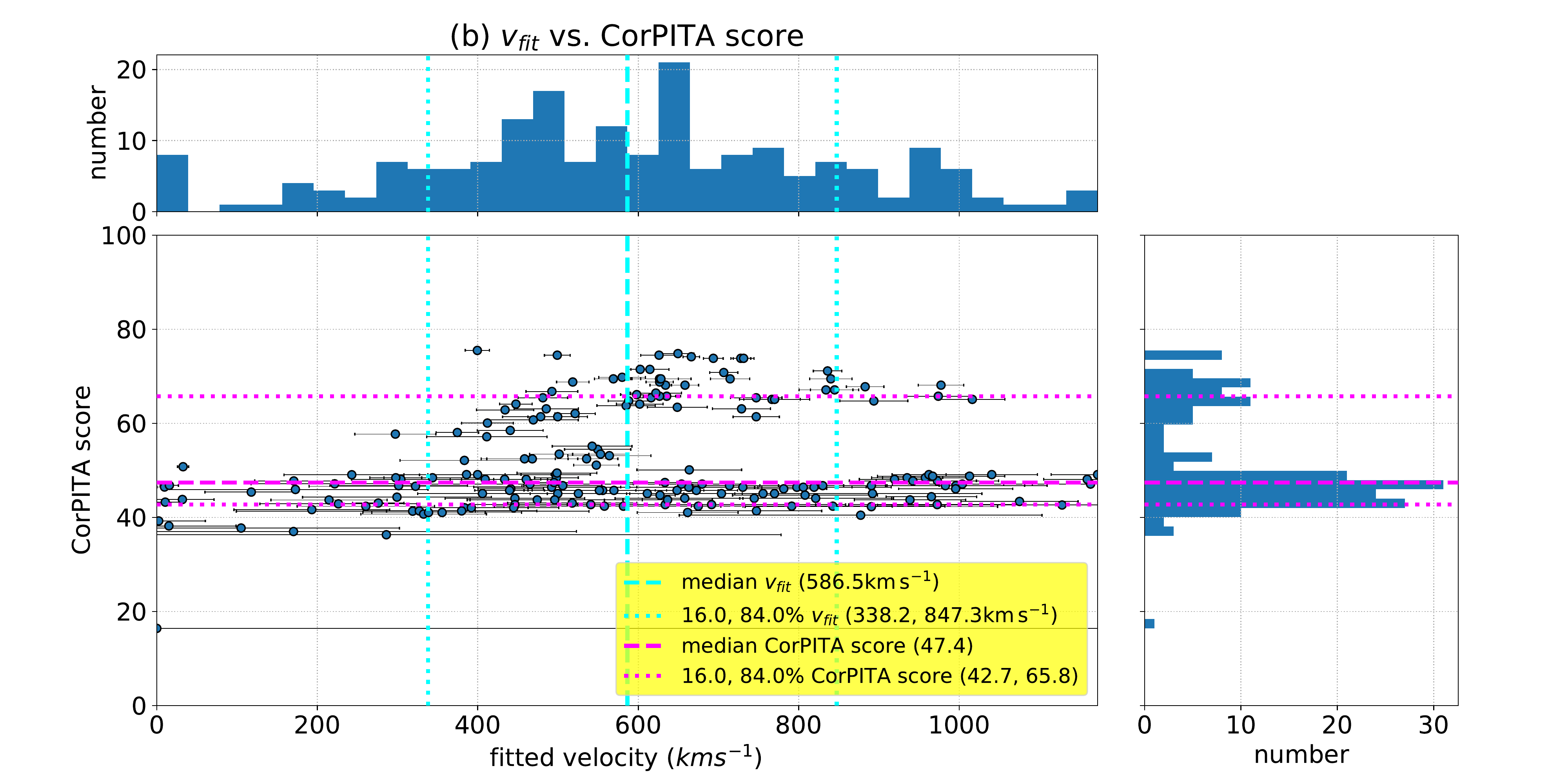}
\caption{Distributions of $\vfit$, $\afit$ and the \longscore\ for the fitted arcs of the 15 February 2011 EUV wave (see Figure \ref{fig:longetal8a}). See the caption of Figure \ref{fig:syntheticwave:dynamics} for more details.}
\label{fig:longetal8adynamics}
\end{center}
\end{figure*}

\begin{figure*}
\begin{center}
\includegraphics[width=6.0cm]{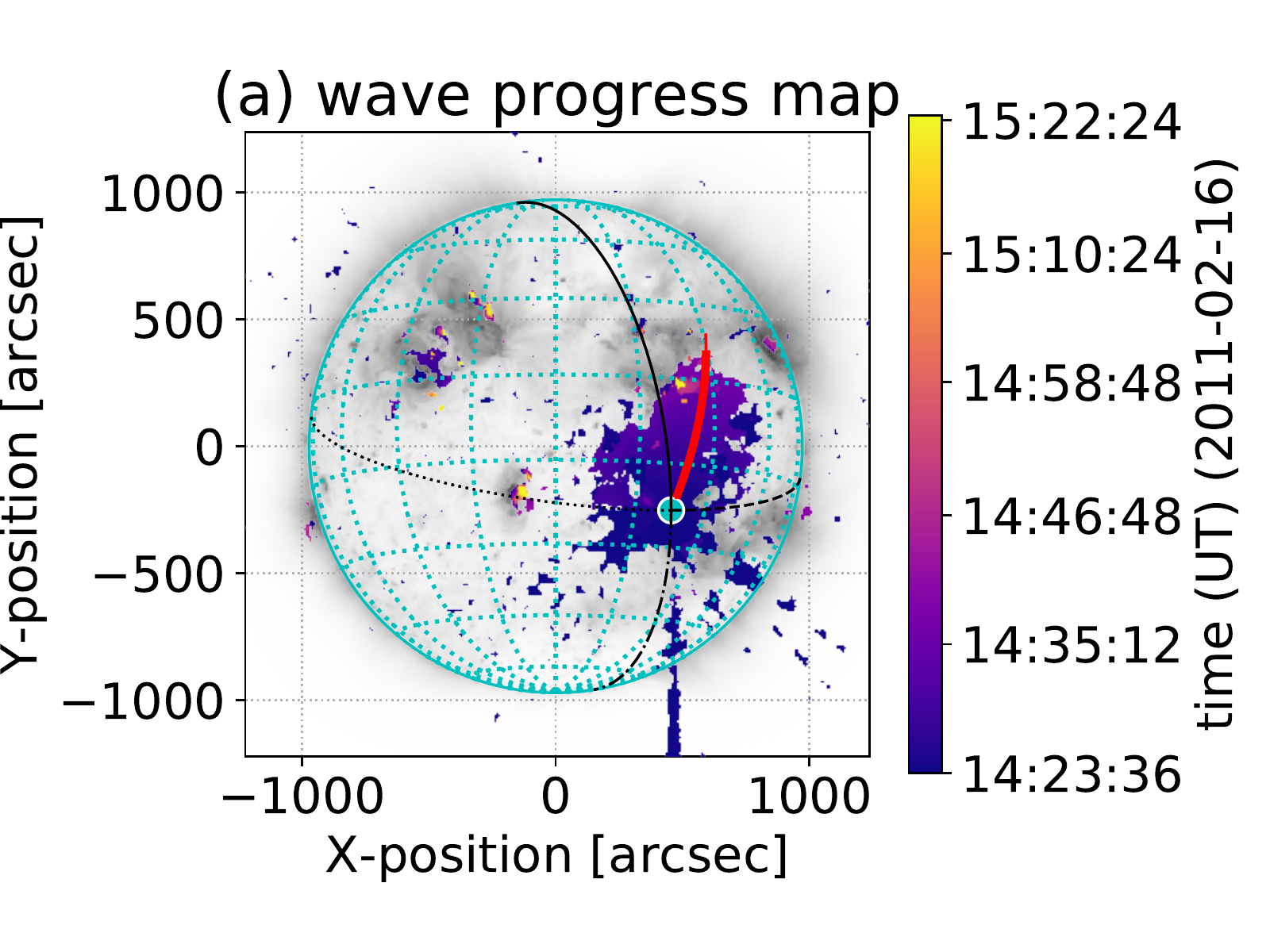}
\includegraphics[width=6.0cm]{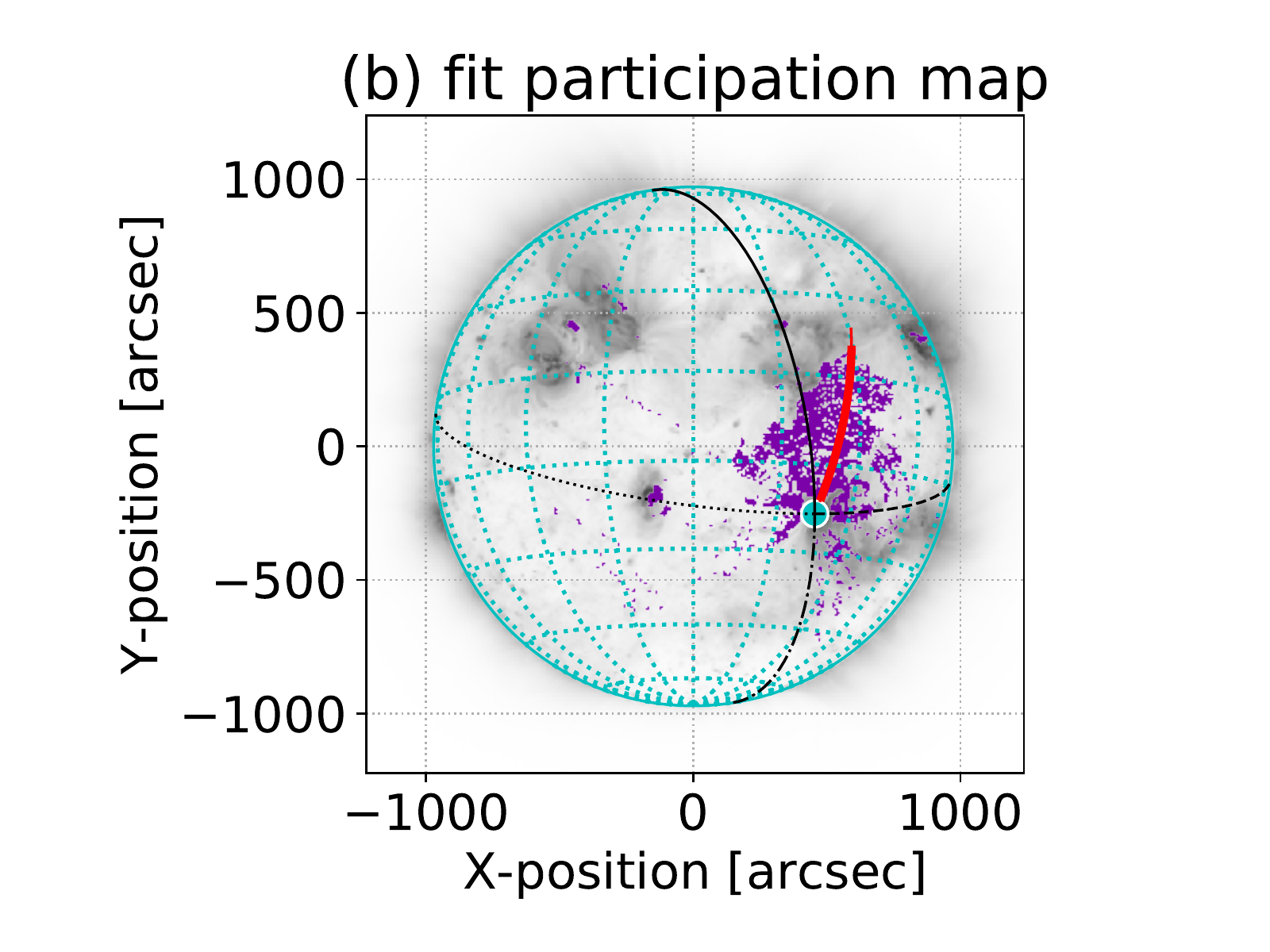}
\includegraphics[width=6.0cm]{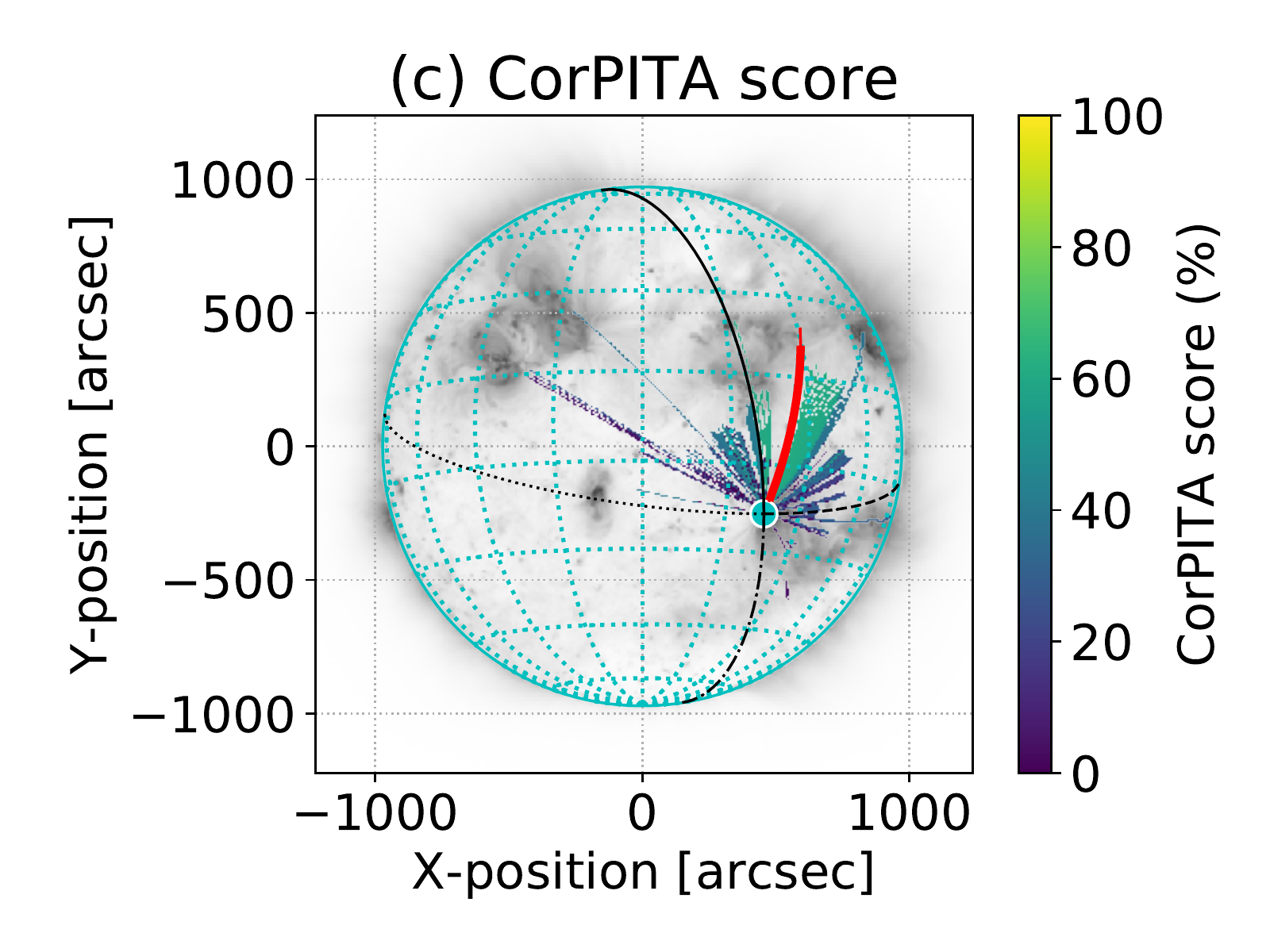}
\includegraphics[width=6.0cm]{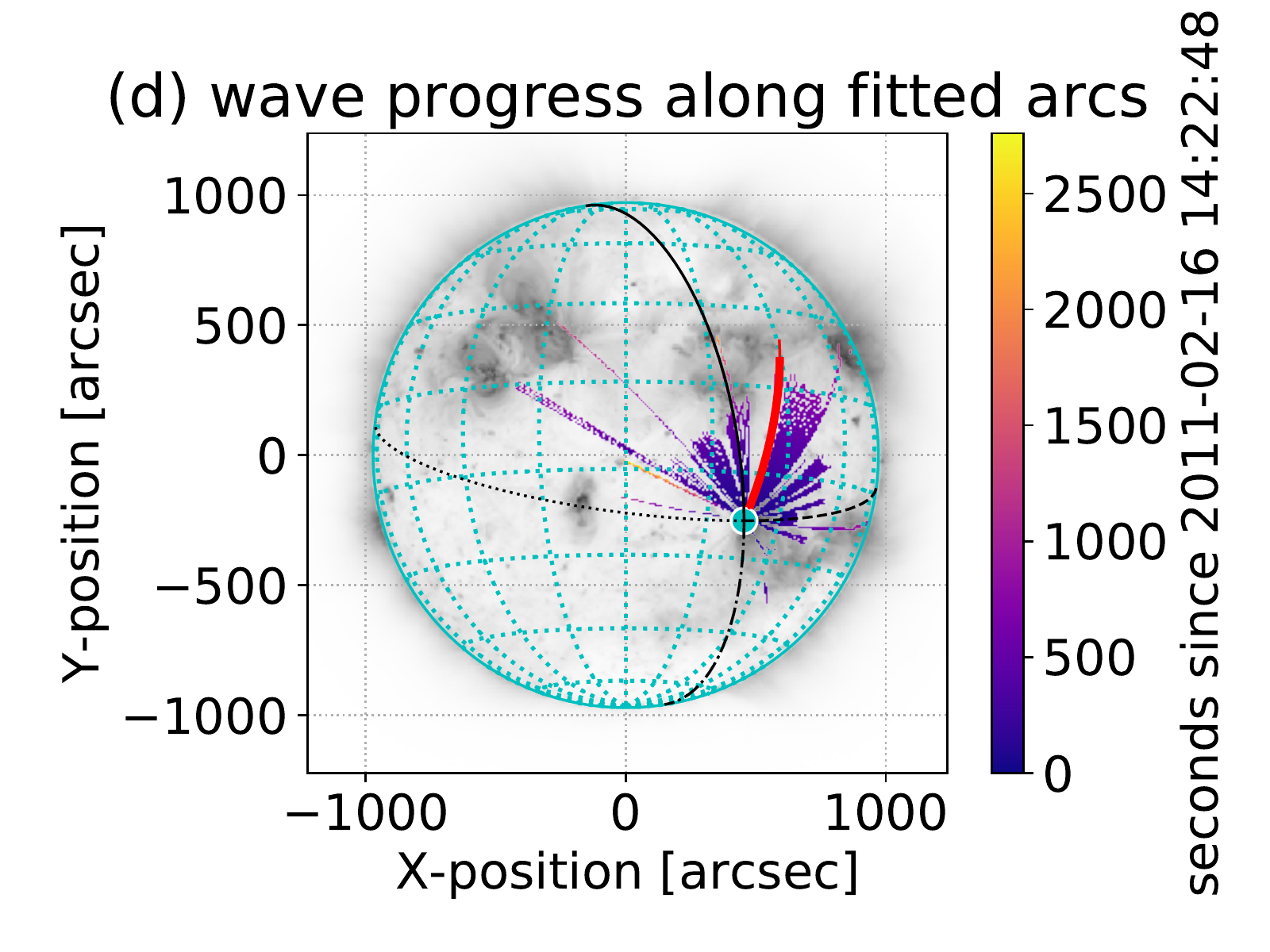}
\includegraphics[width=6.0cm]{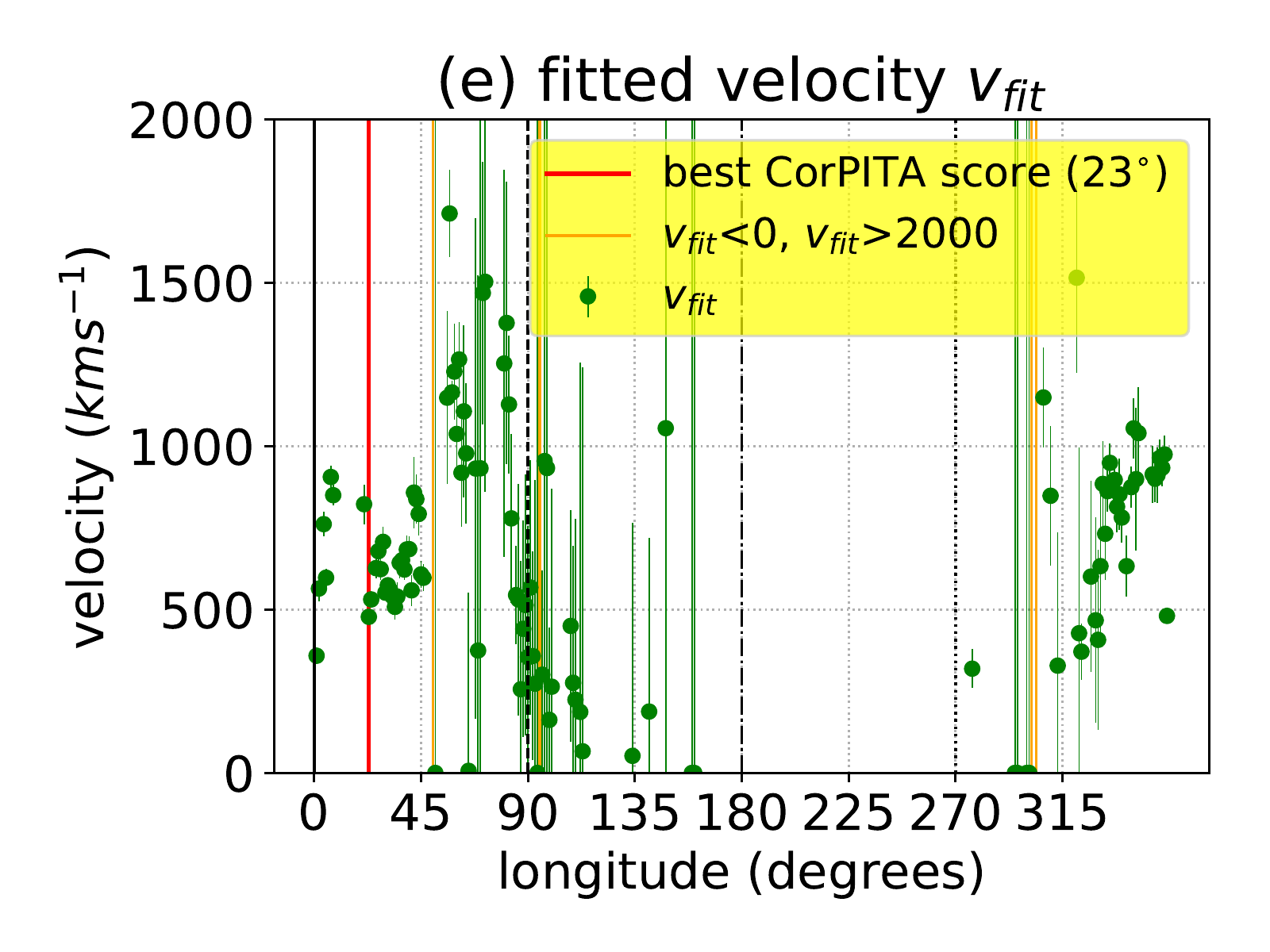}
\includegraphics[width=6.0cm]{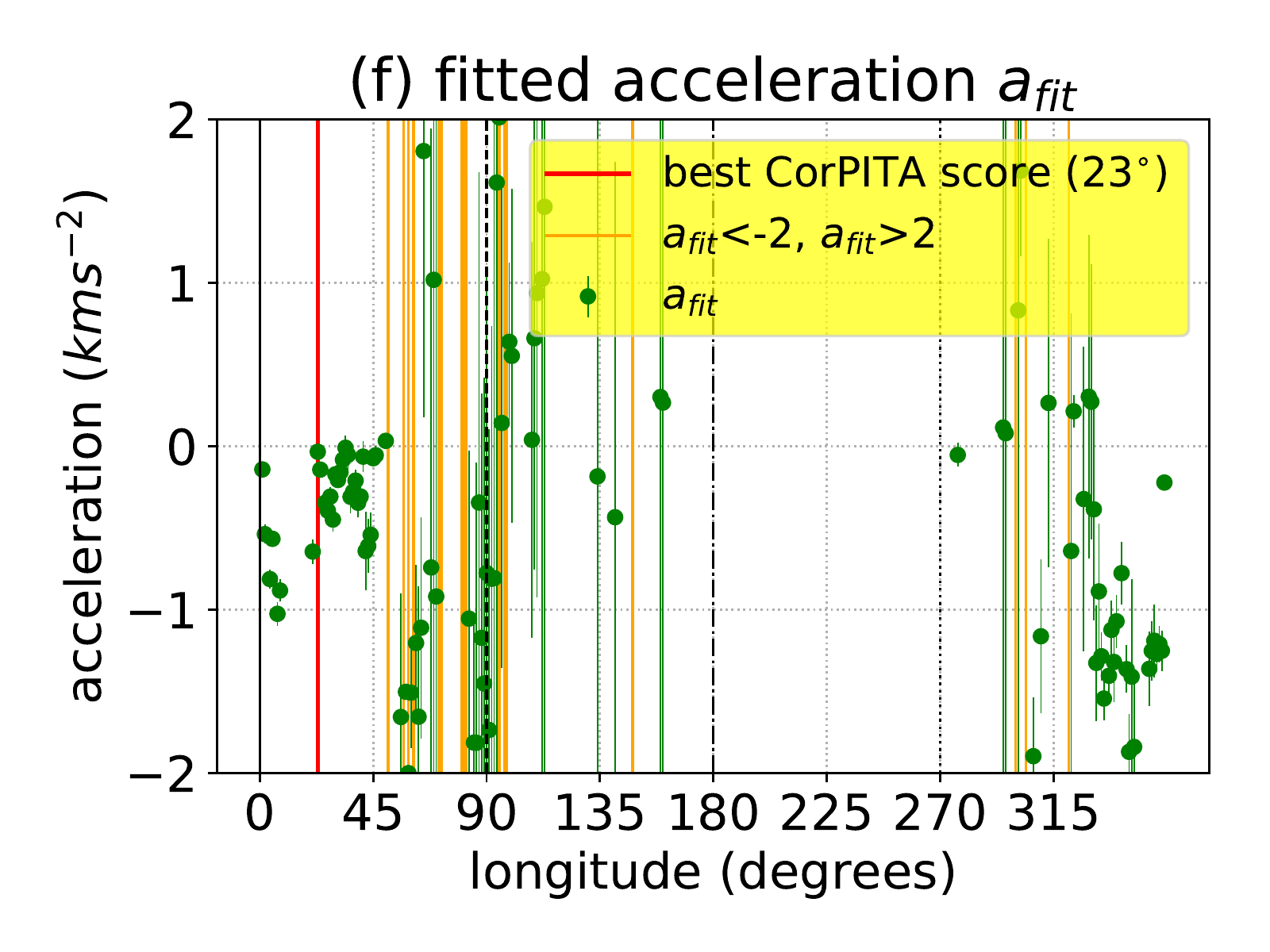}
\includegraphics[width=6.0cm]{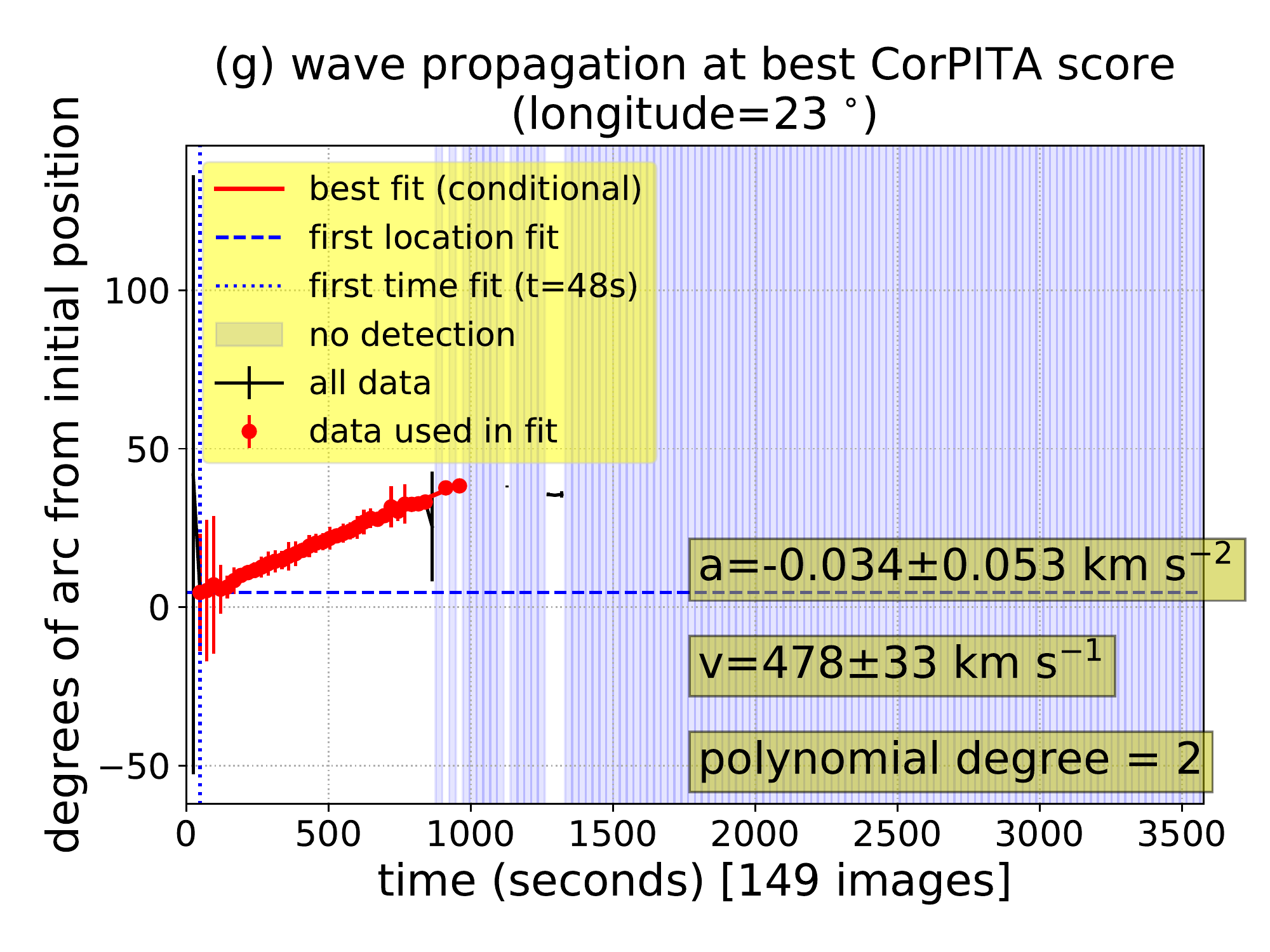}
\caption{AWARE performance for the EUV wave of 16 February 2011. See the caption of Figure \ref{fig:syntheticwave:results} for more details.}
\label{fig:longetal8e}
\end{center}
\end{figure*}

\begin{figure*}
\begin{center}
\includegraphics[width=12.0cm]{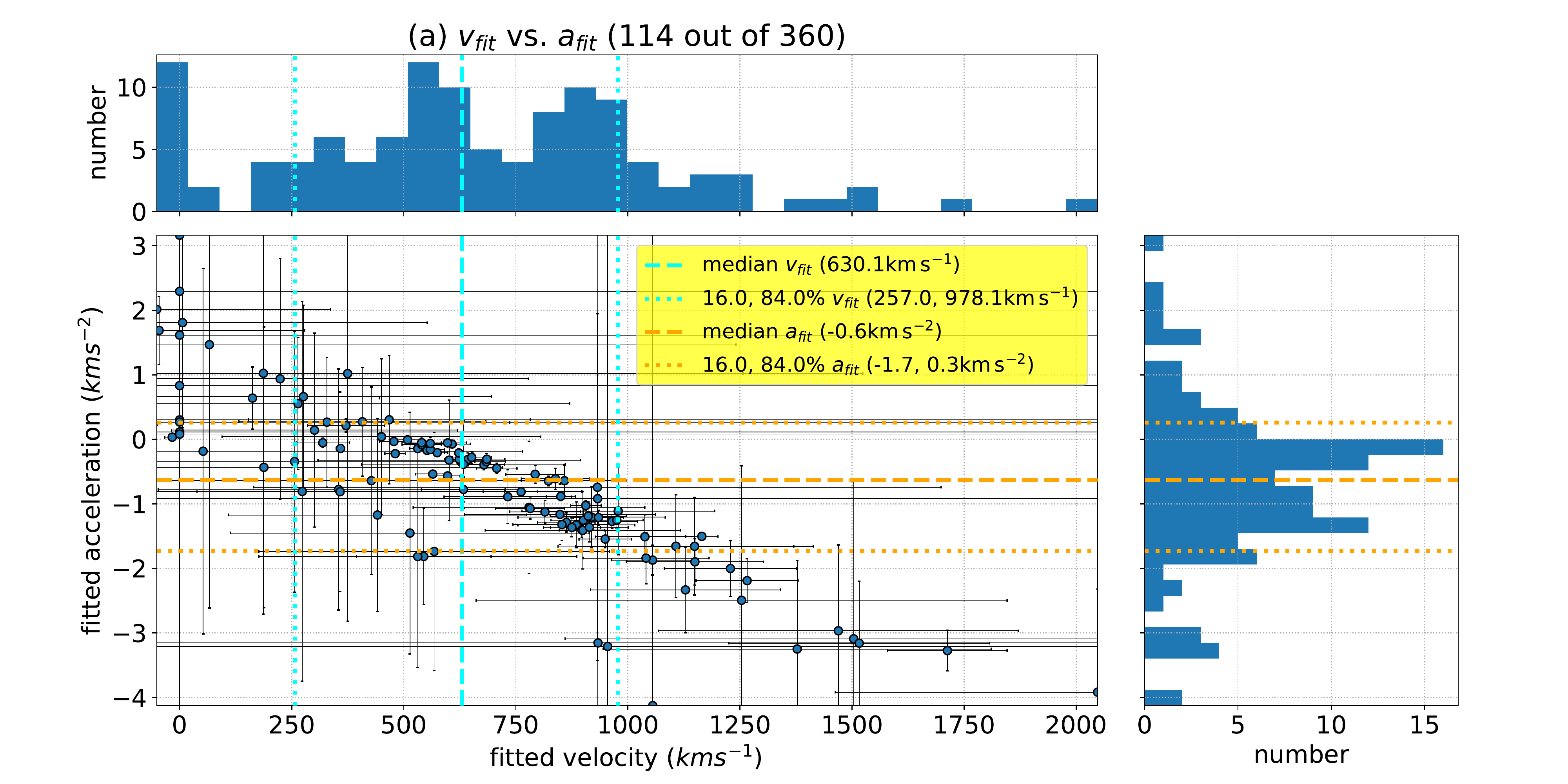}
\includegraphics[width=12.0cm]{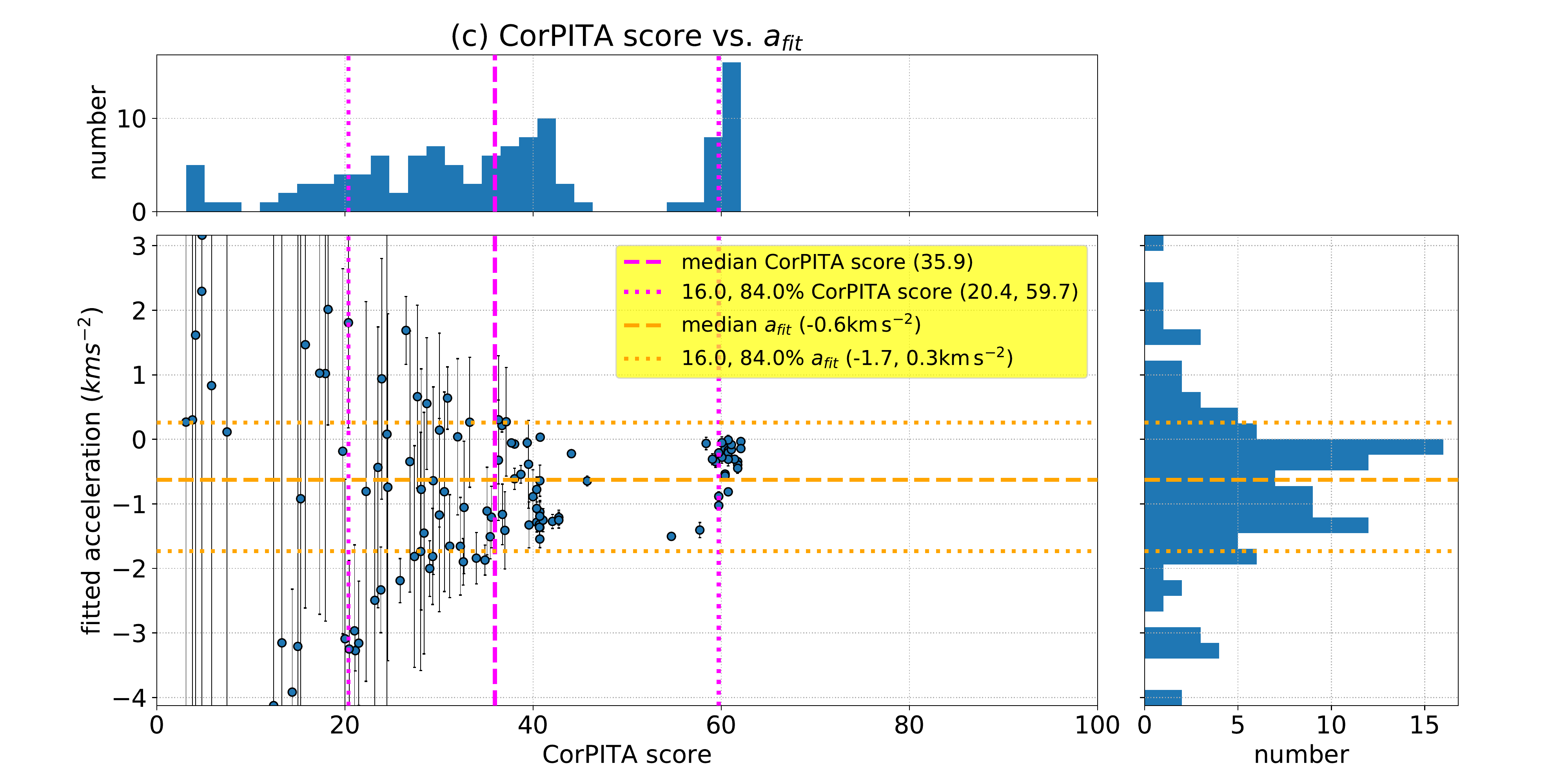}
\includegraphics[width=12.0cm]{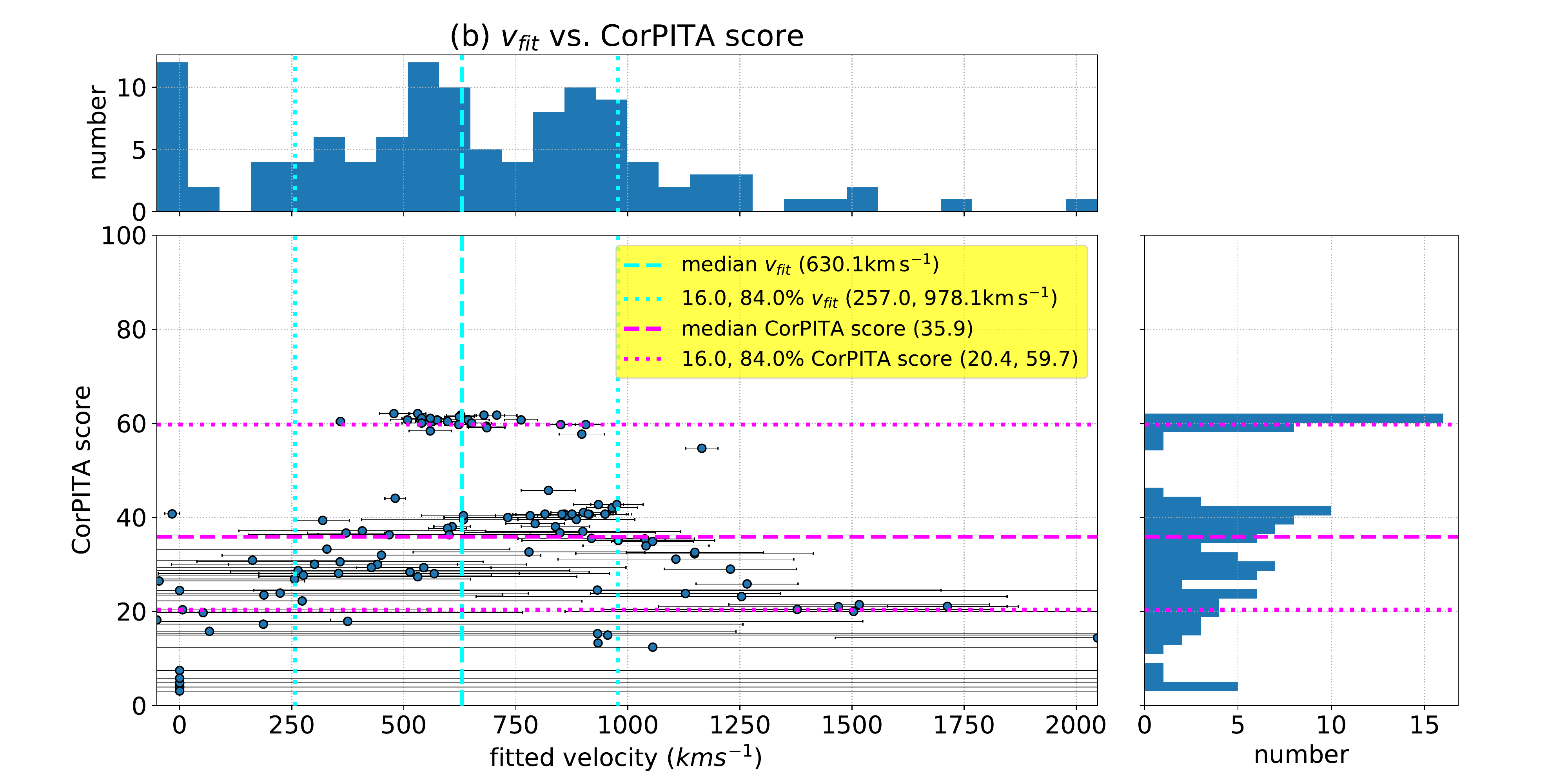}
\caption{Distributions of $\vfit$, $\afit$ and the \longscore\ for the fitted arcs of the 15 February 2011 EUV wave (see Figure \ref{fig:longetal8a}). See the caption of Figure \ref{fig:syntheticwave:dynamics} for more details.}
\label{fig:longetal8edynamics}
\end{center}
\end{figure*}

\rtr{
It is difficult to exactly compare the AWARE results to the results of \citet{2014SoPh..289.3279L}.  The wave progress maps (Figures \ref{fig:longetal4}a, \ref{fig:longetal6}a, \ref{fig:longetal7}a, \ref{fig:longetal8a}a, and \ref{fig:longetal8e}a) are broadly similar to the wave pulse propagation maps of \citet{2014SoPh..289.3279L} (although it may be more appropriate to compare the fitted arcs maps of Figures \ref{fig:longetal4}d, \ref{fig:longetal6}d, \ref{fig:longetal7}d, \ref{fig:longetal8a}d, and \ref{fig:longetal8e}d).  The locations of fitted wave propagation found are broadly similar, but without doing an actual numerical comparison, it is difficult to accurately assess if the AWARE maps show more successfully fitted arcs than any other algorithm.  Reassuringly, running AWARE on the null result data suggested by \citet{2014SoPh..289.3279L} and shown here in Figure \ref{fig:longetal6} produces a similar null result.
}
\rtr{
The values of $\afit$ and $\vfit$ found (Figures \ref{fig:longetal4}e
and f, \ref{fig:longetal4dynamics}, \ref{fig:longetal7}e and f,
\ref{fig:longetal7dynamics}, \ref{fig:longetal8a}e and f,
\ref{fig:longetal8adynamics}, \ref{fig:longetal8e}e and f,
\ref{fig:longetal8edynamics}) for all EUV waves appear reasonable
compared to the values obtained in the literature. Firstly,
\citet{2014SoPh..289.3279L} in analyzing the 15 February 2011 event
(their Figure 8a – d, Figures \ref{fig:longetal8a} and
\ref{fig:longetal8adynamics} here), CorPITA returned an average
$\vfit$ of $406\pm1\speedunit$ in the range
$0\rightarrow1000\speedunit$.  Calculating over all the measured arcs,
AWARE returns a mean $\vfit$  value of $584\pm258 \speedunit$, a
median value of $551 \speedunit$, a central 68\% width
$199\rightarrow877 \speedunit$, with all values falling in the range
of $0\rightarrow1200 \speedunit$ approximately.  For the 16 February 2011
event, \citet{2014SoPh..289.3279L} quote an average $\vfit$ of
$331\pm6 \speedunit$ in a range varying $\approx 100 \rightarrow 975
\speedunit$(their Figure 8e – h, Figures \ref{fig:longetal8e} and
\ref{fig:longetal8edynamics} here).  AWARE returns a mean $\vfit$
value of $592\pm453 \speedunit$ (over all arcs), a median $\vfit$
value of $442 \speedunit$, a central 68\% width of $13\rightarrow820 \speedunit$,
with all values falling in the range of $0\rightarrow2000 \speedunit$ approximately.  Although these values are not the same, they are comparable in magnitude.
}
\rtr{
Secondly, \citet{2017SoPh..292..185L} applied the CorPITA algorithm to
362 EUV events originally identified by \citet{2013ApJ...776...58N}.
Median values to the spread of velocities and accelerations found in
each event are quoted; CorPITA finds median velocities in the range of
$0\rightarrow950 \speedunit$ and median accelerations in the range
of $-0.75\rightarrow0.1 \accunit$ \citet{2013ApJ...776...58N} quote a
velocity range of $200\rightarrow1400 \speedunit$, and an acceleration
range of $-0.7\rightarrow0.3 \accunit$.  For the events analyzed here,
AWARE finds a range of velocities $0\rightarrow2000 \speedunit$, and
accelerations in the approximate range of $-1.7\rightarrow1.9
\accunit$, overlapping with the results of \citet{2017SoPh..292..185L}
and \citet{2013ApJ...776...58N}.  All detections show a correlation
between $\afit$ and $\vfit$. Figure \ref{fig:longetal8adynamics}a
differs from similar plots in that there appears to be two separate
branches of $\afit$ and $\vfit$ dependence above the median value of
$\vfit$.  Interpreting this using the discussion of Section
\ref{sss:fitbias} suggests weak evidence for either two different
distinctive durations of the wavefront or that the ratios of
$\vtrue/\atrue$ differ significantly in at least two locations.  Given
that the \longscore\ values (Figure \ref{fig:longetal8adynamics}b, c)
do not show evidence of two populations (which could be caused by
having two populations that have a significant difference in their
duration, as measured by the existence component $E_{score}$ in
Equation \ref{e:longscore}) then this suggests an interpretation in terms of differing $\vtrue/\atrue$ ratios.}

\section{Summary and Future Work}
\label{s:conclusions}
AWARE uses the novel running difference persistence images to isolate
faint, propagating bright features in coronal channel AIA data.  The
algorithm then applies noise removal and morphological image
operations to better isolate the wave structure.  Great arcs are
traced from the suspected wave launch location and the RANSAC method
is used to find when and where the propagating wavefront intensity
approximately fits an accelerating profile.  These points are fit
using a least-squares fitting algorithm to find value and error
estimates of the initial velocity and acceleration of the
wave. \rtr{As noted above, one hour's worth of AIA 211 \AA\ data are
  used to determine the wave location and dynamics.  The only
  criterion for determining the time-range of data to be downloaded is
  that it be long enough to include the possible lifetime of the wave.
  Stages 1 and 2 of AWARE are sufficient to estimate the wave lifetime
  within the time-range of data downloaded.  Stage 1 of AWARE isolates
  propagating fronts that brighten emission as they propagate;
  therefore, if they are not present they will not be available for
  characterization in stage 2.  In stage 2, the wave location as a
  function of time along each arc is determined. The RANSAC algorithm
  is used to find the inlier wave locations that satisfy the proximity criteria (see Section \ref{sec:aware:dynamics}), eliminating outliers that can cause bad fits.}

AWARE has been tested on both simulated and real datasets, and was found to perform well and produce results comparable with other established methods. The AWARE code repository comes with codes to generate simulated waves.  The mean behavior of AWARE shows that the initial velocity and acceleration of simulated EUV waves can be recovered, subject to error.  Fitting the simulated waves also demonstrates that the fitting algorithm introduces a systematic negative correlation between the fit value of the initial velocity and the acceleration.  A negative correlation persists even when a diverse population of simulated wavefront propagations are considered (Section \ref{sss:fitbias}).  Therefore, the inherent correlation between the initial fit velocity and the acceleration must be taken into account when attempting to determine which physical mechanism is operating that creates EUV waves.

Several future refinements are possible that will improve AWARE and
maximize its effectiveness and use to the solar community. On
comparing the wave progress plots (Section
\ref{sec:aware:segment:persistence}) to the fitted wave progress plots
in Figures \ref{fig:longetal4}, \ref{fig:longetal7},
\ref{fig:longetal8a}, and \ref{fig:longetal8e}, it seems clear that
AWARE loses information about the wavefront. AWARE assumes purely
radial propagation; future versions of AWARE could use the wave
propagation maps to derive the propagation instead of assuming a
radial propagation.  For example, the Huygens principle method
employed by \citet{1999SoPh..190..467W} and
\citet{2006ApJ...645..757W} captured non-radial propagation allowing
the determination of wavefront curvature, variable speed, and
acceleration.  The analysis of Section \ref{sss:fitbias} shows that
reducing the error in locating the wave is crucial to improving
knowledge of wave kinematics.  The noise removal algorithm used in
this article sums the results of the median filter applied at multiple
length-scales. However, there are a number of other noise removal
algorithms. For example, \citet{2002dip..book.....G} describe an
adaptive median filter algorithm that varies the size scale of the area over which the median is calculated based on local greyscale values in the image. Two-dimensional wavelets, along with thresholding algorithms could be used to determine global (or local) noise levels, segmenting the wavefront from the data.  Improved wave segmentation leading to smaller errors on the position of the wavefront would greatly improve our knowledge of the kinematics of EUV waves.  

\rtr{An operational version of AWARE will monitor the Heliophysics Event Knowledgebase (HEK) event database for new events in event types such as as solar flares and eruptions.  The appearance of a new event of one of these types will trigger subsequent data download and analysis}. The HEK event information will be used to determine parameters of the analysis, such as start time of the interval to be analyzed, and potential locations on the Sun of a wave origin. AWARE will perform the image processing and wave assessment steps described above, finally writing a record into a database that is easily accessed through existing solar feature and event databases. The availability of such a catalog will allow large-scale statistical studies of EUV waves and their properties. There are several additional properties that could be stored about each EUV wave in such a catalog.  The total number of degrees around the starting point at which a propagating wavefront was fit would give an assessment of its total angular extent.  For example, if $\theta_{wave}=360$, then there was a successful detection of the wave completely encircling the starting point.  Also, the number of distinct angular regions of wave propagation would yield information as to how discontinuous the wavefront is compared to a single wave.  Median values of populations are less sensitive to outliers and so the median \longscore\ of all the arcs extending from the starting point would give a useful assessment of the entire wave.

\rtr{AWARE is the fourth published EUV wave detection algorithm at the time of writing.  Along with NEMO, CorPITA and Solar Demon, EUV wave detection is now at the stage that it should now be possible to compare the effectiveness of these algorithms on the same simulated and observational data with the goal of obtaining useful insight on the performance of each of them (at the very least making the comparisons in Section \ref{sec:obs} easier).  Such comparison work has been useful in understanding the performance of automated loop tracing algorithms \citep{2008SoPh..248..359A} and magnetic field extrapolation \citep{2006SoPh..235..161S, 2008SoPh..247..269M}.  Approaches similar in style to these works would create a useful benchmark, and guide the development of improved algorithms to better determine the dynamics, and ultimately the physics, of EUV waves in the solar atmosphere.}

\begin{acks}
We are grateful to the developers of SSWIDL \citep{ssw}, IPython \citep{ipython}, SunPy \citep{2015CSD....8a4009S}, matplotlib \citep{Hunter:2007:matplotlib}, Scikit-Learn \citep{scikit-learn}, SymPy \citep{sympy} and the Scientific Python stack for providing data preparation, manipulation, analysis, and display packages.  LH and JI acknowledge the support of the Solar Data Analysis Center (SDAC) through the SESDA grant 80GSFC17C0003.
\end{acks}


\bibliographystyle{spr-mp-sola}

\bibliography{eitwave-paper.bib}

\end{article} 

\end{document}